\documentclass[]{aastex631}
\usepackage{bm}
\usepackage{enumerate}
\usepackage{soul}
\usepackage{multirow}
\usepackage{amsmath}
\usepackage{natbib}

\newcommand{\qa}{$q_{\bm{\phi}}(\Delta a\,|\,\bm{\theta})$}
\newcommand{\qr}{$q_{\bm{\phi}}(\Delta r\,|\,\bm{\theta})$}
\newcommand{\qv}{$q_{\bm{\phi}}(\Delta v\,|\,\bm{\theta})$}

\newcommand{\qvs}{$q_{\bm{\phi}}(\Delta v_{\rm sat}\,|\,\bm{\theta})$}
\newcommand{\qrrc}{$q_{\bm{\phi}}(\Delta r_{\rm cen}/R_{\rm vir}\,|\,\bm{\theta})$}
\newcommand{\qrrs}{$q_{\bm{\phi}}(\Delta r_{\rm sat}/R_{\rm vir}\,|\,\bm{\theta})$}
\newcommand{\qvvc}{$q_{\bm{\phi}}(\Delta v_{\rm cen}/V_{\rm vir}\,|\,\bm{\theta})$}
\newcommand{\qvvs}{$q_{\bm{\phi}}(\Delta v_{\rm sat}/V_{\rm vir}\,|\,\bm{\theta})$}

\newcommand{\qrr}{$q_{\bm{\phi}}(\Delta r/R_{\rm vir}\,|\,\bm{\theta})$}
\newcommand{\qvv}{$q_{\bm{\phi}}(\Delta v/V_{\rm vir}\,|\,\bm{\theta})$}

\newcommand{\qac}{$q_{\bm{\phi}}(\Delta a_{\rm cen}\,|\,\bm{\theta})$}
\newcommand{\qas}{$q_{\bm{\phi}}(\Delta a_{\rm sat}\,|\,\bm{\theta})$}

\begin{document}

\title{Modeling the Kinematics of Central and Satellite Galaxies Using Normalizing Flows}

\correspondingauthor{James Kwon}
\email{jameskwon@ucsb.edu}

\author[0000-0001-9802-362X]{K. J. Kwon}
\affiliation{Department of Physics, University of California, Santa Barbara, Broida Hall, Santa Barbara, CA 93106, USA}

\author[0000-0003-1197-0902]{ChangHoon Hahn}
\affiliation{Department of Astrophysical Sciences, Princeton University, Peyton Hall, Princeton, NJ 08544, USA}

\begin{abstract}
Galaxy clustering contains information on cosmology, galaxy evolution, and the relationship between galaxies and their dark matter hosts. On small scales, the detailed kinematics of galaxies within their host halos determines the galaxy clustering. In this paper, we investigate the dependence of the central and satellite galaxy kinematics on $\bm{\theta}$, the intrinsic host halo properties (mass, spin, concentration), cosmology ($\Omega_{\textrm{m}}$,~$\sigma_8$), and baryonic feedback from active galactic nuclei and supernovae~($A_{\textrm{AGN1}}$,~$A_{\textrm{AGN2}}$,~$A_{\textrm{SN1}}$,~$A_{\textrm{SN2}}$). We utilize 2,000 hydrodynamic simulations in CAMELS run using IllustrisTNG and SIMBA galaxy formation models. Focusing on central and satellite galaxies with $M>10^9M_\ast$, we apply neural density estimation~(NDE) with normalizing flows to estimate their $p(\Delta r\,|\,\bm{\theta})$ and $p(\Delta v\,|\,\bm{\theta})$, where $\Delta r$ and $\Delta v$ are the magnitudes of the halo-centric spatial and velocity offsets. With NDE, we accurately capture the dependence of galaxy kinematics on each component of $\bm{\theta}$. For central galaxies, we identify significant spatial and velocity biases dependent on halo mass, concentration, and spin. For satellite distributions, we find significant deviations from an NFW profile and evidence that they consist of distinct orbiting and infalling populations. However, we find no significant dependence on $\bm{\theta}$ besides a weak dependence on host halo spin. For both central and satellite galaxies, there is no significant dependence on cosmological parameters and baryonic feedback. These results provide key insights for improving the current halo occupation distribution~(HOD) models. This work is the first in a series that will re-examine and develop HOD frameworks for improved modeling of galaxy clustering at smaller scales.
\end{abstract}

\section{Introduction}
Galaxy clustering is one of the most important probes for large-scale structure \citep[e.g.,][]{percival_01, peacock_01, tegmark_04, cole_05, zehavi_11, white_11_hod_clustering, tinker_12, reid_14, alam_17, ivanov_2020, kobayashi_23_sdss3}. It encodes key cosmological information about e.g., the growth and expansion history of the Universe~(\citealt{ivanov2023_H0chapter}; see \citealt{huterer2023_growth} for a recent review). It also offers a unique opportunity to study the connection between galaxies and their host halos as well as the complex physics of galaxy formation and evolution (e.g., \citealt{abazajian_05, masjedi_06, zheng+_2007, zheng_07_degeneracy, white_07, vdB_13_small_scale, guo_15a, Molino, yuan_22_fs8}). Numerous galaxy redshift surveys have thus been conducted to analyze galaxy clustering: e.g., Sloan Digital Sky Survey~\citep[SDSS;][]{sdss}, the Baryon Oscillation Spectroscopic Survey~\citep[BOSS;][]{boss}, and the Extended Baryon Oscillation Spectroscopic Survey~\citep[eBOSS;][]{eboss}. Many more upcoming surveys are planned for the near future, e.g., the Dark Energy Spectroscopic  Instrument~\citep[DESI;][]{desi}, the Subaru Prime Focus Spectrograph~\citep[PFS;][]{takada2014}, the ESA {\em Euclid} satellite mission~\citep{laureijs2011}, and the Nancy Grace Roman Space Telescope~\citep[Roman;][]{spergel2015, wang2022a}, to probe galaxies over even larger cosmic volumes.

Simulations, used to analyze galaxy clustering, are most commonly modeled using the halo occupation framework, where galaxies occupy dark matter halos predicted by cosmological $N$-body simulations~(\citealt{benson_00, peacock_00, scoccimarro_2001, yang_2003_clf, reid_12_boss_hod, zhai2019_aemulus}; see \citealt{wechsler_tinker_18} for a recent review). The Halo Occupation Distribution (HOD) model, one of the main halo occupation models, provides an empirical prescription of populating halos with galaxies, without computationally modeling the physics of galaxy formation~\citep[e.g.,][]{seljak_2000, peacock_2002, berlind_2002, zheng_2005, zheng+_2007,watson_hod_mass_loss,hearin_13_age}. It consists of two procedures: determining the number of galaxies within a halo (occupation number) and assigning them positions and velocities relative to the halo. In earlier HOD models, the occupation number of central and satellite galaxies was determined solely by the host halo mass. More recent models, based on observational evidence, now include additional dependence on other halo properties or their environment~\citep[e.g.,][]{croton_07, zentner_14, hearin_16, zehavi_18}.

For the positions and velocities of galaxies, the simplest model places central galaxies at the center of the host halos with zero relative velocity. The positions of satellite galaxies are sampled to trace the Navarro-Frenk-White~\citep[NFW;][]{nfw} profiles of host halos and their velocities are sampled using the velocity dispersion profiles of host halos, derived from solving the Jean's equation using the assumed halo density profile. State-of-the-art HOD models include additional flexibility through galaxy concentration bias and velocity bias~\citep[e.g.,][]{guo_15a, guo_15b, hearin_16}. Concentration bias allows the concentration of satellite galaxies to deviate from the dark matter halo~\citep[e.g.,][]{nagai_05, croton_07} and velocity bias allows for the systematic discrepancy between the velocities of galaxies and their host halos~\citep[e.g.,][]{berlind_03_hydro_velbias, reid_14, guo_15a, guo_15b}. 

Despite the continual developments, there are still major open questions in HOD modeling. For instance, there is no consensus as to how closely satellites trace the host density profiles (e.g., \citealt{nagai_05, tal_2012, watson_12, budzynski_12, guo_15a,  McDonough_2022}). In addition, given the evidence of galaxy assembly bias \citep[e.g.,][]{zentner_14, hearin_16, zehavi_18, xu_2021, hadzhiyska_21_gal_assembly} in the occupation number, it remains to be investigated whether halo properties and environment also impact the kinematics. Furthermore, studies have consistently demonstrated that the presence of baryons can leave an imprint on matter or halo clustering in small scales (e.g., \citealt{masjedi_06,white_07,van_daalen_2011_agn_small, chan_2015, hellwing_2016, chisari2019_baryons, schneider2019_baryons,springel_18,foreman_2020, van_daalen_2020_gal_form_power_spectrum}). In particular, \cite{van_daalen_2011_agn_small,van_daalen_2020_gal_form_power_spectrum} and \cite{foreman_2020} used hydrodynamic simulations to show that baryonic processes such as feedback from active galactic nuclei (AGN) or supernovae (SN) can alter the matter power spectrum and bispectrum from what is predicted by dark-matter-only (DMO) simulations by $> 1$\% at $k\gtrsim 1 \ h\,{\rm Mpc}^{-1}$. Beyond the matter distribution, it remains to be seen whether these baryonic processes also impact the kinematics of galaxies within the halo.

Whether and to what extent the kinematics of galaxies depend on secondary halo properties, cosmology, or baryonic processes remains to be investigated in detail. Quantifying these dependencies will enable us to develop HOD models that more accurately model galaxy clustering on small scales. This is particularly timely as recent works have now firmly established the advantages of analyzing galaxy clustering down to small scales~\citep[e.g.,][]{verde2002, sefusatti2005, hahn2021, massara_2021, gualdi_21, eickenberg_22_wavelet,  wang2022a, yuan2022, massara_2023, hou_2023,
porth_23}. Simulation-based analyses with emulators or simulation-based inference using the HOD model have now been successfully applied to observations~\citep{zhai2019_aemulus, simbig2022_plk, valogiannis_22_wst, storey-fisher_23_aemulus_VI, zhai_23_aemulus, paillas2023, lange2023, thiele2023, simbig_wave2, simbig_bk, simbig_cnn, valogiannis2023_wst}. Improved HOD models and a better understanding of the galaxy-halo connection will enable us to more robustly exploit this additional information on small scales.

In this paper, we investigate the dependence of central and satellite galaxy kinematics on internal halo properties (mass, concentration, spin), cosmology ($\Omega_{\rm m}$, $\sigma_8$), and baryonic feedback processes by AGN and SNe. We use galaxies from 2,000 hydrodynamic simulations of the Cosmology and Astrophysics with Machine Learning Simulations~(CAMELS; \citealt{camels_2021, camels_2022}) suite: 1,000 simulations using the IllustrisTNG~\citep{tng_weinberger_17, tng_pillepich_18, nelson_2019} and 1,000 from the SIMBA~\citep{dave_2019} galaxy formation models. CAMELS simulations are constructed using a wide range of cosmology and baryonic processes. This provides an ideal dataset for examining the kinematic properties of galaxies and their dependence on cosmological and baryonic feedback parameters. To quantify the dependencies between the galaxy kinematics and the parameters, we use neural density estimation (NDE) with normalizing flows, which can accurately estimate complex high-dimensional conditional probability distributions. With NDE, we quantify the impact of internal halo properties, cosmological parameters, and baryonic feedback on galaxy kinematics. Furthermore, we examine deviations in galaxy kinematics from the standard HOD models.

The structure of this paper is as follows. In Section~\ref{sec:data}, we describe CAMELS and halo catalogs that we use to extract information on galaxy kinematics and halo properties. Section~\ref{sec:nde} summarizes how we use NDEs to estimate central and satellite galaxy kinematics from CAMELS. We present our results in Section~\ref{sec:result}. Based on our findings, we discuss the implications for informing and improving HOD models in Section~\ref{sec:disc}. We summarize our analysis in Section~\ref{sec:summary}.

\section{Data\label{sec:data}}
In this section, we briefly describe CAMELS and the halo catalogs constructed from CAMELS that we use to extract the galaxy kinematics and the intrinsic properties of the host halos. 

\subsection{CAMELS\label{subsec:camels}}
CAMELS is a suite of cosmological simulations designed for machine-learning-based cosmological analyses. It consists of 4,033 simulations, each with a comoving volume of (25 $h^{-1}{\rm Mpc}$)$^3$. 2,184 of these are the (magneto-)hydrodynamic simulations that implement the subgrid physics models of IllustrisTNG or SIMBA. Each hydrodynamic simulation evolves $256^3$ dark matter particles of mass $6.49\times 10^7 (\Omega_{\rm m}-\Omega_{\rm b})/0.251\ h^{-1} M_\odot$ with $256^3$ fluid elements of mass $1.27\times 10^{7}\ h^{-1} M_\odot$ from $z=127$ to $z=0$ for given values of the cosmological and hydrodynamic parameters.  

For our study, we use the 2,000 hydrodynamic simulations at $z=0$ constructed using the IllustrisTNG and SIMBA models (1,000 each). These simulations vary two cosmological parameters, $\Omega_{\rm m}$ and $\sigma_8$, uniformly across the range $[0.1,0.5]$ and $[0.6,1.0]$, respectively. The other cosmological parameters are fixed to $\Omega_{\rm b}=0.049$, $h=0.6711$, $n_s=0.9624$, $M_\nu=0.0$ eV, $w=-1$, and $\Omega_K=0$. The simulations also vary four hydrodynamic parameters, $A_{\rm hydro}=\{A_{\rm SN1}, A_{\rm SN2}, A_{\rm AGN1}, A_{\rm AGN2}\}$, that control the energy scale and the injection rate of SN and AGN feedback. $A_{\rm SN1}$ and $A_{\rm SN2}$ provide normalization factors for the galactic wind flux and speed, respectively. $A_{\rm AGN1}$ and $A_{\rm AGN2}$ control the energy output normalization and specific energy of AGN feedback. They are all log-uniformly sampled across the range $A_{\rm SN1}$, $A_{\rm AGN1}\in[0.25,4.0]$ and $A_{\rm SN2}$, $A_{\rm AGN2}\in[0.5,2.0]$. We note that although they share the $A_{\rm hydro}$ parameterization, each subgrid physics model implements $A_{\rm hydro}$ differently~\citep[see][for more details]{camels_2021}. In total, six parameters---$\Omega_{\rm m}$, $\sigma_8$, $A_{\rm SN1}, A_{\rm SN2}, A_{\rm AGN1}, A_{\rm AGN2}$---of our CAMELS simulations are varied and arranged in a Latin Hypercube~(LH).

\subsection{\textsc{Rockstar} Halos\label{subsec:rockstar}}
For each snapshot of the simulations, CAMELS provides multiple halo catalogs constructed using different halo-finding algorithms. In this work, we use halo catalogs generated by the Robust Overdensity Calculation using K-Space Topologically Adaptive Refinement~\citep[\textsc{Rockstar};][]{behroozi_2013}, a 6D phase space temporal halo-finding method. {\sc Rockstar} implements a variant of the 3D friends-of-friends (FoF) algorithm to find overdense regions. Within each FoF group, it identifies the hierarchy of FoF subgroups by adaptively establishing a 6D phase-space metric at each depth. This process terminates when the number of particles in a subgroup falls below a minimum threshold. Once all 6D FoF subgroups have been identified, a seed halo is assigned to each subgroup at the deepest level of the hierarchy. Subsequently, particle membership is determined based on the phase-space distance to seed halos to determine the halo substructures. \textsc{Rockstar} then calculates subhalo and halo properties using the particle membership. In this work, we only include host halos and subhalos from IllustrisTNG and SIMBA suites that host galaxies with stellar mass, $M_*$, greater than $10^9M_\odot$. This is to focus on intermediate- to high-mass galaxies that are typically probed by cosmological galaxy surveys~\citep[e.g.,][]{sdss_lrg_2001, dawson_13, desi_bgs_target, desi_lrg_23}. We also remove a small fraction ($\ll 1\%$) of central galaxies that are identified outside of their host halos. These are mostly galaxies that have recently experienced major mergers, for which determining the host halos is ambiguous.

From the galaxy/subhalo properties, we derive the kinematics of central and satellite galaxies using their halo-centric radial position and velocity. For central galaxies, we use $X_{\rm off}$ and $V_{\rm off}$ of host halos provided by {\sc Rockstar}. $X_{\rm off}$ is the comoving distance between the density peak, the position of the particle with the minimum gravitational potential, and the center of mass of a halo, calculated using all particles within the virial radius, $R_{\rm vir}$\footnote{\textsc{Rockstar} determines the halo center using $n$ innermost particles such that it minimize the Poisson error of the center.}. $V_{\rm off}$ is the magnitude of the velocity offset between the bulk (all particles within~$R_{\rm vir}$) and the core (particles within~0.1$R_{\rm vir}$) of the halo. For satellite galaxies, we derive $\Delta r$ by taking the magnitude of the difference between the comoving position vectors of its subhalo and its host halo, $|\bm{r}_{\rm sat}- \bm{r}_{\rm host}|$. Similarly, $\Delta v$ is taken to be the magnitude of the difference between the bulk velocities of the satellite and host halos, $|\bm{v}_{\rm sat}-\bm{v}_{\rm host}|$.

\begin{figure}
    \centering
    \includegraphics[width=0.5\textwidth]{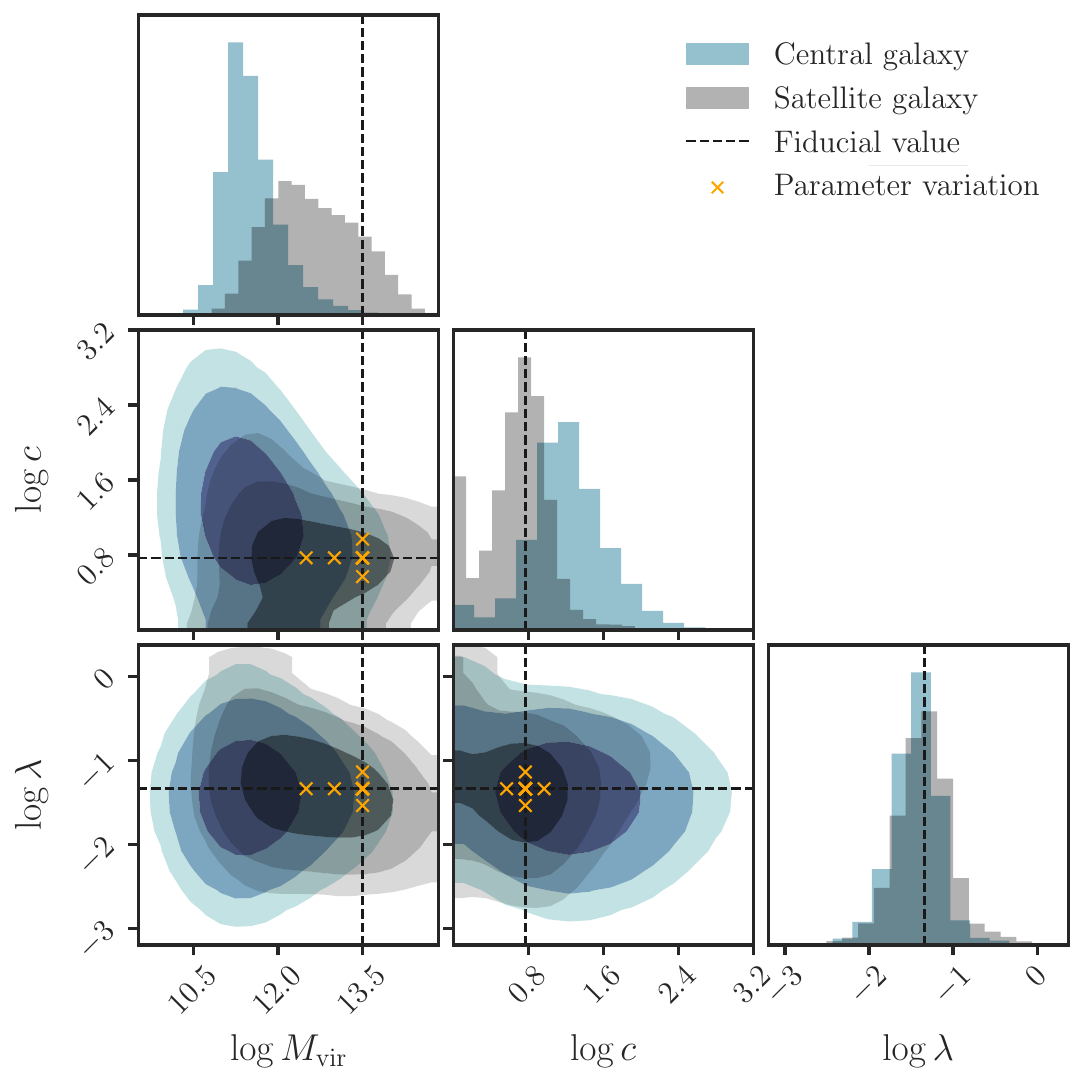}
    \caption{The distribution of host halo properties, $M_{\rm vir}, \lambda, c$, for central galaxies (blue) and satellite galaxies (gray) in the SIMBA hydrodynamic suite. The contours are marked for 68$^{\rm th}$, 95$^{\rm th}$, and 99$^{\rm th}$ percentiles. In total, we have 359,494 central galaxies and 22,244 satellite galaxies with $M_* > 10^9 M_\odot$ in SIMBA. IllustrisTNG has a similar distribution, with 276,527 central and 34,695 satellite galaxies. The black dashed lines mark the fiducial values of halo properties. We examine the distribution of galaxy kinematics at the points marked in orange (`x') to identify the dependence on each halo property.}
    \label{fig:ill_train}
\end{figure}

In addition to the kinematic properties, we also use three internal properties of host halos, which we collectively refer to as $\bm{\theta}_{\rm h}$. We include halo virial mass ($M_{\rm vir}$), spin ($\lambda$), and concentration ($c$). $\lambda$ is given by $J\sqrt{|E|}/GM_{\rm vir}^{2.5}$, where $J$ is the magnitude of the halo angular momentum and $|E|$ is the magnitude of the total energy of the halo~\citep{peebles_1969}. We derive $c$, which has been shown to be a proxy of halo assembly history~\citep{castorina2013, mao2015_concentration}, as the ratio of the fitted NFW scale radius and $R_{\rm vir}$. In Figure~\ref{fig:ill_train}, we present the distribution of $\bm{\theta}_{\rm h}$ for the central (blue) and satellite galaxies (gray) in SIMBA. The contours are smoothed using a Gaussian kernel with 1.5$\sigma$. We only show the $\bm{\theta}_{\rm h}$ distribution of SIMBA galaxies for clarity, as IllustrisTNG galaxies have a similar distribution. 

In total, our dataset includes 359,494 central galaxies and 22,244 satellites from SIMBA, and 276,527 central galaxies and 34,695 satellites from IllustrisTNG. We randomly select 10\% of the galaxies and reserve them for testing; we use the remaining galaxies for training\footnote{Each of 2,000 simulations is not specifically set aside for training or testing; i.e., different galaxies from the same simulation can be used in training, validation, and testing}. For each galaxy $i$, we compile $(\Delta r_i, \Delta v_i, \bm{\theta}_{{\rm h},i}, \Omega_{{\rm m}, i}, \sigma_{8,i}, A_{{\rm hydro},i})$.

\section{Neural Density Estimation with Normalizing Flows\label{sec:nde}}
Our objective is to identify how the host halo properties, cosmological parameters, and baryonic processes affect $\Delta r$ and $\Delta v$ of galaxies within a halo. We also seek to quantify any deviations in the $\Delta r$ and $\Delta v$ distributions from the standard HOD prescription. To do this, we estimate $p(\Delta r\,|\,\bm{\theta})$ and $p(\Delta v\,|\,\bm{\theta})$, where $\bm{\theta} = \{\bm{\theta}_{\rm h}, \Omega_{\rm m}, \sigma_8, A_{\rm hydro}\}$ for CAMELS using neural density estimation (NDE), which has been used extensively in the literature to accurately estimate complex high-dimensional distributions~\citep[e.g.,][]{pydelfi_2019, hiflow, gw_prop, sedflow_2022, dai_seljak_22, lovell_23}.

With NDE, we train neural networks ($q$) with free parameters $\bm{\phi}$ such that $q_{\bm{\phi}}(\Delta a\,|\,\bm{\theta})\approx p(\Delta a\,|\,\bm{\theta})$, where $\Delta a$ refers to either $\Delta r$ and $\Delta v$. Once trained, \qa{} can be evaluated and sampled at different values of $\bm{\theta}$. This means that we can isolate the effect of a single parameter by comparing \qa{} to $q_{\bm{\phi}}(\Delta a\,|\,\bm{\theta}')$, where we only vary a single parameter in $\bm{\theta}$ to obtain $\bm{\theta}'$. In this work, we use NDE based on normalizing flows~\citep{tabak_vanden_10, tabak_13}, which estimates a target distribution by mapping it to a base distribution that is simple and easy to evaluate (e.g., a multivariate Gaussian distribution). The mapping between the target and base distributions is set to a series of invertible and differentiable transformations. This enables us to calculate and sample the target distribution using the base distribution with a change of variables.

Among various normalizing flow architectures, e.g., Masked Autoregressive Flow \citep{papam_17}, Neural Spline Flow~\citep{durkan_19}, we use Masked Autoencoder for Distribution Estimation~\citep[MADE;][]{made_2015, nflows}, based on experimentation. MADE decomposes $p(\bm{x}\,|\,\bm{\theta})$ into a product of nested conditional probabilities, $\prod_{i=1}^{\text{dim}(\bm{x})}p(x_i\,|\,\bm{x}_{<i},\bm{\theta})$, where $\bm{x}_{<i}=(x_1,x_2,\cdots,x_{i-1})$. Given the ordering of input $\bm{x}$, an autoregressive model parameterizes each $p(x_i\,|\,\bm{x}_{<i},\bm{\theta})$ with $\bm{x}_{<i}$ and $\bm{\theta}$ as the only input. MADE preserves the autoregressive property by using binary masks to drop out connections between $p(x_i\,|\,\bm{x}_{<i},\bm{\theta})$ and $\bm{x}_{\geq i}$. We use the MADE implementation from the 
\texttt{sbi}\footnote{\href{https://github.com/sbi-dev/sbi}{https://github.com/sbi-dev/sbi}} \texttt{python} package~\citep{greenberg_19, sbi_2020}.

To train our flow, we first divide our dataset $\{(\Delta a_i, \bm{\theta}_i)\}$ from Section~\ref{sec:data} into a training and validation set with a 90/10 split. Then, we maximize the total log-likelihood $\sum_{i=1}^n\log q_{\bm{\phi}}(\Delta a_i\,|\,\bm{\theta}_i)$, for galaxies in our training set. This is equivalent to minimizing the Kullback-Leibler divergence between $q_{\bm{\phi}}(\Delta a\,|\,\bm{\theta})$ and $p(\Delta a\,|\,\bm{\theta})$. For the maximization, we use the \textsc{Adam} optimizer~\citep{kingma2017adam} with a learning rate of $5\times10^{-4}$. We impose early stopping to prevent overfitting by stopping training when the log-likelihood evaluated on the validation set fails to increase after 20 consecutive epochs. We determine the number of hidden layers and units of $q_{\bm \phi}$ through experimentation to achieve the lowest validation losses. In total, we train 8 separate flows to estimate $p(\Delta r\,|\,\bm{\theta})$ and $p(\Delta v\,|\,\bm{\theta})$\footnote{In practice, we estimate $p(\log \Delta r\,|\,\bm{\theta})$ and $p(\log\Delta v\,|\,\bm{\theta})$ to reduce the dynamic range of the kinematics data. However, we find that the choice between training NDEs on linear or logarithmic scales does not have a significant impact on their accuracy.}, of central and satellite galaxies, and for IllustrisTNG and SIMBA. In Appendix~\ref{app:nde_valid}, we validate the accuracy of our trained NDEs using the 10\% of the CAMELS galaxies reserved for testing.

\section{Results\label{sec:result}}
In this section, we present the dependence of the kinematics of central and satellite galaxies on their host halo properties, cosmology, and baryonic processes using \qr{} and \qv{}. To focus on the effect of individual parameters, we first select a fiducial $\bm{\theta}_{\rm fid}$:
\begin{equation} \label{eq:theta_fid}
\bm{\theta}_{\rm fid} = (\log M_{\rm vir}=13.5, \log c=0.77, \log \lambda=-1.34, \Omega_{\rm m}=0.3, \sigma_8=0.8, A_{\rm hydro}=1),
\end{equation}
where $M_{\rm vir}$ is in units of $h^{-1}M_\odot$. We then compare $q_{\bm{\phi}}(\Delta r\,|\,\bm{\theta}_{\rm fid})$ to $q_{\bm{\phi}}(\Delta r\,|\,\bm{\theta}')$, where $\bm{\theta}'$ has the same parameter values as $\bm{\theta}_{\rm fid}$, except for a single parameter.

We set the fiducial $M_{\rm vir}$ based on the typical host halo mass of luminous red galaxies~\citep[LRG; e.g.,][]{lrg_09}. We choose the fiducial concentration near the center of the training data (Figure~\ref{fig:ill_train}) and the fiducial $\lambda$ based on the typical spin of halos in numerical simulations~\citep{spin_dist}. The fiducial $\Omega_{\rm m}$ and $\sigma_8$ values are at the center of the CAMELS prior range, which is consistent with \cite{planck_2018} constraints. The fiducial $A_{\rm hydro}$ are set to unity, which corresponds to the original feedback settings of IllustrisTNG and SIMBA. We note that although we do not include $R_{\rm vir}$ in $\bm{\theta}_{\rm h}$, it varies based on $M_{\rm vir}$ and $\Omega_{\rm m}$\footnote{$M_{\rm vir}=4\pi\Delta(x)\,\rho_{\rm crit}/3 R_{\rm vir}^3$, where $x=1-\Omega_{\rm m}$ and $\Delta (x) = 18\pi^2 + 82x -39x^2$~\citep{bryan_1998}.}. In this work, we present results at $\bm{\theta}_{\rm fid}$ specified in Equation~\ref{eq:theta_fid}. However, our results do not vary significantly for different $\bm{\theta}_{\rm fid}$. 

\begin{figure}
    \centering   \gridline{\fig{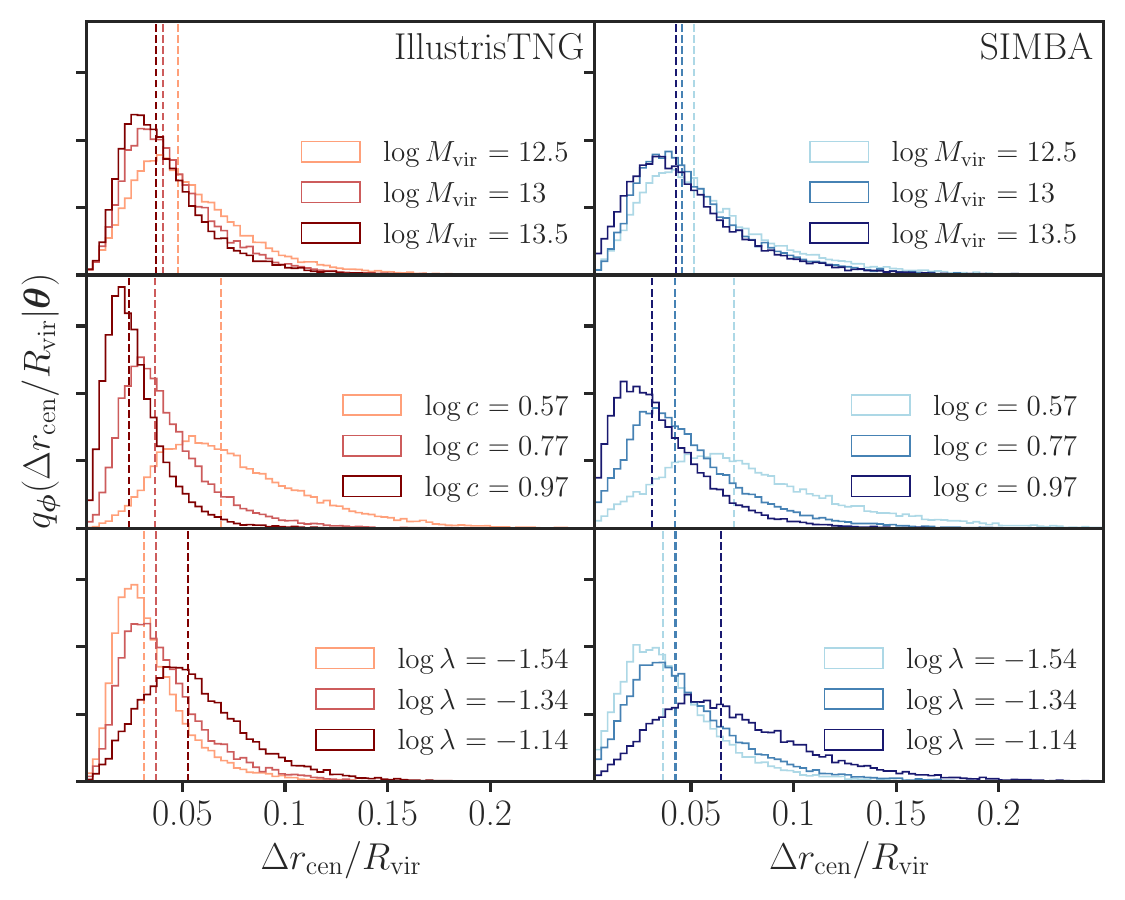}{0.5\textwidth}{(a)}
    \fig{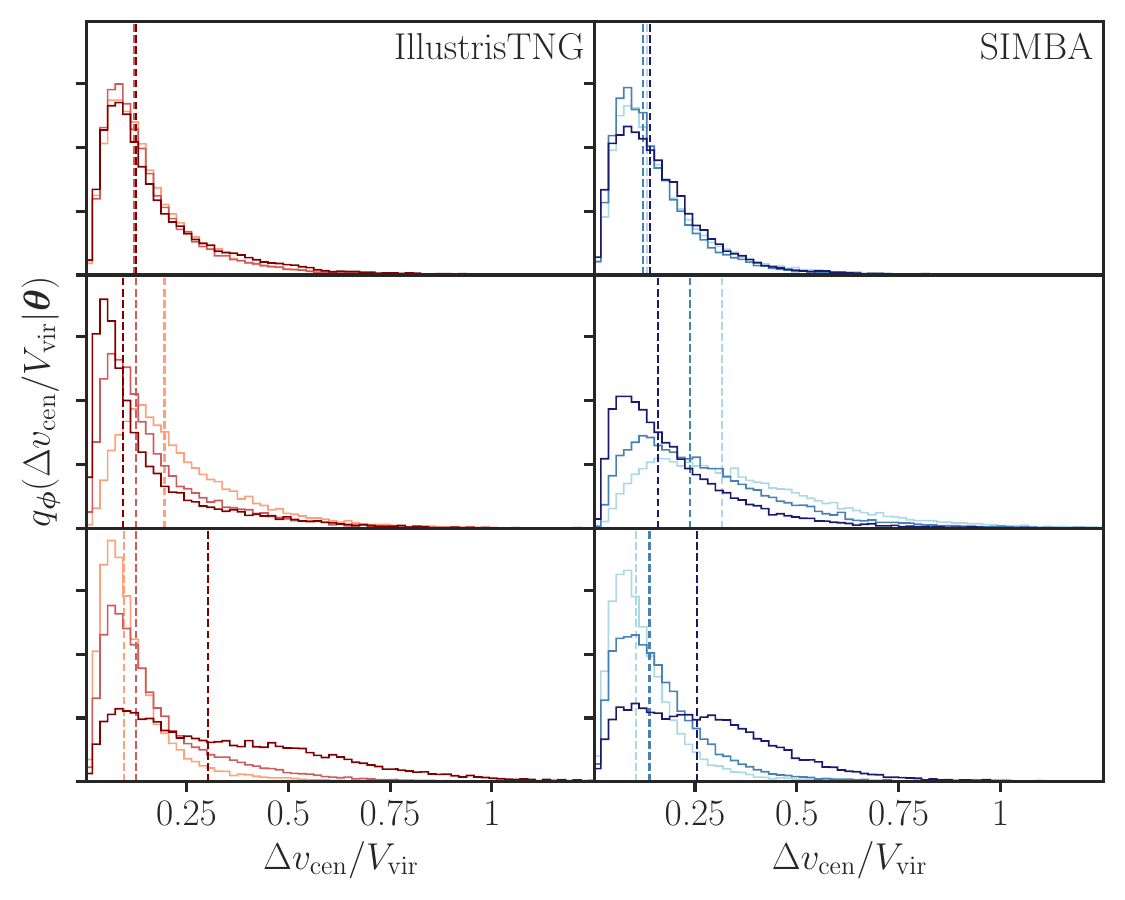}{0.5\textwidth}{(b)}}
    \caption{Dependence of the central galaxy kinematics, \qrrc{} and \qvvc{}, on $\bm{\theta}_{\rm h}$ ($M_{\rm vir}$, $c$, and $\lambda$, from top to bottom) in IllustrisTNG (red) and SIMBA (blue). The $y$-axis denotes the probability density on a linear scale. The parameters that do not appear in the legends are kept at the fiducial values. The dashed lines mark the medians of the distributions. We find a significant dependence of \qrrc{} not only on $M_{\rm vir}$ but also on $c$ and $\lambda$ for both IllustrisTNG and SIMBA: \qrrc{} is skewed towards lower $\Delta r_{\rm cen}/R_{\rm vir}$ for higher $M_{\rm vir}$ and $c$ while the trend is reversed for $\lambda$. The trends in both simulation suites are consistent with each other; however, \qrrc{} is broader in SIMBA than in IllustrisTNG.}
    \label{fig:cen_internal}
\end{figure}

\subsection{Central Galaxies\label{subsec:cen_kin}}
\noindent\ul{\emph{Host halo properties}}: 
In Figure~\ref{fig:cen_internal}, we present the dependence of central galaxy kinematics, \qrrc{}~(a) and \qvvc{}~(b), on $\log M_{\rm vir}$ (top), $\log c$ (center), and $\log \lambda$ (bottom) for IllustrisTNG (red) and SIMBA (blue). For clarity we present $\Delta r$ and $\Delta v$ in units of $R_{\rm vir}$ and $V_{\rm vir}$. The value of the varied parameter increases from light to dark. The dashed vertical lines indicate the medians of the distributions to highlight any trends. Overall, we find that the dependence of the central kinematics on each halo property is qualitatively consistent between IllustrisTNG and SIMBA. Furthermore, we find that central galaxies are radially displaced from the center of the halos and have non-zero relative velocities. This validates the use of velocity bias of central galaxies in current state-of-the-art HOD models.

Figure~\ref{fig:cen_internal} also confirms a number of expected dependencies of $\Delta r_{\rm cen}$ and $\Delta v_{\rm cen}$ with halo properties. For instance, the medians of \qrrc{}~and \qvvc{}~decrease as $M_{\rm vir}$ or $c$ increase. The gravitational potential is deeper for higher $M_{\rm vir}$ and steeper towards the center for higher $c$. In contrast, the medians of the distributions increase with increasing $\lambda$. At fixed fiducial $M_{\rm vir}$ and $R_{\rm vir}$, increasing $\lambda$ is equivalent to increasing $J$, i.e., increased rotational support against the gravitational infall, which would increase $\Delta r_{\rm cen}$ and $\Delta v_{\rm cen}$. Furthermore, \qrrc{} is slightly broader in SIMBA than IllustrisTNG, most noticeably in the $\log c$ and $\log \lambda$ panels.
\vspace{10pt}

\begin{figure}
    \centering
    \gridline{\fig{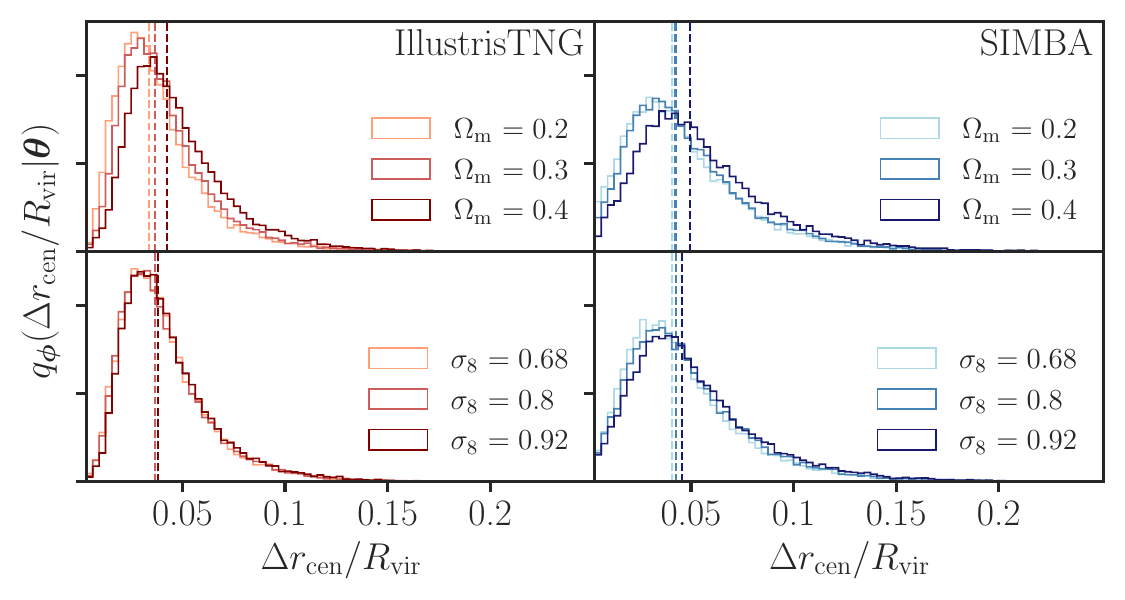}{0.5\textwidth}{(a)}
          \fig{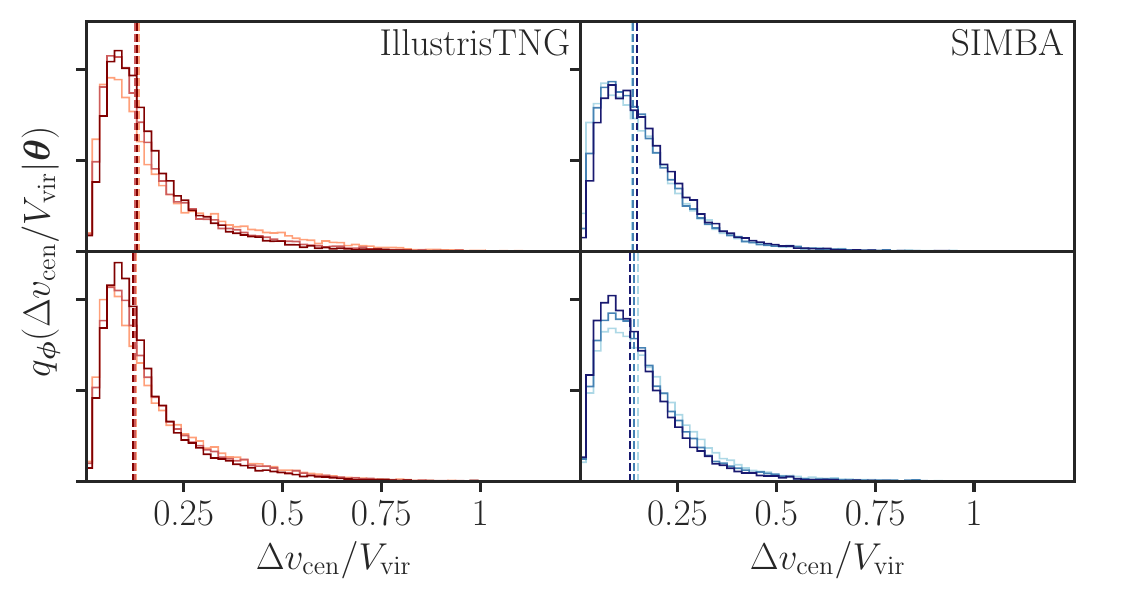}{0.5\textwidth}{(b)}}
    \caption{Dependence of the central galaxy kinematics, \qrrc{} and \qvvc{}, on $\Omega_{\rm m}$ (top) and $\sigma_8$ (bottom) in IllustrisTNG (red) and SIMBA (blue). The parameters that do not appear in the legends are kept at the fiducial values. The dashed lines mark the medians of the distributions. In both IllustrisTNG and SIMBA, \qrrc{} is slightly skewed to higher $\Delta r_{\rm cen}/R_{\rm vir}$ at higher $\Omega_{\rm m}$. Overall, we find no significant dependence of galaxy kinematics on the cosmological parameters.}
    \label{fig:cen_cosmo}
\end{figure}

\noindent\ul{\emph{Cosmological parameters}}: In Figure~\ref{fig:cen_cosmo}, we present the dependence of central galaxy kinematics, \qrrc{}~(a) and \qvvc{}~(b), on $\Omega_{\rm m}$ (top) and $\sigma_8$ (bottom) for IllustrisTNG (red) and SIMBA (blue). Overall, the cosmological parameters have a negligible effect on the central galaxy kinematics. For both IllustrisTNG and SIMBA, there is a marginal dependence of $\Omega_{\rm m}$ on \qrrc{}, where central galaxies are slightly farther away from the halo centers at higher $\Omega_{\rm m}$. This trend remains even when we vary halo mass with $\Omega_{\rm m}$, based on the fixed number density of the halo mass function (Appendix~\ref{app:hmf}). Nevertheless, the effect of $\Omega_{\rm m}$ on \qrrc{} and \qvvc{} is small. \vspace{10pt}

\begin{figure}
    \centering
    \gridline{\fig{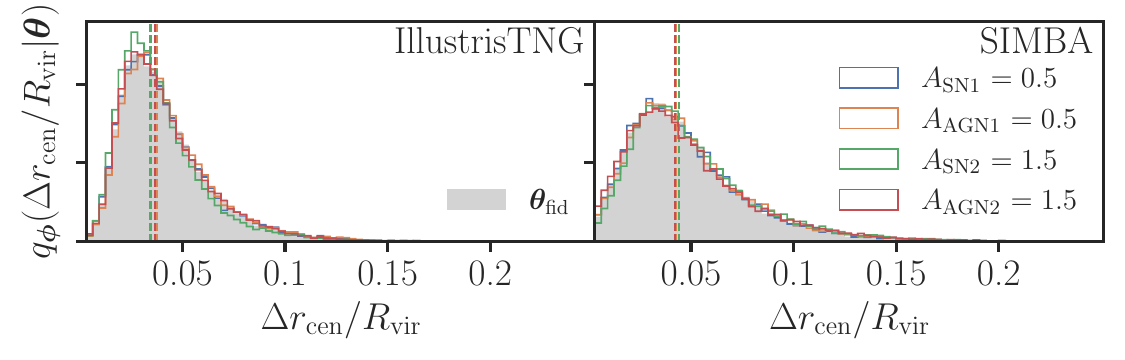}{0.5\textwidth}{(a)}
          \fig{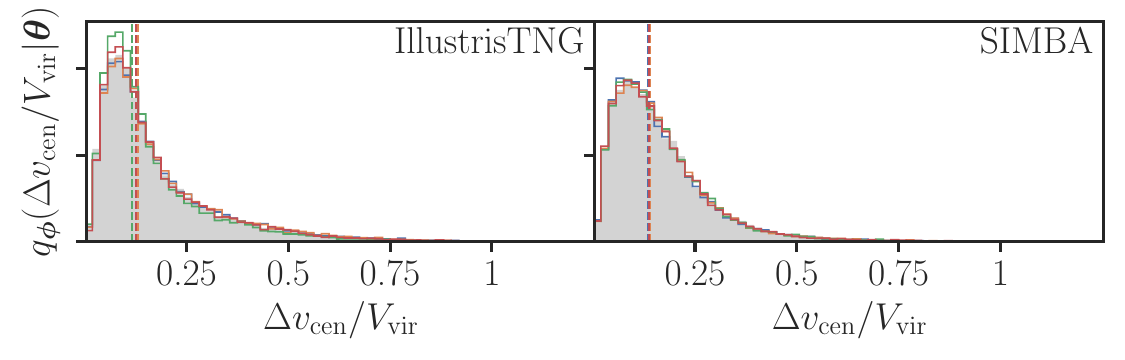}{0.5\textwidth}{(b)}}
    \caption{Dependence of the central galaxy kinematics, \qrrc{} and \qvvc{}, on the baryonic feedback parameters. The distributions vary a single feedback parameter as indicated in the legend (blue, orange, green, and red), while the rest are kept at the fiducial values. We include the distributions at $\bm{\theta}_{\rm fid}$ for reference (gray). The dashed lines mark the medians of the distributions. The baryonic feedback parameters do not significantly impact central galaxy kinematics.}
    \label{fig:cen_hydro}
\end{figure}

\noindent\ul{\emph{Baryonic feedback}}: In Figure~\ref{fig:cen_hydro}, we present the dependence of central galaxy kinematics, \qrrc{}~(a) and \qvvc{}~(b), on $A_{\rm hydro}$ for IllustrisTNG (left) and SIMBA (right). In each panel, we include the fiducial distributions (gray-shaded) for comparison. For the other distributions, we vary a single feedback parameter: $A_{\rm SN1}=0.5$ (blue), $A_{\rm AGN1}=0.5$ (orange), $A_{\rm SN2}=1.5$ (green), and $A_{\rm AGN2}=1.5$ (red). We set the parameters to either 0.5 or 1.5 to ensure that we sufficiently vary baryonic feedback. For reference, in SIMBA, setting the parameters to 0.5 or 1.5 modifies the stellar mass function (SMF) by $\gtrsim 25\%$ at $10^{11}M_\odot$. Despite the significant impact they have on the galaxy population, the baryonic feedback parameters overall do not significantly impact the kinematics of central galaxies.

\begin{figure}
    \centering
    \gridline{\fig{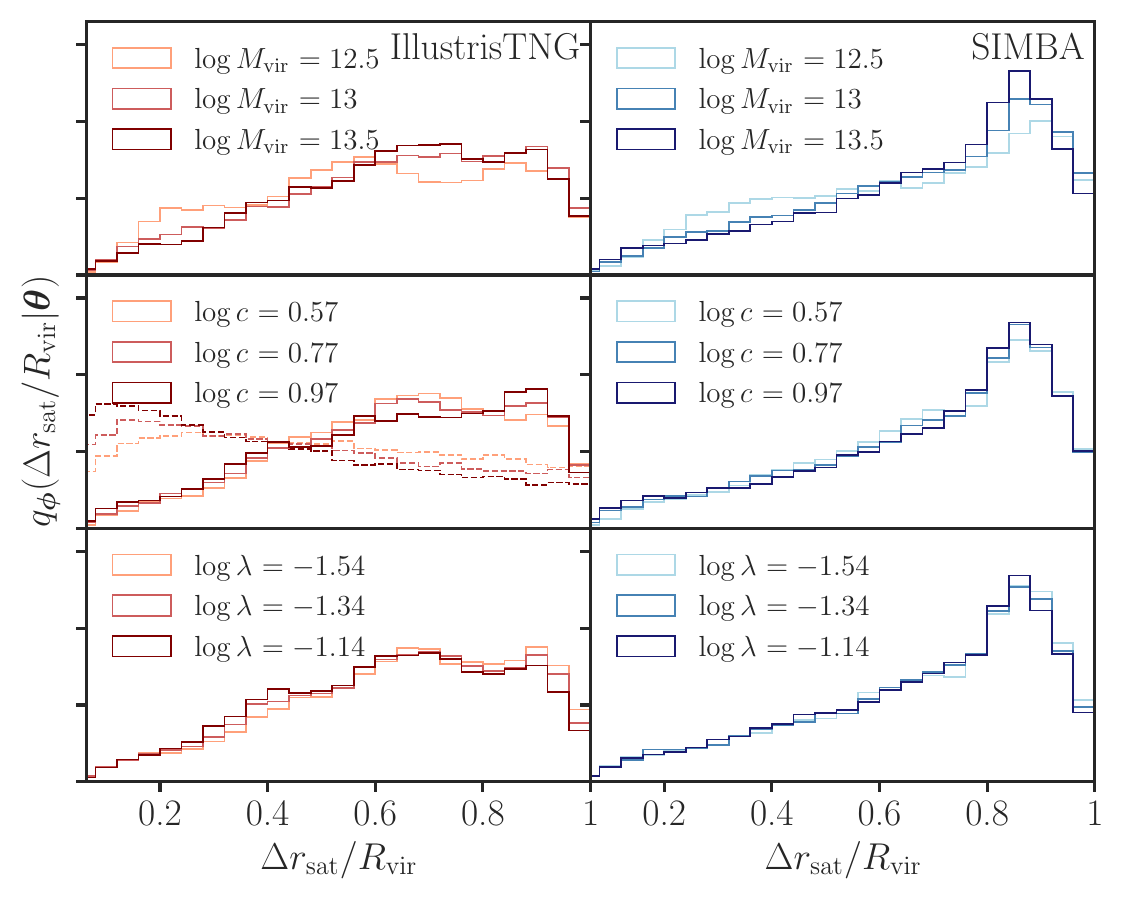}{0.5\textwidth}{(a)}
    \fig{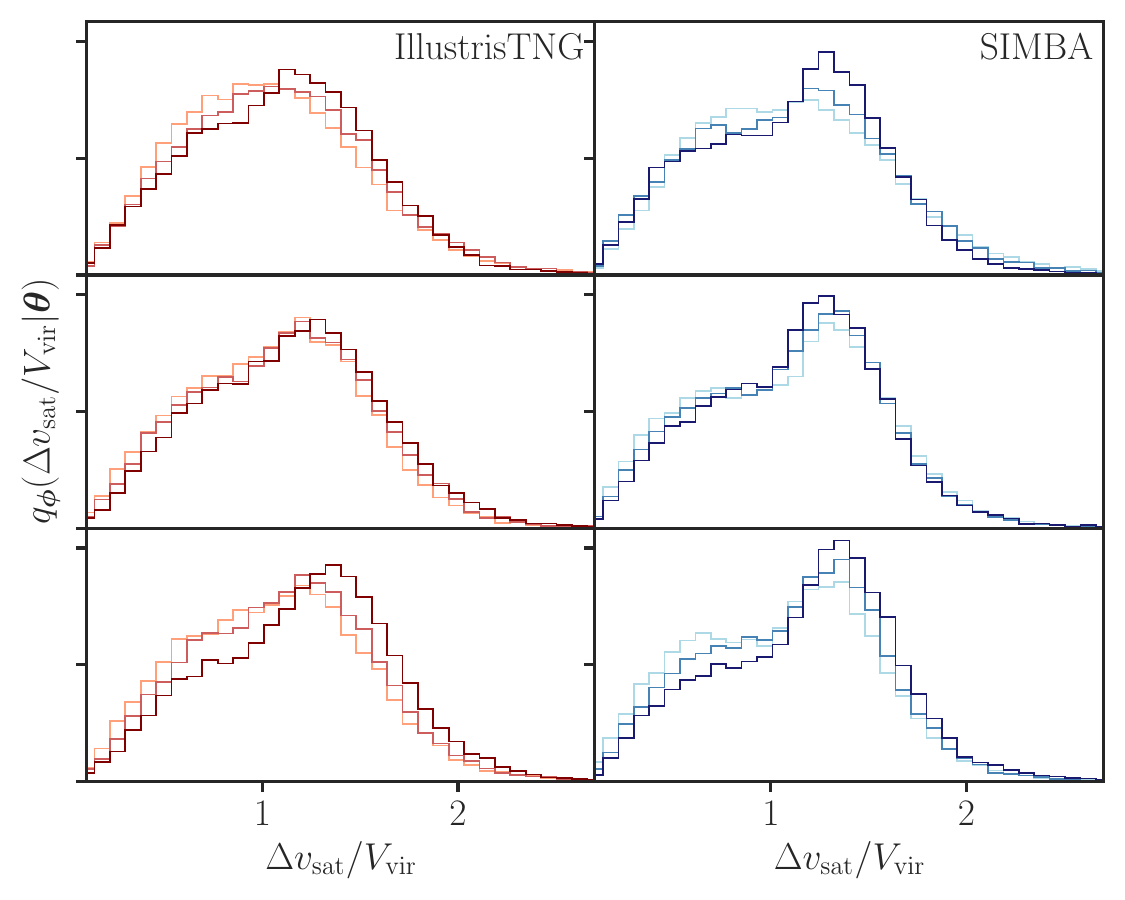}{0.5\textwidth}{(b)}}
    \caption{Dependence of the satellite galaxy kinematics, \qrrs{} and \qvvs{}, on $\bm{\theta}_{\rm h}$ ($M_{\rm vir}$, $c$, and $\lambda$, from top to bottom) in IllustrisTNG (red) and SIMBA (blue). The parameters that do not appear in the legends are kept at the fiducial values. For reference, we include NFW profiles using the three values of $c$ (dashed; $\log c$ panel). \qrrs{} differs between IllustrisTNG and SIMBA. They also differ from the NFW profiles (Section~\ref{subsec:sub_strip}). In addition, \qvvs{} exhibits bimodality, which is due to infalling and orbiting populations of satellite galaxies (Appendix~\ref{app:infall}). Overall, the effect of $\bm{\theta}_{\rm h}$ on the satellite kinematics is not significant, except for the effect of $\lambda$ on $\Delta v_{\rm sat}/V_{\rm vir}$, which changes the infalling fraction.}
    \label{fig:sat_internal}
\end{figure}

\subsection{Satellite Galaxy Kinematics\label{subsec:sat_kin}}
\noindent\ul{\emph{Host halo properties}}: 
In Figure~\ref{fig:sat_internal}, we present the dependence of satellite galaxy kinematics, \qrrs{}~(a) and \qvvs{}~(b), on $\log M_{\rm vir}$ (top), $\log c$ (center), and $\log \lambda$ (bottom) for IllustrisTNG (red) and SIMBA (blue). We note that for \qrrs{}, we impose an upper limit of $\Delta r_{\rm sat}/R_{\rm vir}=1$, since {\sc Rockstar} only identifies satellites within $R_{\rm vir}$. This only discards a small fraction of satellites and does not affect our overall results. We find significant discrepancies in \qrrs{} between IllustrisTNG and SIMBA. Compared to IllustrisTNG, the satellite distribution in SIMBA is skewed toward $R_{\rm vir}$. Despite the discrepancy, we find no strong dependence of \qrrs{} on any $\bm{\theta}_{\rm h}$ for either subgrid model. The lack of dependence on $M_{\rm vir}$ is consistent with \cite{hadzhiyska_2022_one_halo}, where they examined the radial distribution of satellite subhalos in MilleniumTNG~\citep[MTNG;][]{pakmor_mtng_23} as a function of host halo mass. 

Beyond the lack of $\bm{\theta}_{\rm h}$ dependence, we find that \qrrs{}~deviates significantly from the NFW profile (dashed). This can be driven by numerical effects that unphysically disrupt subhalos and make them unidentifiable by halo-finding algorithms at the inner radii of host halos~\citep[e.g.,][]{van_kampen_95, kampen_2000, moore_96, vKampen_99, klypin_99_overmerging, vdB_17, vdB_18_2, diemer_23}. We discuss this in more detail in Section~\ref{subsec:sub_strip}. 

For \qvvs{}, we first note that the distribution is significantly bimodal. This bimodality comes from separate orbiting and infalling populations of satellites and the latter approximately accounts for $10\%$ of our satellites in IllustrisTNG and SIMBA (see Appendix~\ref{app:infall}). The infalling populations are distributed around $\Delta v_{\rm sat}/V_{\rm vir}\approx 1.5$. In addition, the bimodality becomes slightly more pronounced with higher $M_{\rm vir}$ in SIMBA, while the effect of $M_{\rm vir}$ is negligible in IllustrisTNG. There is also a slight effect of $\lambda$ on \qvs{}, where the bimodality becomes stronger with higher $\lambda$ in both IllustrisTNG and SIMBA.

\begin{figure}
    \centering
   \gridline{\fig{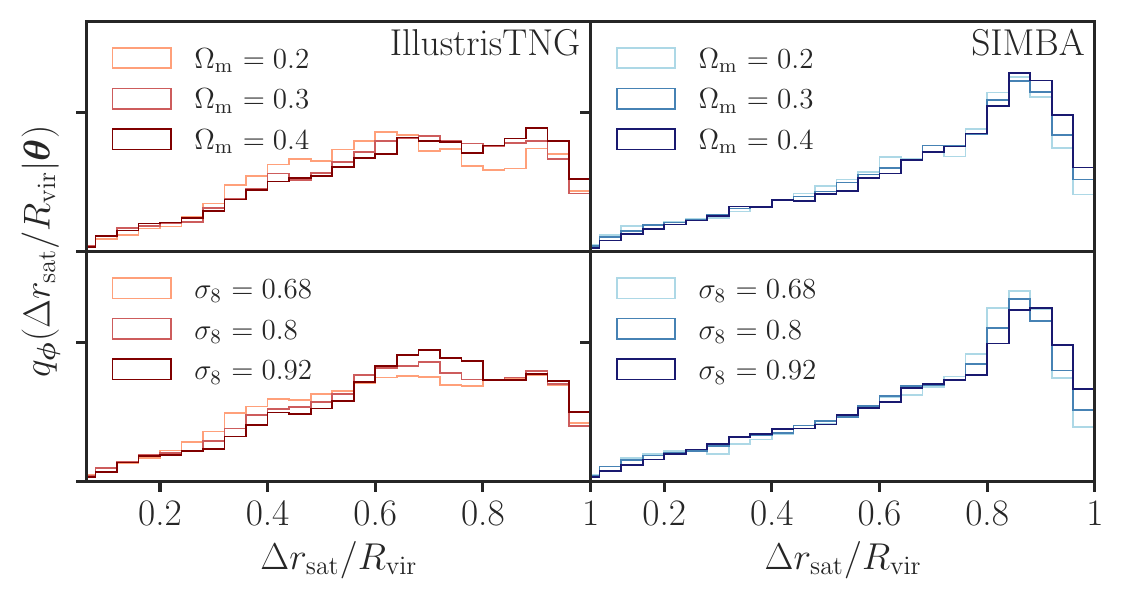}{0.5\textwidth}{(a)}
          \fig{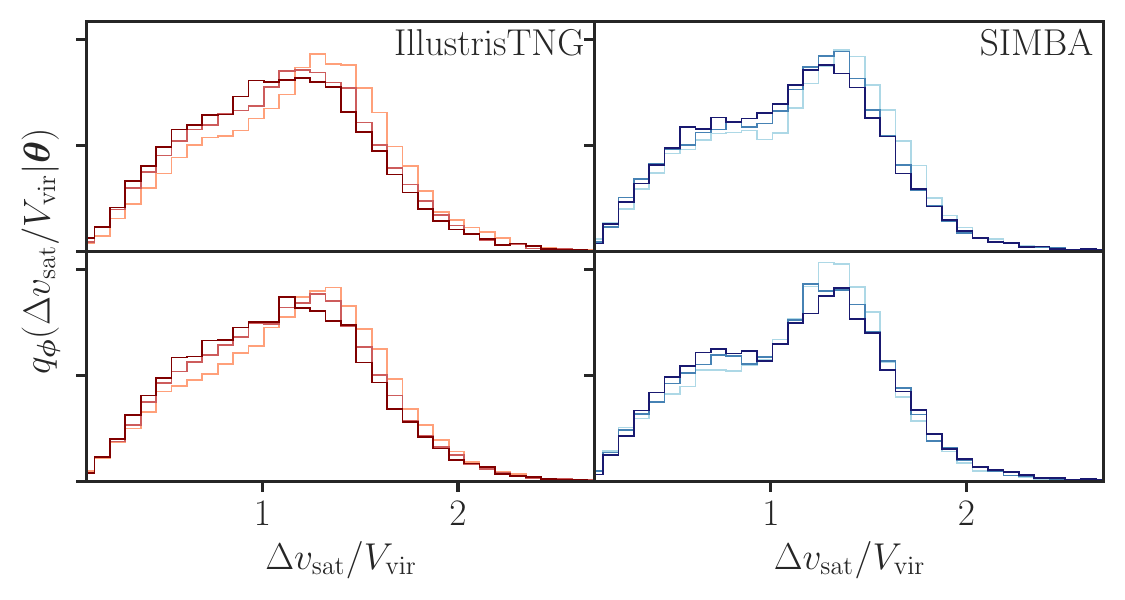}{0.5\textwidth}{(b)}}
    \caption{Dependence of the satellite galaxy kinematics, \qrrs{} and \qvvs{}, on $\Omega_{\rm m}$ (top) and $\sigma_8$ (bottom) in IllustrisTNG (red) and SIMBA (blue). The parameters that do not appear in the legends are kept at the fiducial values. The effects of the cosmological parameters are small in both hydrodynamic suites.}
    \label{fig:sat_cosmo}
\end{figure}
\vspace{10pt}
\noindent\ul{\emph{Cosmological parameters}}: In Figure~\ref{fig:sat_cosmo}, we present the dependence of satellite galaxy kinematics, \qrrs{}~(a) and \qvvs{}~(b), on $\Omega_{\rm m}$ (top) and $\sigma_8$ (bottom) for IllustrisTNG (red) and SIMBA (blue). For \qrrs{}, we find no significant dependence on $\Omega_{\rm m}$ and $\sigma_8$. For \qvvs{}, the distribution slightly shifts to lower $\Delta v_{\rm sat}$ with higher $\Omega_{\rm m}$ in both IllustrisTNG and SIMBA. Additionally, exclusively in IllustrisTNG, the velocity distribution shifts to lower $\Delta v_{\rm sat}$ with higher $\sigma_8$. In Appendix~\ref{app:hmf}, we vary $M_{\rm vir}$ with $\Omega_{\rm m}$ using the fixed number density in the halo mass function. We find a similar weak dependence. Overall, varying $\Omega_{\rm m}$ and $\sigma_8$ does not significantly affect the satellite kinematics as well. 

\begin{figure}
    \centering
    \gridline{\fig{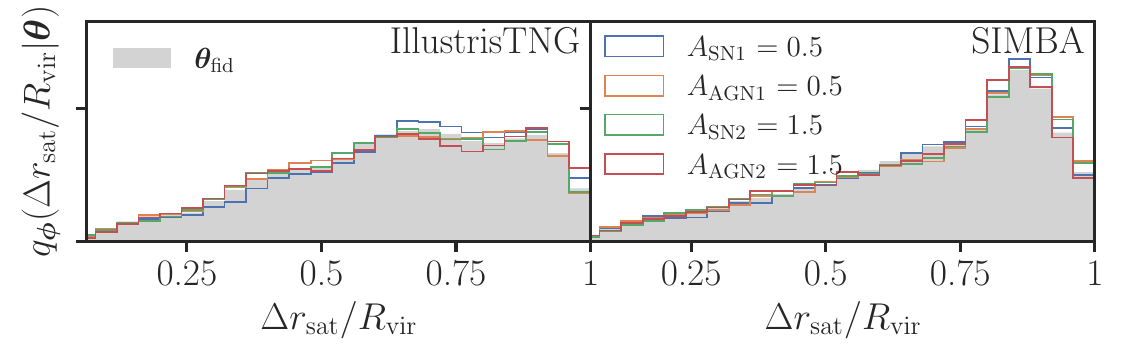}{0.5\textwidth}{(a)}
          \fig{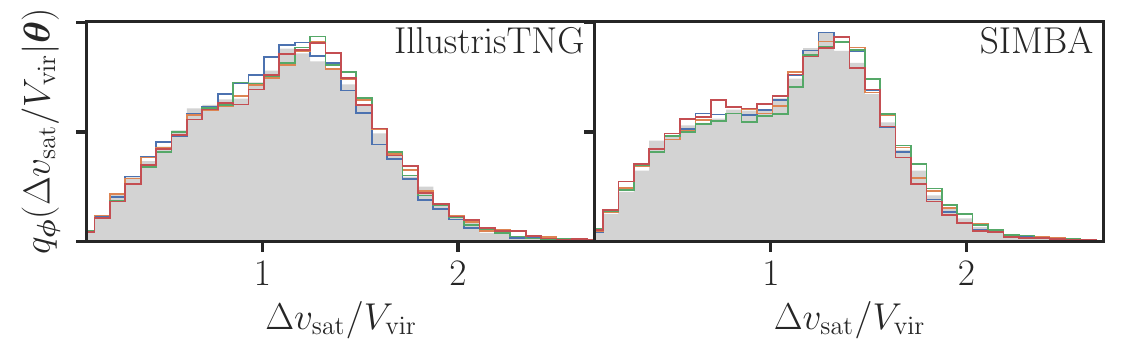}{0.5\textwidth}{(b)}}
    \caption{Dependence of the satellite galaxy kinematics, \qrrs{} and \qvvs{}, on the baryonic feedback parameters. The distributions vary a single feedback parameter as indicated in the legend (blue, orange, green, and red), while the rest are kept at the fiducial values. We include the distributions at $\bm{\theta}_{\rm fid}$ for reference (gray). The effect of baryonic feedback is negligible for satellite galaxy kinematics.}
    \label{fig:sat_hydro}
\end{figure}

\vspace{10pt}
\noindent\ul{\emph{Baryonic feedback}}: In Figure~\ref{fig:sat_hydro}, we present the dependence of satellite galaxy kinematics, \qrrs{}~(a) and \qvvs{}~(b), on $A_{\rm hydro}$ for IllustrisTNG (left) and SIMBA (right). We find no significant dependence of \qrrs{} or \qvvs{} on the change of the baryonic feedback parameters. This is in spite of the fact that we vary $A_{\rm hydro}$ sufficiently to modify the SMF (Section~\ref{subsec:cen_kin}). While studies find that baryonic processes can affect small-scale matter clustering, we find that they alone do not significantly impact the kinematics of satellite galaxies.

\section{Discussion\label{sec:disc}}
\subsection{Galaxy Kinematics\label{subsec:cond_dep}}
Galaxy kinematics estimated by our NDEs have several implications for HOD models. First, central galaxies are not aligned with the centers of their host halos with zero halo-centric velocity. Instead, on average, they are located at $\Delta r_{\rm cen}\sim 0.03R_{\rm vir}$ with $\Delta v_{\rm cen}\sim 0.1V_{\rm vir}$\footnote{This spatial bias is not due to gravitational softening, as the halo density profile at $r=0.03R_{\rm vir}$ is weakly affected by the choice of the softening length~\citep{power_03, zhang_19_opt}.}. Our spatial and velocity biases of central galaxies are consistent with observational evidence. \cite{vdb_05_ps} found that brightest halo galaxies in the Two-Degree Field Galaxy Redshift Survey \citep[2dFGRS;][]{cole_05} and SDSS have non-zero specific kinetic energy in the coordinate frame defined with respect to the satellites and the velocity bias is larger in more massive halos. \cite{guo_15b} reached similar conclusions using SDSS-III BOSS central galaxies. Additionally, \cite{zitrin_12} showed that bright cluster galaxies in 10,000 SDSS clusters have a spatial bias from the halo center. In addition to spatial and velocity biases, we also find that the kinematics of central galaxies depend significantly on $c$ and $\lambda$. Incorporating these biases and dependencies into HOD models will enable more accurate modeling of central galaxy kinematics. 

Beyond these effects, we find little evidence that there are additional dependencies that need to be included in HOD models for central galaxies. In particular, we find that AGN and SN feedback do not significantly impact the central galaxy kinematics. This is consistent with previous works. For instance, \cite{hellwing_2016} matched halos in a hydrodynamic simulation and its dark matter only counterpart in the Evolution and Assembly of GaLaxies and their Environments~\citep[EAGLE;][]{eagle_15} simulation. They found peculiar velocity offsets between galaxies are consistent with zero.

For satellite galaxies, we find that the radial distribution differs significantly between IllustrisTNG and SIMBA. The distributions of both models also deviate from the NFW profile, with distributions skewed towards higher $\Delta r_{\rm sat}/R_{\rm vir}$ than NFW. Similar studies by \cite{yuan_22_lrg_elg} and \cite{hadzhiyska_2022_one_halo} find LRG satellites in IllustrisTNG and MTNG, selected using sliding color-magnitude cuts and stellar mass cuts respectively, have consistent shapes of radial distributions as our \qrrs{}. In this work, we confirm that the deviation from NFW is not unique to IllustrisTNG but is also evident in SIMBA. However, as mentioned above, the deviation from NFW may be driven by known numerical limitation of $N$-body simulations, where satellite subhalos near the center of host halos can experience unphysical mass stripping and disruption. We discuss this limitation further in the following section.

In addition, we find signatures of accreted, infalling satellite galaxies in \qvvs{} (see also Appendix~\ref{app:infall}). The velocity distribution of the satellite galaxies exhibits bimodality due to the coexistence of infalling and orbiting populations. We estimate in Appendix~\ref{app:infall} that the infalling population accounts for $\sim 10\%$ of satellite galaxies with $M_\ast > 10^9 M_\odot$. The distinct populations of infalling and orbiting satellites pose the question of whether the current HOD models should explicitly include an infalling population. Since their kinematics are distinct from orbiting satellites, the infalling population can significantly impact small-scale galaxy clustering, especially in redshift space. Furthermore, including infalling satellites may enable us to probe galaxy formation and galaxy-halo connection~\citep[e.g.,][]{mihos_94, gill_2005, more_11, rocha_12, lange_19_sat_kin, wright_22} or test general relativity \citep[e.g.,][]{zw_13, zw_14}. If one were to include infalling satellites in HOD modeling, our NDEs would provide a direct way of including them by using the dynamics of simulations. 

Aside from the shape of the satellite galaxy kinematics distributions, we do not find any significant parameter dependence on $\bm{\theta}_{\rm h}$, $\Omega_{\rm m}$, $\sigma_8$, or $A_{\rm hydro}$. We find a dependence of \qvvs{}~on $\lambda$; however, the effect is small and the impact is primarily on the relative amplitude between the orbiting and infalling fraction. Interestingly, baryonic feedback does not significantly impact satellite positions or velocities, even when we vary $A_{\rm hydro}$ enough to induce $\gtrsim 25\%$ changes to the SMF at $10^{11}M_\odot$.

Overall, our results indicate that HOD models can assign the kinematics of satellite galaxies without considering a detailed dependence on $\bm{\theta}_{\rm h}$, $\Omega_{\rm m}$, $\sigma_8$, and AGN and SN feedback. However, for both central and satellite galaxies, we find differences in $q_{\bm{\phi}}(\Delta r/R_{\rm vir}\,|\,\bm{\theta})$ between IllustrisTNG and SIMBA, which indicates that hydrodynamic simulations have yet to converge in their modeling of galaxy kinematics. This underscores the necessity for HOD models to incorporate more flexible parameterizations of galaxy kinematics that can describe the range of predictions in simulations. 

\subsection{Satellite Incompleteness\label{subsec:sub_strip}}
The radial distribution of our satellites significantly deviates from NFW in both IllustrisTNG and SIMBA. The lack of satellites near the host halo centers compared to the NFW profile is partly anticipated by known numerical limitations in $N$-body simulations. The disruption of subhalos by dynamical friction and tidal stripping can be unrealistically enhanced by gravitational softening~\citep{moore_96, klypin_99_overmerging, per_10, vdB_17, vdB_18, vdB_18_2}. Additional numerical limitations in the two-body relaxation, particle-halo heating, and particle-halo tidal heating can also quickly disintegrate subhalos \citep{van_kampen_95, kampen_2000, moore_96, klypin_99_overmerging}. Even when subhalos are not fully disrupted, numerical effects may influence the recovered subhalo properties~\citep{mansfield_21} and halo-finding algorithms may lose subhalos experiencing tidal deformation~\citep{diemer_23}. These causes of satellite incompleteness are relevant even to subhalos with a large number of particles~\citep[e.g.,][]{vdB_17, vdB_18, diemer_23}. 

To assess the degree of subhalo incompleteness, we compare the galaxy kinematics derived from CAMELS with the \textsc{UniverseMachine}~\citep[UM;][]{behroozi_2019}. UM is a semi-empirical model that employs analytic prescription to account for both the numerical and physical disruption of satellite subhalos. It is built on halo merger trees of Bolshoi-Planck $N$-body simulations~\citep{klypin_16, puebla_16} of comoving box size equal to 250 ~$h^{-1}$Mpc. Once satellite subhalos become unidentifiable by \textsc{Rockstar}, UM mitigates satellite incompleteness by analytically integrating softened force-law to track ``orphan'' satellites until they are considered disrupted or merged~\citep[see Appendix~B of][]{behroozi_2019}. 

In Figure~\ref{fig:UM}, we compare the halo-centric radial and velocity distributions of satellite galaxies in UM (gray shade), IllustrisTNG (red), and SIMBA (blue). From UM, we take satellite galaxies at $z=0$ with $M_\ast > 10^{9}M_\odot$ and host halo mass around our fiducial value, ${13.45}\leq \log M_{\rm vir}\leq {13.55}$. For IllustrisTNG and SIMBA, we use our NDEs with $\bm{\theta}_{\rm fid}$, which has consistent cosmology with the UM. Compared to our \qrrs{} and \qvvs{} from CAMELS, the satellite $\Delta r$ and $\Delta v$ distributions are skewed toward the center of the halo in UM. The contrast between the satellite distributions of UM and the CAMELS simulations suggests that numerical disruption may indeed affect satellites in hydrodynamic simulations. Moreover, this indicates that we cannot take hydrodynamic simulations as the ground truth and that HOD models should incorporate additional flexibility to account for our current lack of understanding regarding the kinematics of satellites near the halo center.

\begin{figure}
    \centering
    \gridline{\fig{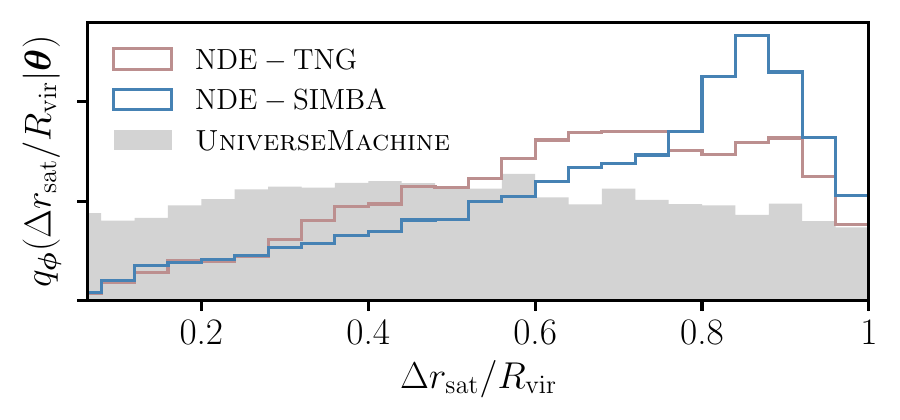}{0.45\textwidth}{(a)}
    \fig{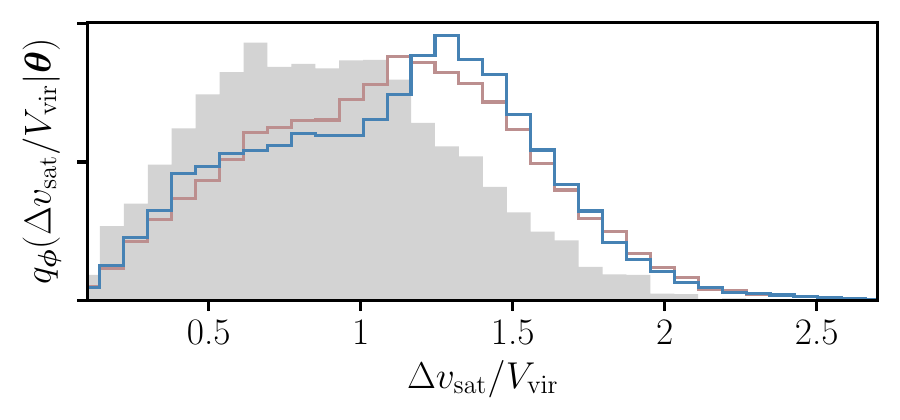}{0.45\textwidth}
{(b)}}
    \caption{Comparison of the $\Delta r_{\rm sat}/R_{\rm vir}$ and $\Delta v_{\rm sat}/V_{\rm vir}$ of satellite galaxies in IllustrisTNG (red), and SIMBA (blue), and the \textsc{UniverseMachine} (UM, gray shade). The IllustrisTNG and SIMBA distributions are derived from our NDEs with $\bm{\theta}_{\rm fid}$ while the UM distribution is derived from $M_*>10^9M_\odot$ UM satellite galaxies at $z=0$ with ${13.45}\leq \log M_{\rm vir}\leq {13.55}$. UM uses analytic prescriptions to mitigate satellite incompleteness. The comparison suggests that IllustrisTNG and SIMBA may experience satellite incompleteness, which significantly reduces the fraction of satellite galaxies close to the host halo center.}
    \label{fig:UM}
\end{figure}

\subsection{Limitations and Outlook\label{subsec:lim}}
The primary goal of this study is to investigate how the positions and velocities of galaxies depend on the internal halo properties, cosmological parameters, and feedback parameters. Currently, our NDEs focus solely on the magnitudes of halo-centric position and velocity offsets, $\Delta r$ and $\Delta v$, without considering correlations between them. We also do not include information on anisotropies in satellite galaxy kinematics. 
Some studies using observations from 2dFGRS and SDSS~\citep[e.g.,][]{sales_04, bailin_08_sdss} 
as well as simulations~\citep[e.g.,][]{zentner_05, hadzhiyska_2022_one_halo} find evidence for anisotropies among satellite galaxies. While these effects may affect how galaxies should be placed within a halo in practice, they do not impact our results on the parameter dependence.
We also note that, in principle, our NDE-based approach can incorporate correlations between $\Delta r$ and $\Delta v$ as well as satellite anisotropies.

Additionally, while this work focuses solely on the internal properties of host halos, many studies have demonstrated the important role of environmental properties and assembly history of halos in the galaxy-halo connection~\citep[e.g.,][]{sheth_tormen_2004, shen_2006, croton_07, hahn_2007, hahn_2009, borzyszkowski_17, paranjape_18, xu_2021, delgado_22, hadzhisyksa_22_two_halo}. The shear field defined by the tidal tensor of ambient mass distribution, for instance, may play a pivotal role in shaping the satellite kinematics by affecting the orbits and survival of satellites within halos \citep[e.g.,][]{hahn_2007, hahn_2009, paranjape_18}. In future works, we plan to incorporate environment information at both the dark matter particle level and halo catalog level using, e.g., $k^{\rm th}$ nearest neighbor~\citep[e.g.,][]{banerjee2021_knn} and graph neural networks~\citep[e.g.,][]{cranmer2019, jespersen2022_gnn, villanueva-domingo2022_gnn, gnn_wu_23}.

This work is the first in a series of re-evaluating the HOD using CAMELS to develop a more accurate framework for modeling galaxy clustering at small scales. In subsequent work, we will incorporate additional parameters, such as the environmental properties of halos and angular information of galaxy kinematics, into our NDE-based approach. This will provide a more complete understanding of the galaxy-halo connection and whether the HOD model needs to be expanded for additional parameter dependency.

Once the importance of additional parameters have been explored, we will derive new analytic HOD models using symbolic regression~\citep[e.g.,][]{pysr}. We will construct these models to also incorporate current modeling uncertainties in the satellite kinematics near the halo center by using a functional form that can describe satellites in hydrodynamic simulations and with corrections for satellite incompleteness (Section~\ref{subsec:sub_strip}). This will provide a more flexible and comprehensive description of galaxy kinematics within halos than current HOD models, effectively capturing any additional parameter dependencies. Finally, we will demonstrate that our new HOD models can more accurately describe
higher-order galaxy clustering down to small scales and test whether they can meet the 
precision level required to analyze upcoming galaxy surveys.

In the upcoming work, we also plan to extend our work to additional simulations. We extensively validated our NDEs and demonstrated their robustness for our analysis (see Appendix~\ref{app:nde_valid}). However, incorporating more simulations will further enhance their accuracy, especially by increasing the number of satellites and massive host halos. 
We will also expand the comparison of galaxy kinematics to other hydrodynamic simulations in CAMELS that employ different subgrid prescriptions for modeling baryonic processes, such as Astrid~\citep{bird_2022}, EAGLE, Ramses~\citep{teyssier02_ramses}, and Magneticum~\citep{hirschmann2014_magneticum}. Future CAMELS suites will also increase the number of simulation parameters to 28, including $\Omega_{\rm b}$, $h$, and additional stellar and galactic wind parameters. 
By using additional simulations with more cosmological and astrophysical parameters, we will 
further validate our findings and develop an even more robust and flexible HOD model. 

\section{summary\label{sec:summary}}
 A better understanding of galaxy kinematics will enable us to model galaxy clustering more accurately and, thus, extract more constraining power from galaxy surveys for cosmology, galaxy evolution, and galaxy-halo connection studies. In this paper, we investigate the dependence of the central and satellite galaxy kinematics on $\bm{\theta}$, the intrinsic host halo properties (mass, spin, concentration), cosmology ($\Omega_{\rm m}$, $\sigma_8$), and baryonic feedback from AGN and SNe ($A_{\rm SN1}, A_{\rm SN2}, A_{\rm AGN1}, A_{\rm AGN2}$).

We use simulated galaxies from 2,000 CAMELS hydrodynamic simulations constructed using the IllustrisTNG and SIMBA galaxy formation models. We focus on central and satellite galaxies with $M_* > 10^9 M_\odot$, identified using the \textsc{Rockstar} halo finder. We apply NDE based on normalizing flows to estimate the $p(\Delta r\,|\,\bm{\theta})$ and $p(\Delta v\,|\,\bm{\theta})$ of central and satellite galaxies. This enables us to accurately capture the dependence of galaxy kinematics on $\bm{\theta}$ and quantify the impact of each parameter on $p(\Delta r\,|\,\bm{\theta})$ and $p(\Delta v\,|\,\bm{\theta})$.

We summarize our main findings below: 
\begin{enumerate}[(i)]
    \item Central galaxies have significant spatial bias and velocity bias with respect to their host halos, consistent with previous observational evidence. We find significant dependence in these biases on host halo mass, concentration, and spin. In contrast, we find no significant dependence of central galaxy kinematics on $\Omega_{\rm m}$, $\sigma_8$, and AGN and SN feedback.
    \item Satellite galaxies in CAMELS have radial distributions that strongly deviate from an NFW profile, skewing away from the halo center. We find a deviation from NFW in both IllustrisTNG and SIMBA satellites. This discrepancy, which is most significant near the halo center, suggests that hydrodynamic simulations may experience satellite incompleteness. Additionally, the significant differences in the radial distributions of satellites in IllustrisTNG and SIMBA indicate that HOD models should incorporate additional flexibility when modeling satellite positions.
    \item The velocity distribution of satellite galaxies is bimodal due to orbiting and infalling satellites (Appendix~\ref{app:infall}). The infalling population accounts for $\sim 10\%$ of $M_\ast > 10^9 M_\odot$ satellites.
    \item The velocity distribution of satellites has a slight dependence on host halo spin which affects the bimodality. Besides this weak dependence, the kinematics of satellite galaxies do not depend strongly on other host halo properties, cosmology, or feedback parameters. 
\end{enumerate}

Our results provide timely insights into improving HOD models and our understanding of the galaxy-halo connection. Galaxy clustering analyses are currently pushing to small scales to exploit additional constraining power. At the same time, upcoming galaxy surveys like DESI, PFS, and Euclid, with their unprecedented statistical power, will have significantly higher accuracy requirements for modeling galaxy clustering. Our efforts seek to develop a new framework for HOD that will meet the upcoming requirements in modeling galaxy clustering on small scales.

This paper is the first of a series that will re-examine the HOD using CAMELS. In subsequent works, we will incorporate additional parameters, such as the environmental properties of halos and angular information of galaxy kinematics to reveal any additional parameter dependencies. We will then derive new analytic HOD models that account for current uncertainties in the satellite kinematics near the halo center due to satellite incompleteness in simulations. Finally, we will demonstrate that our new HOD model can more accurately describe the higher-order galaxy clustering down to small scales at the precision level required by upcoming galaxy surveys. Ultimately, these works will develop a new HOD framework that can more accurately, flexibly, and robustly model galaxy clustering down to small scales.

\appendix
\section{NDE Validation\label{app:nde_valid}}
\begin{figure}
    \centering
    \gridline{\fig{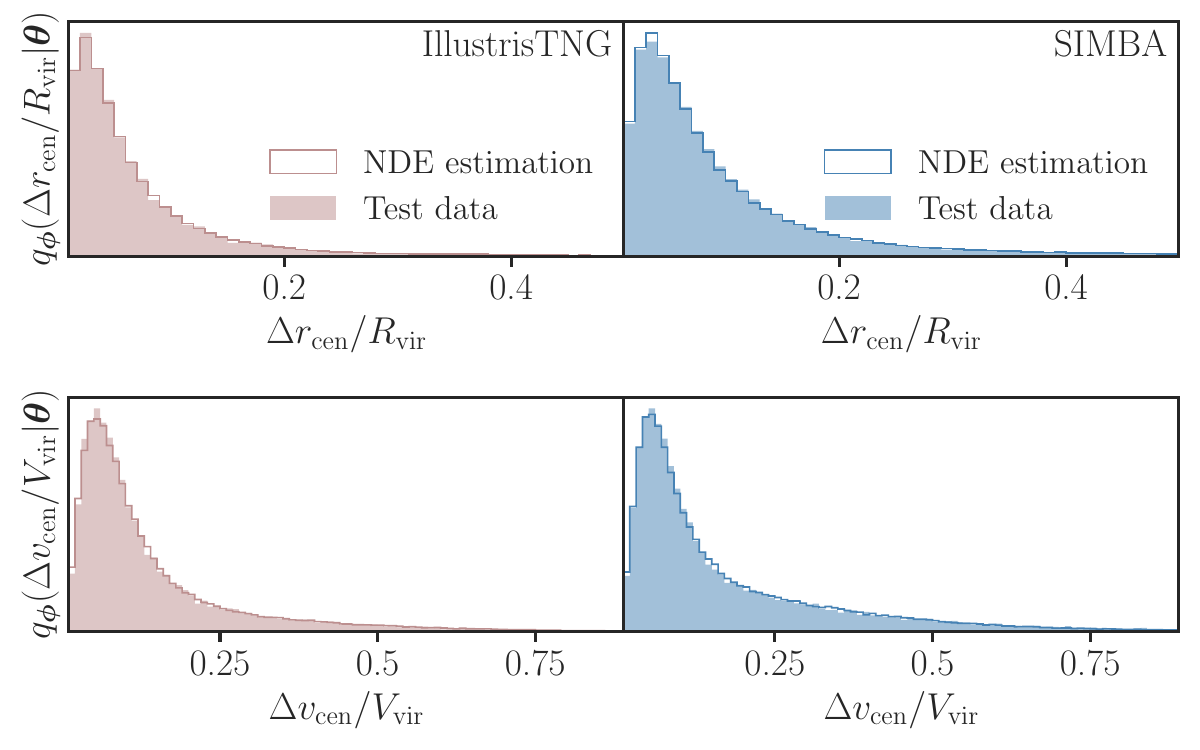}{0.5\textwidth}{(a)}
    \fig{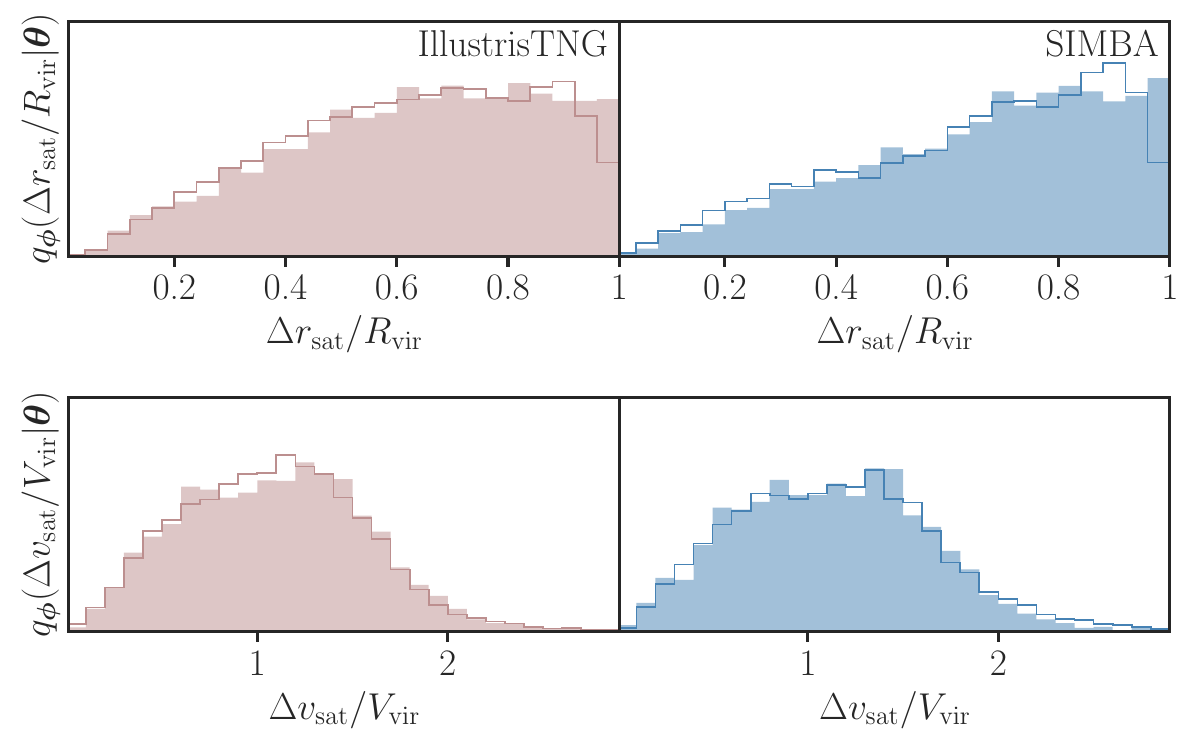}{0.5\textwidth}
{(b)}}
    \caption{
    Comparison between the $\Delta r/R_{\rm vir}$ (top) and $\Delta v/V_{\rm vir}$ (bottom) distributions drawn from our NDEs (solid line) and from the test sample (shaded). We compare the distributions for central (a) and satellite (b) galaxies in IllustrisTNG (red) and SIMBA (blue). The NDE distributions are constructed using $\Delta r'/R_{\rm vir}$ and $\Delta v'/V_{\rm vir}$ drawn from the NDEs at the $\bm{\theta}$ values of every test galaxy. We find excellent agreement between the NDE and test sample distributions, which demonstrates the overall accuracy of our NDEs. 
    }
    \label{fig:cen_nde_valid}
\end{figure}
In this section, we validate the accuracy of the NDEs, \qrrc{}, \qrrs{}, \qvvc{}, and \qvvs{}, using the randomly selected 10\% of CAMELS galaxies that we reserve for testing (Section~\ref{sec:data}). This test sample consists of 27,652 central galaxies and 3,469 satellites for IllustrisTNG and 35,949 central galaxies and 2,224 satellites for SIMBA. For each test galaxy~$i$, we compile its parameters: $\bm{\theta}_i=(M_{{\rm vir}, i}$, $c_i$, $\lambda_i$, $\Omega_{{\rm m}, i}$, $\sigma_{8, i}$, $A_{{\rm hydro}, i})$. We then use the NDEs to draw $\Delta r_{{\rm cen},i}'$, $\Delta r_{{\rm sat},i}'$, $\Delta v_{{\rm cen},i}'$, and $\Delta v_{{\rm sat},i}'$ for the given $\bm{\theta}_i$. Afterwards, we compare the distributions of the kinematic quantities drawn from 
our NDEs to $p(\Delta r_{{\rm cen},i}/R_{\rm vir})$, $p(\Delta r_{{\rm sat},i}/R_{\rm vir})$, $p(\Delta v_{{\rm cen},i}/V_{\rm vir})$, and $p(\Delta v_{{\rm sat},i}/V_{\rm vir})$ of the test galaxies.

In Figure~\ref{fig:cen_nde_valid}, we present the comparison between the $p(\Delta r'/R_{\rm vir})$ and $p(\Delta r/R_{\rm vir})$ (top) and $p(\Delta v'/V_{\rm vir})$ and $p(\Delta v/V_{\rm vir})$ (bottom) for IllustrisTNG (red) and SIMBA (blue), and for central (a) and satellite (b) galaxies. $p(\Delta r'/R_{\rm vir})$ and $p(\Delta v'/V_{\rm vir})$ correspond to the distributions drawn from our NDEs (solid line). We draw 15 samples from our NDEs for each test galaxy to reduce the noise in the distribution. Meanwhile, $p(\Delta r/R_{\rm vir})$ and  $p(\Delta v/V_{\rm vir})$ correspond to the distribution of all test galaxies (shaded). For both central and satellite galaxies, and for both subgrid models, the distributions from our NDEs are in excellent agreement with the test sample distributions. This demonstrates the overall accuracy of our NDEs.

\begin{figure}
    \centering
    \gridline{\fig{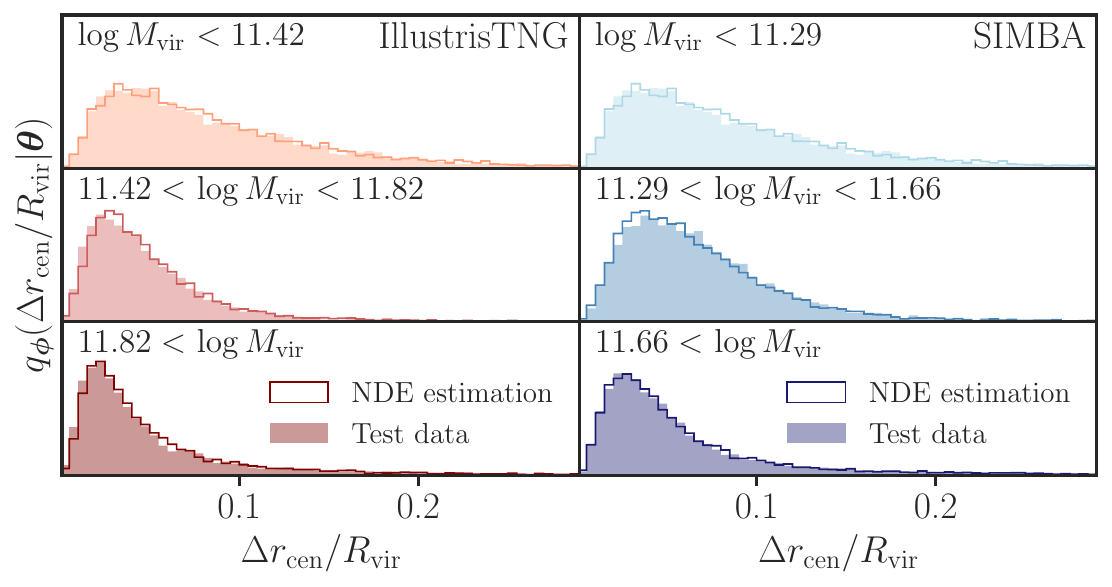}{0.5\textwidth}{(a)}\fig{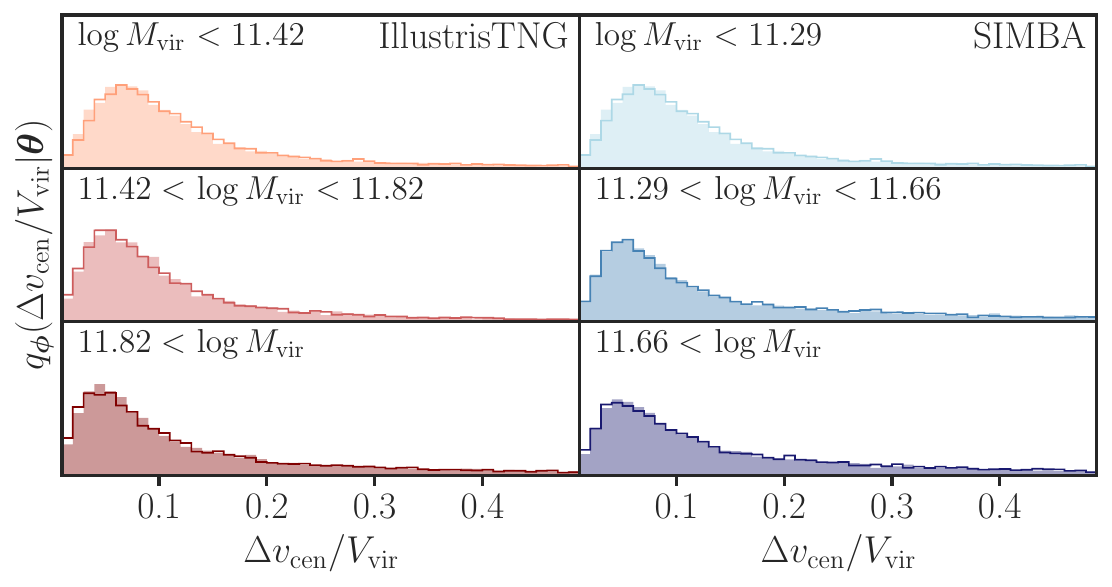}{0.5\textwidth}{(b)}}
    \caption{
    Comparison between the $\Delta r_{\rm cen}/R_{\rm vir}$ (a) and $\Delta v_{\rm cen}/V_{\rm vir}$ (b) distributions drawn from our NDEs (solid line) and from the test sample (shaded) divided into low, intermediate, and high $M_{\rm vir}$ bins. We compare the distributions for IllustrisTNG (red) and SIMBA (blue). The NDE distributions are constructed in the same fashion as in Figure~\ref{fig:cen_nde_valid} using the test galaxies in each $M_{\rm vir}$ bin. By splitting the test sample by $M_{\rm vir}$, our comparison examines the accuracy of our NDEs, while accounting for parameter dependencies. We find good agreement between the NDE and test sample distributions in every panel. We similarly find good agreements for other components of $\bm{\theta}$ and for satellites. 
    }
    \label{fig:cen_Mvir_subreimge}
\end{figure}

We further validate the NDEs, this time focusing on the parameter dependence. Instead of comparing $p(\Delta r/R_{\rm vir})$ and $p(\Delta v/V_{\rm vir})$ over the full test sample, we split the sample into thirds along each parameter of $\bm{\theta}$. Then, for each of the test subsamples, we repeat the validation above. In Figure~\ref{fig:cen_Mvir_subreimge}, we present the validation results for central galaxies divided using $\log M_{\rm vir}$ for IllustrisTNG~(red) and SIMBA~(blue). As in Figure~\ref{fig:cen_nde_valid}, we mark the distributions of our NDE draws in solid and the distributions from the test sample in the shaded regions. From top to bottom, we show the comparison in low, intermediate, and high $M_{\rm vir}$ ranges. In all of the $M_{\rm vir}$ ranges, for both IllustrisTNG and SIMBA, the distributions are in excellent agreement. We similarly find good agreements for all other parameters in $\bm{\theta}$ and for satellites. 

Finally, we validate our NDEs using the Test of Accuracy with Random Points~\citep[TARP;][]{lemos_23_tarp} coverage test. TARP assesses the accuracy of NDEs by estimating the expected coverage probabilities (ECPs) for given credibility levels. For the test, we randomly select 10,000 central galaxies and use all satellites from the IllustrisTNG and SIMBA test datasets. 
For each selected test galaxy $i$ with $(\Delta a_{i},\bm{\theta}_i)$, we draw 1,000 kinematics estimates, $\{\Delta a_i'\}$, using our NDEs. We sample a random point $\Delta a_{i,r}$ uniformly 
from the range of $\Delta a$ spanned by the selected test galaxies. For a given credibility level, we estimate the ECP from the fraction of $\{\Delta a_i'\}$ with $|\Delta a'_i  -\Delta a_{i,r}|<|\Delta a_{i}-\Delta a_{i,r}|$. For an optimal NDE, ECP is equal to the credibility level.

\begin{figure}
    \centering
    \gridline{\fig{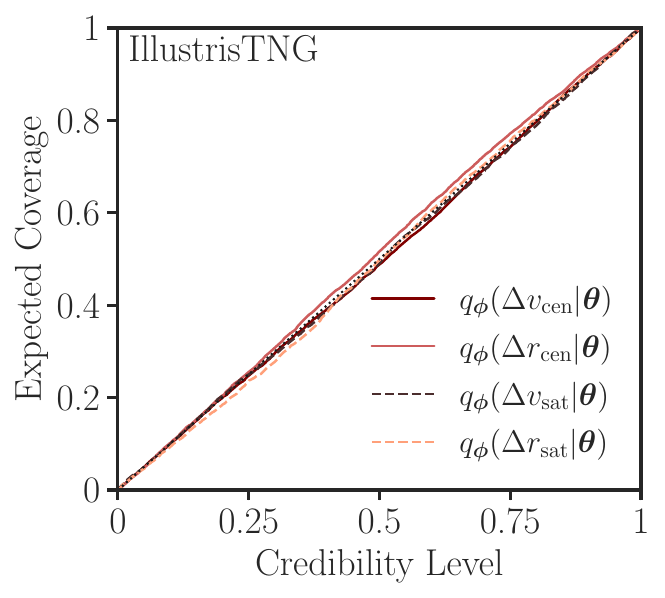}{0.35\textwidth}{(a)}\fig{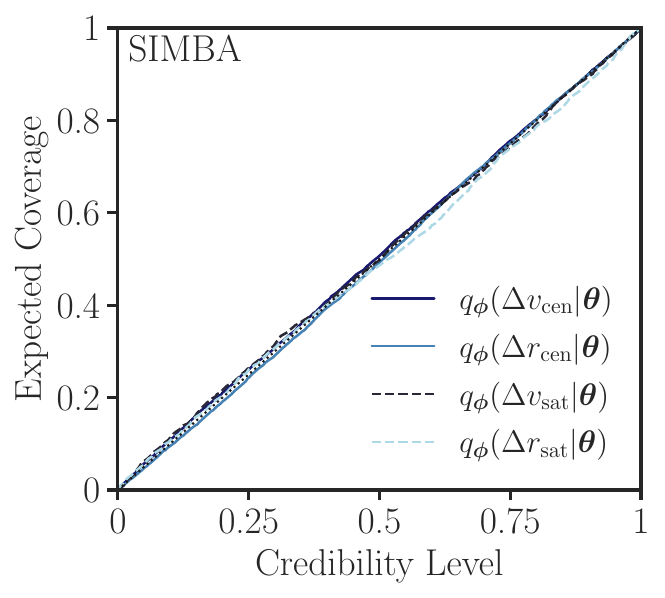}{0.35\textwidth}{(b)}}
    \caption{Comparison between credibility levels and corresponding expected coverage probabilities for \qac{}~(solid lines) and \qas{}~(dashed lines) in IllustrisTNG (a) and SIMBA (b). We compute the expected coverage using randomly selected 10,000 central galaxies and all satellites in test datasets of both hydrodynamic suites. The good agreement between the expected coverage probability and the credibility level demonstrates that our NDEs provide accurate estimates of kinematics distributions.}
    \label{fig:cov}
\end{figure}

In Figure~\ref{fig:cov}, we compare the ECPs to the credibility levels for \qac{}~(solid lines) and \qas{}~(dashed lines) in IllustrisTNG~(a) and SIMBA~(b), derived from TARP. The excellent agreement between the ECPs and credibility levels demonstrates that all of our NDEs provide accurate estimates of kinematics distributions. We therefore conclude that our NDEs estimate $p(\Delta a|\bm{\theta})$ in CAMELS with sufficient accuracy.

\section{$\Omega_{\rm\lowercase{m}}$ dependence\label{app:hmf}}
In Section~\ref{sec:result}, we examine the dependence of central and satellite kinematics on $\Omega_{\rm m}$. However, to estimate the dependence on $\Omega_{\rm m}$ alone, we must account for the fact that the definition of halo mass, $M_{\rm vir}$, changes with $\Omega_{\rm m}$. In this section, we examine whether the kinematics dependence, or lack thereof, on $\Omega_{\rm m}$ is affected if we vary $M_{\rm vir}$ according to $\Omega_{\rm m}$. 

To disentangle the dependence on $\Omega_{\rm m}$ from $M_{\rm vir}$, for a given $\Omega_{\rm m}$ we set $M_{\rm vir}$ to the halo mass that matches the fiducial number density of $M_{\rm vir}=10^{13.5}M_\odot$ halos at the fiducial $\Omega_{\rm m}=0.3$ cosmology. We do this by measuring the halo mass functions (HMF) for IllustrisTNG and SIMBA at $\Omega_{\rm m}=0.2$, 0.3, and 0.4, then matching them to the fiducial halo number density. For $\Omega_{\rm m} = 0.2$ and 0.4 we use $M_{\rm vir} \approx 10^{13.25}M_\odot$ and $10^{13.72}M_\odot$ for IllustrisTNG, and $M_{\rm vir} \approx 10^{13.14}M_\odot$ and $10^{13.67}M_\odot$ for SIMBA.

\begin{figure}
    \centering
    \gridline{\fig{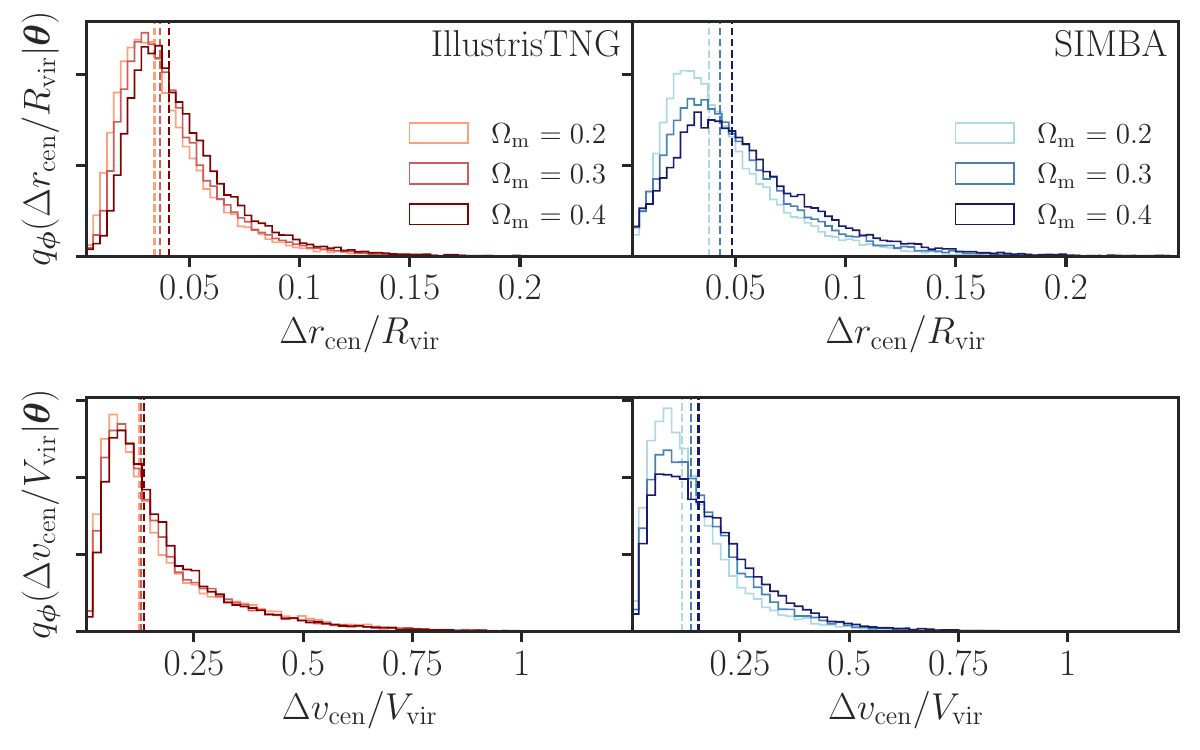}{0.5\textwidth}{(a)}
    \fig{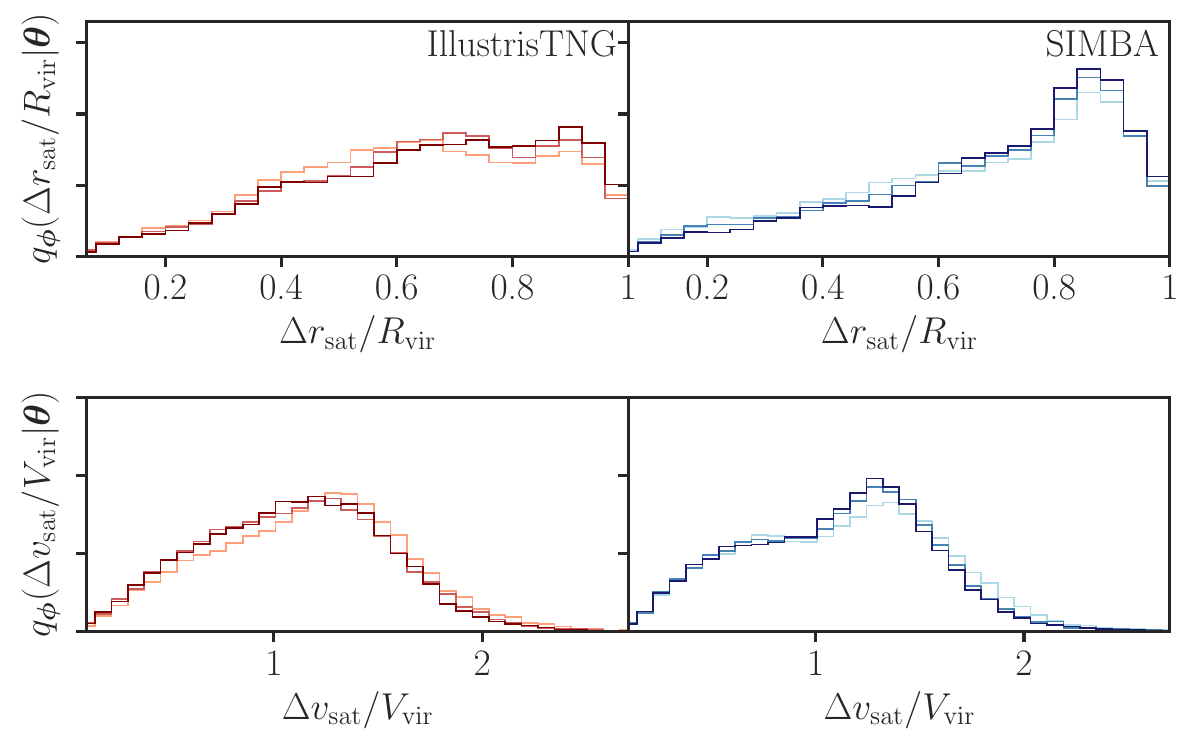}{0.5\textwidth}{(b)}}
    \caption{Dependence of the central galaxy kinematics (a) and satellite kinematics (b), \qrr{} (top) and \qvv{} (bottom), on $\Omega_{\rm m}$ in IllustrisTNG (red) and SIMBA (blue), while varying $M_{\rm vir}$ to a fixed fiducial halo number density. The parameters that do not appear in the legends are kept at the fiducial values. The dashed lines in (a) mark the medians of the distributions. Overall, varying $\Omega_{\rm m}$ does not impact our findings in Section~\ref{sec:result} and we recover the same weak $\Omega_{\rm m}$ dependence for both central and satellite galaxy kinematics.
    } \label{fig:cen_cosmo_hmf_corrected}
\end{figure}

In Figure~\ref{fig:cen_cosmo_hmf_corrected}, we present the central galaxy kinematics (a) and satellite kinematics (b), \qrr{} (top) and \qvv{} (bottom) for different $\Omega_{\rm m}$ while varying $M_{\rm vir}$ to match the fiducial number density. We include the distributions for IllustrisTNG (red) and SIMBA (blue). Even after fixing number density, i.e., disentangling the potential dependence on $M_{\rm vir}$, there is only a slight dependence on $\Omega_{\rm m}$ for both central and satellite galaxy kinematics. This is consistent with our findings in Section~\ref{sec:result} and confirms that $\Omega_{\rm m}$ does not significantly impact the kinematics of both central and satellite galaxies.

\begin{figure}
    \centering
    \gridline{\fig{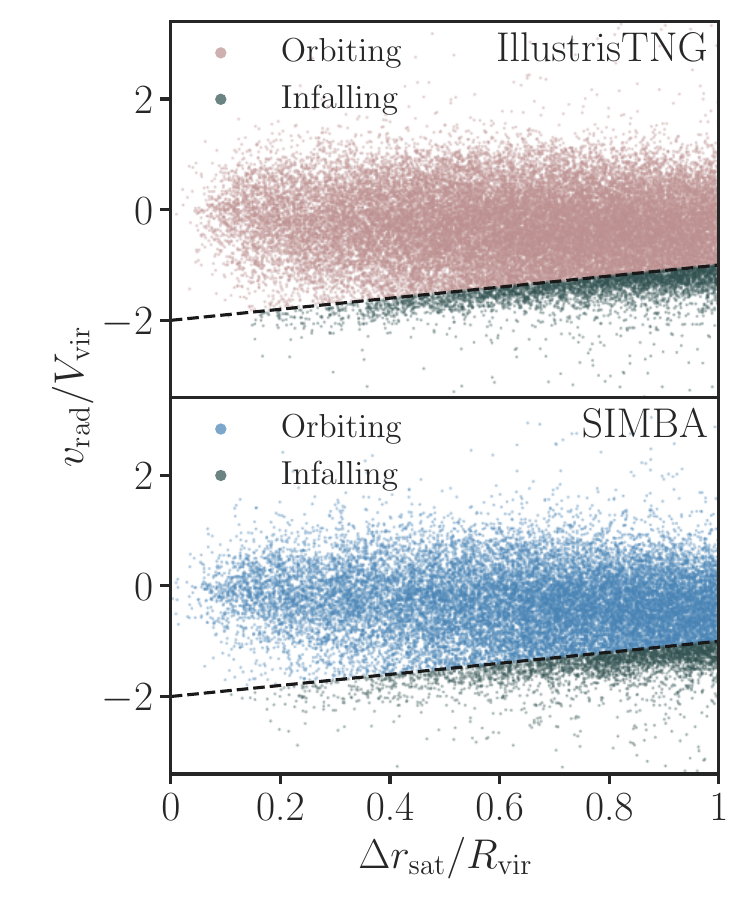}{0.33\textwidth}{(a)}
    \fig{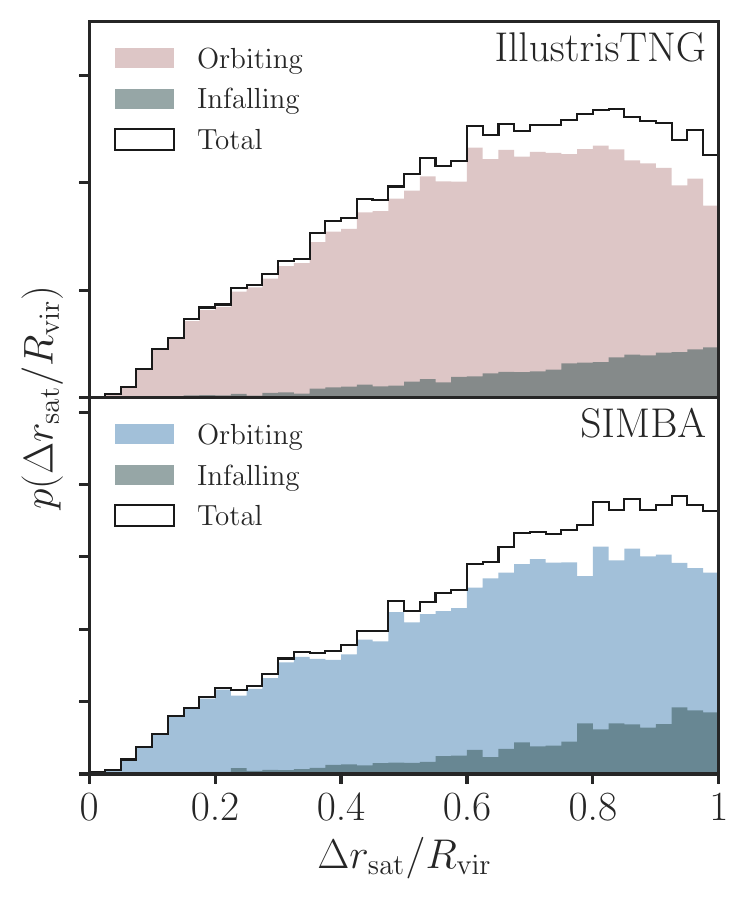}{0.33\textwidth}{(b)}
    \fig{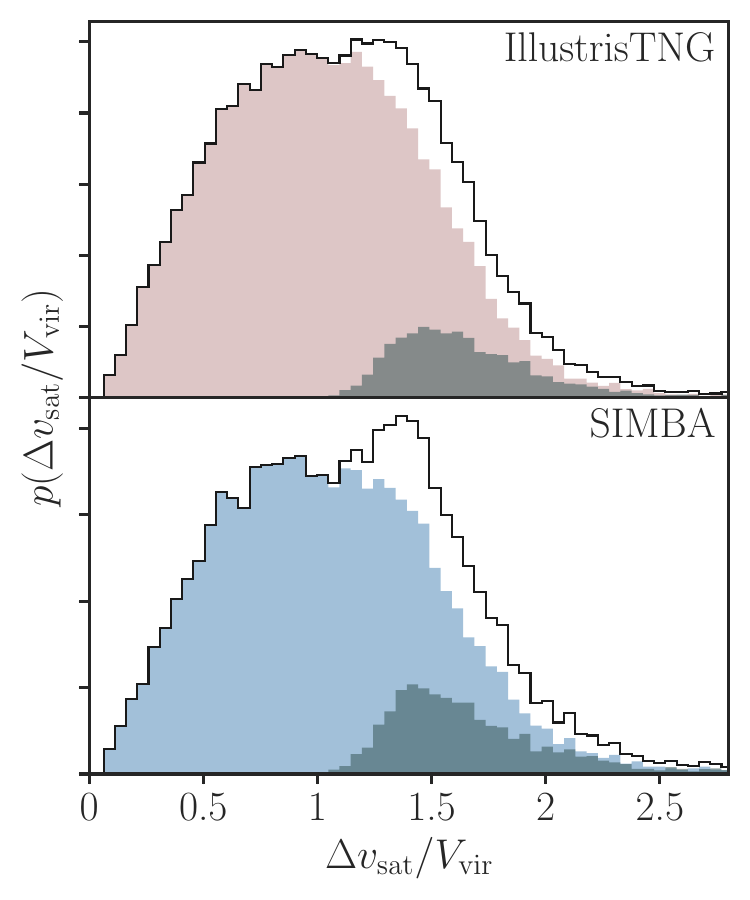}{0.33\textwidth}{(c)}}
    \caption{(a) Satellite radial velocity, $v_{\rm rad}/V_{\rm vir}$ distribution as a function of satellite radial position, $\Delta r_{\rm sat}/R_{\rm vir}$ for all satellites in IllustrisTNG (top) and SIMBA (bottom). Negative $v_{\rm rad}$ is toward the halo center. We classify infalling galaxies from orbiting using an approximate criterion (dashed line) derived from \cite{aung_infall}. (b) $p(\Delta r_{\rm sat}/R_{\rm vir})$ of the satellite galaxies. We present the total satellite population in black and the infalling population in dark green. Infalling satellites contributes significantly near $\Delta r_{\rm sat} = R_{\rm vir}$. (c) $p(\Delta v_{\rm sat}/V_{\rm vir})$  of the satellite galaxies. The bimodality in $p(\Delta v_{\rm sat}/V_{\rm vir})$ is primarily driven by infalling galaxies distributed around $\Delta v_{\rm sat} \approx 1.5 V_{\rm vir}$.
   }
    \label{fig:sat_vrad}
\end{figure}

\section{Infalling Satellites \label{app:infall}}
We consistently find a bimodality in \qvvs{} for both IllustrisTNG and SIMBA (Section~\ref{subsec:sat_kin}). One of the modes is centered at $\Delta {v}_{\rm sat}/V_{\rm vir} \approx 1$ while the other is centered at $\approx 1.5$. In this section, we examine whether these modes come from separate orbiting and infalling populations. 

We begin by selecting infalling galaxies using an approximate cut derived from the position of infalling galaxies on $v_{\rm rad}$--$\Delta r_{\rm sat}$ plane in \cite{aung_infall} (see Figure~2 therein): $v_{\rm rad}/V_{\rm vir} > \Delta r_{\rm sat}/R_{\rm vir} - 2$. We use this approximate selection as CAMELS snapshots have low time resolution so we cannot accurately trace the pericentric passages of our satellite galaxies. In Figure~\ref{fig:sat_vrad}a, we show the distribution of $\Delta r_{\rm sat}$ and $v_{\rm rad}$ for all IllustrisTNG (red) and SIMBA (blue) satellite galaxies with $M_\ast > 10^9M_\odot$. We mark our approximate infall cut (black dashed) and the infalling population in dark green.

In Figure~\ref{fig:sat_vrad}b and c, we show the satellite radial distribution and velocity offset distribution of the same IllustrisTNG (red) and SIMBA (blue) satellite galaxies. We denote the total satellite population in black and the infalling satellites in dark green. For the satellite radial distribution, we find that infalling satellites contribute significantly near $\Delta r_{\rm sat} = R_{\rm vir}$ and account for some of the differences between the IllustrisTNG and SIMBA distributions. Furthermore, for the velocity offset distribution, we find the second mode at $\Delta v_{\rm sat}/V_{\rm vir}\approx 1.5$ is primarily due to the infalling population. 

$\sim 10\%$ of the total satellite population consists of infalling satellites. Despite their significant contribution, infalling satellites are currently not included in standard HOD models. As we show in this section, the kinematics of infalling galaxies are distinct from the orbiting satellites. This suggests that excluding them in HOD models may significantly limit how accurately we can model small-scale galaxy clustering, especially in redshift space.

\section*{acknowledgement}
It is a pleasure to thank Andrew Hearin, Francisco Villaescusa-Navarro, Romain Teyssier, and Tjitske Starkenburg for discussions on the comprehensive aspects of hydrodynamic simulations as well as constructive feedback on the paper. The authors also benefited from detailed discussions with Chirag Modi, Christopher Lovell, John Wu, Peter Behroozi, Shy Genel, and Zixuan Peng. KJK acknowledges support from the National Science Foundation under Grant No. 2012086. CH was supported by the AI Accelerator program of the Schmidt Futures Foundation.

\bibliography{sample631}{}

\begin{thebibliography}{}
\expandafter\ifx\csname natexlab\endcsname\relax\def\natexlab#1{#1}\fi
\providecommand{\url}[1]{\href{#1}{#1}}
\providecommand{\dodoi}[1]{doi:~\href{http://doi.org/#1}{\nolinkurl{#1}}}
\providecommand{\doeprint}[1]{\href{http://ascl.net/#1}{\nolinkurl{http://ascl.net/#1}}}
\providecommand{\doarXiv}[1]{\href{https://arxiv.org/abs/#1}{\nolinkurl{https://arxiv.org/abs/#1}}}

\bibitem[{{Abazajian} {et~al.}(2005){Abazajian}, {Zheng}, {Zehavi}, {Weinberg}, {Frieman}, {Berlind}, {Blanton}, {Bahcall}, {Brinkmann}, {Schneider}, \& {Tegmark}}]{abazajian_05}
{Abazajian}, K., {Zheng}, Z., {Zehavi}, I., {et~al.} 2005, \apj, 625, 613, \dodoi{10.1086/429685}

\bibitem[{{Alam} {et~al.}(2017){Alam}, {Ata}, {Bailey}, {Beutler}, {Bizyaev}, {Blazek}, {Bolton}, {Brownstein}, {Burden}, {Chuang}, {Comparat}, {Cuesta}, {Dawson}, {Eisenstein}, {Escoffier}, {Gil-Mar{\'\i}n}, {Grieb}, {Hand}, {Ho}, {Kinemuchi}, {Kirkby}, {Kitaura}, {Malanushenko}, {Malanushenko}, {Maraston}, {McBride}, {Nichol}, {Olmstead}, {Oravetz}, {Padmanabhan}, {Palanque-Delabrouille}, {Pan}, {Pellejero-Ibanez}, {Percival}, {Petitjean}, {Prada}, {Price-Whelan}, {Reid}, {Rodr{\'\i}guez-Torres}, {Roe}, {Ross}, {Ross}, {Rossi}, {Rubi{\~n}o-Mart{\'\i}n}, {Saito}, {Salazar-Albornoz}, {Samushia}, {S{\'a}nchez}, {Satpathy}, {Schlegel}, {Schneider}, {Sc{\'o}ccola}, {Seo}, {Sheldon}, {Simmons}, {Slosar}, {Strauss}, {Swanson}, {Thomas}, {Tinker}, {Tojeiro}, {Maga{\~n}a}, {Vazquez}, {Verde}, {Wake}, {Wang}, {Weinberg}, {White}, {Wood-Vasey}, {Y{\`e}che}, {Zehavi}, {Zhai}, \& {Zhao}}]{alam_17}
{Alam}, S., {Ata}, M., {Bailey}, S., {et~al.} 2017, \mnras, 470, 2617, \dodoi{10.1093/mnras/stx721}

\bibitem[{{Alsing} {et~al.}(2019){Alsing}, {Charnock}, {Feeney}, \& {Wandelt}}]{pydelfi_2019}
{Alsing}, J., {Charnock}, T., {Feeney}, S., \& {Wandelt}, B. 2019, \mnras, 488, 4440, \dodoi{10.1093/mnras/stz1960}

\bibitem[{{Aung} {et~al.}(2023){Aung}, {Nagai}, {Rozo}, {Wolfe}, \& {Adhikari}}]{aung_infall}
{Aung}, H., {Nagai}, D., {Rozo}, E., {Wolfe}, B., \& {Adhikari}, S. 2023, \mnras, 521, 3981, \dodoi{10.1093/mnras/stad601}

\bibitem[{{Bailin} {et~al.}(2008){Bailin}, {Power}, {Norberg}, {Zaritsky}, \& {Gibson}}]{bailin_08_sdss}
{Bailin}, J., {Power}, C., {Norberg}, P., {Zaritsky}, D., \& {Gibson}, B.~K. 2008, \mnras, 390, 1133, \dodoi{10.1111/j.1365-2966.2008.13828.x}

\bibitem[{{Banerjee} \& {Abel}(2021)}]{banerjee2021_knn}
{Banerjee}, A., \& {Abel}, T. 2021, \mnras, 500, 5479, \dodoi{10.1093/mnras/staa3604}

\bibitem[{{Behroozi} {et~al.}(2019){Behroozi}, {Wechsler}, {Hearin}, \& {Conroy}}]{behroozi_2019}
{Behroozi}, P., {Wechsler}, R.~H., {Hearin}, A.~P., \& {Conroy}, C. 2019, \mnras, 488, 3143, \dodoi{10.1093/mnras/stz1182}

\bibitem[{{Behroozi} {et~al.}(2013){Behroozi}, {Wechsler}, \& {Wu}}]{behroozi_2013}
{Behroozi}, P.~S., {Wechsler}, R.~H., \& {Wu}, H.-Y. 2013, \apj, 762, 109, \dodoi{10.1088/0004-637X/762/2/109}

\bibitem[{{Benson} {et~al.}(2000){Benson}, {Cole}, {Frenk}, {Baugh}, \& {Lacey}}]{benson_00}
{Benson}, A.~J., {Cole}, S., {Frenk}, C.~S., {Baugh}, C.~M., \& {Lacey}, C.~G. 2000, \mnras, 311, 793, \dodoi{10.1046/j.1365-8711.2000.03101.x}

\bibitem[{{Berlind} \& {Weinberg}(2002)}]{berlind_2002}
{Berlind}, A.~A., \& {Weinberg}, D.~H. 2002, \apj, 575, 587, \dodoi{10.1086/341469}

\bibitem[{{Berlind} {et~al.}(2003){Berlind}, {Weinberg}, {Benson}, {Baugh}, {Cole}, {Dav{\'e}}, {Frenk}, {Jenkins}, {Katz}, \& {Lacey}}]{berlind_03_hydro_velbias}
{Berlind}, A.~A., {Weinberg}, D.~H., {Benson}, A.~J., {et~al.} 2003, \apj, 593, 1, \dodoi{10.1086/376517}

\bibitem[{{Bird} {et~al.}(2022){Bird}, {Ni}, {Di Matteo}, {Croft}, {Feng}, \& {Chen}}]{bird_2022}
{Bird}, S., {Ni}, Y., {Di Matteo}, T., {et~al.} 2022, \mnras, 512, 3703, \dodoi{10.1093/mnras/stac648}

\bibitem[{{Borzyszkowski} {et~al.}(2017){Borzyszkowski}, {Porciani}, {Romano-D{\'\i}az}, \& {Garaldi}}]{borzyszkowski_17}
{Borzyszkowski}, M., {Porciani}, C., {Romano-D{\'\i}az}, E., \& {Garaldi}, E. 2017, \mnras, 469, 594, \dodoi{10.1093/mnras/stx873}

\bibitem[{{Bryan} \& {Norman}(1998)}]{bryan_1998}
{Bryan}, G.~L., \& {Norman}, M.~L. 1998, \apj, 495, 80, \dodoi{10.1086/305262}

\bibitem[{{Budzynski} {et~al.}(2012){Budzynski}, {Koposov}, {McCarthy}, {McGee}, \& {Belokurov}}]{budzynski_12}
{Budzynski}, J.~M., {Koposov}, S.~E., {McCarthy}, I.~G., {McGee}, S.~L., \& {Belokurov}, V. 2012, \mnras, 423, 104, \dodoi{10.1111/j.1365-2966.2012.20663.x}

\bibitem[{{Castorina} \& {Sheth}(2013)}]{castorina2013}
{Castorina}, E., \& {Sheth}, R.~K. 2013, \mnras, 433, 1529, \dodoi{10.1093/mnras/stt824}

\bibitem[{{Chan} {et~al.}(2015){Chan}, {Kere{\v{s}}}, {O{\~n}orbe}, {Hopkins}, {Muratov}, {Faucher-Gigu{\`e}re}, \& {Quataert}}]{chan_2015}
{Chan}, T.~K., {Kere{\v{s}}}, D., {O{\~n}orbe}, J., {et~al.} 2015, \mnras, 454, 2981, \dodoi{10.1093/mnras/stv2165}

\bibitem[{{Chisari} {et~al.}(2019){Chisari}, {Mead}, {Joudaki}, {Ferreira}, {Schneider}, {Mohr}, {Tr{\"o}ster}, {Alonso}, {McCarthy}, {Martin-Alvarez}, {Devriendt}, {Slyz}, \& {van Daalen}}]{chisari2019_baryons}
{Chisari}, N.~E., {Mead}, A.~J., {Joudaki}, S., {et~al.} 2019, The Open Journal of Astrophysics, 2, 4, \dodoi{10.21105/astro.1905.06082}

\bibitem[{{Cole} {et~al.}(2005){Cole}, {Percival}, {Peacock}, {Norberg}, {Baugh}, {Frenk}, {Baldry}, {Bland-Hawthorn}, {Bridges}, {Cannon}, {Colless}, {Collins}, {Couch}, {Cross}, {Dalton}, {Eke}, {De Propris}, {Driver}, {Efstathiou}, {Ellis}, {Glazebrook}, {Jackson}, {Jenkins}, {Lahav}, {Lewis}, {Lumsden}, {Maddox}, {Madgwick}, {Peterson}, {Sutherland}, \& {Taylor}}]{cole_05}
{Cole}, S., {Percival}, W.~J., {Peacock}, J.~A., {et~al.} 2005, \mnras, 362, 505, \dodoi{10.1111/j.1365-2966.2005.09318.x}

\bibitem[{{Cranmer}(2023)}]{pysr}
{Cranmer}, M. 2023, arXiv e-prints, arXiv:2305.01582, \dodoi{10.48550/arXiv.2305.01582}

\bibitem[{{Cranmer} {et~al.}(2019){Cranmer}, {Xu}, {Battaglia}, \& {Ho}}]{cranmer2019}
{Cranmer}, M.~D., {Xu}, R., {Battaglia}, P., \& {Ho}, S. 2019, arXiv e-prints, arXiv:1909.05862, \dodoi{10.48550/arXiv.1909.05862}

\bibitem[{{Croton} {et~al.}(2007){Croton}, {Gao}, \& {White}}]{croton_07}
{Croton}, D.~J., {Gao}, L., \& {White}, S. D.~M. 2007, \mnras, 374, 1303, \dodoi{10.1111/j.1365-2966.2006.11230.x}

\bibitem[{{Dai} \& {Seljak}(2022)}]{dai_seljak_22}
{Dai}, B., \& {Seljak}, U. 2022, \mnras, 516, 2363, \dodoi{10.1093/mnras/stac2010}

\bibitem[{Dav\'e {et~al.}(2019)Dav\'e, Angl\'es-Alc\'azar, Narayanan, Li, Rafieferantsoa, \& Appleby}]{dave_2019}
Dav\'e, R., Angl\'es-Alc\'azar, D., Narayanan, D., {et~al.} 2019, Mon. Not. Roy. Astron. Soc., 486, 2827, \dodoi{10.1093/mnras/stz937}

\bibitem[{{Dawson} {et~al.}(2013{\natexlab{a}}){Dawson}, {Schlegel}, {Ahn}, {Anderson}, {Aubourg}, {Bailey}, {Barkhouser}, {Bautista}, {Beifiori}, {Berlind}, {Bhardwaj}, {Bizyaev}, {Blake}, {Blanton}, {Blomqvist}, {Bolton}, {Borde}, {Bovy}, {Brandt}, {Brewington}, {Brinkmann}, {Brown}, {Brownstein}, {Bundy}, {Busca}, {Carithers}, {Carnero}, {Carr}, {Chen}, {Comparat}, {Connolly}, {Cope}, {Croft}, {Cuesta}, {da Costa}, {Davenport}, {Delubac}, {de Putter}, {Dhital}, {Ealet}, {Ebelke}, {Eisenstein}, {Escoffier}, {Fan}, {Filiz Ak}, {Finley}, {Font-Ribera}, {G{\'e}nova-Santos}, {Gunn}, {Guo}, {Haggard}, {Hall}, {Hamilton}, {Harris}, {Harris}, {Ho}, {Hogg}, {Holder}, {Honscheid}, {Huehnerhoff}, {Jordan}, {Jordan}, {Kauffmann}, {Kazin}, {Kirkby}, {Klaene}, {Kneib}, {Le Goff}, {Lee}, {Long}, {Loomis}, {Lundgren}, {Lupton}, {Maia}, {Makler}, {Malanushenko}, {Malanushenko}, {Mandelbaum}, {Manera}, {Maraston}, {Margala}, {Masters}, {McBride}, {McDonald}, {McGreer}, {McMahon}, {Mena}, {Miralda-Escud{\'e}},
  {Montero-Dorta}, {Montesano}, {Muna}, {Myers}, {Naugle}, {Nichol}, {Noterdaeme}, {Nuza}, {Olmstead}, {Oravetz}, {Oravetz}, {Owen}, {Padmanabhan}, {Palanque-Delabrouille}, {Pan}, {Parejko}, {P{\^a}ris}, {Percival}, {P{\'e}rez-Fournon}, {P{\'e}rez-R{\`a}fols}, {Petitjean}, {Pfaffenberger}, {Pforr}, {Pieri}, {Prada}, {Price-Whelan}, {Raddick}, {Rebolo}, {Rich}, {Richards}, {Rockosi}, {Roe}, {Ross}, {Ross}, {Rossi}, {Rubi{\~n}o-Martin}, {Samushia}, {S{\'a}nchez}, {Sayres}, {Schmidt}, {Schneider}, {Sc{\'o}ccola}, {Seo}, {Shelden}, {Sheldon}, {Shen}, {Shu}, {Slosar}, {Smee}, {Snedden}, {Stauffer}, {Steele}, {Strauss}, {Streblyanska}, {Suzuki}, {Swanson}, {Tal}, {Tanaka}, {Thomas}, {Tinker}, {Tojeiro}, {Tremonti}, {Vargas Maga{\~n}a}, {Verde}, {Viel}, {Wake}, {Watson}, {Weaver}, {Weinberg}, {Weiner}, {West}, {White}, {Wood-Vasey}, {Yeche}, {Zehavi}, {Zhao}, \& {Zheng}}]{boss}
{Dawson}, K.~S., {Schlegel}, D.~J., {Ahn}, C.~P., {et~al.} 2013{\natexlab{a}}, \aj, 145, 10, \dodoi{10.1088/0004-6256/145/1/10}

\bibitem[{{Dawson} {et~al.}(2013{\natexlab{b}}){Dawson}, {Schlegel}, {Ahn}, {Anderson}, {Aubourg}, {Bailey}, {Barkhouser}, {Bautista}, {Beifiori}, {Berlind}, {Bhardwaj}, {Bizyaev}, {Blake}, {Blanton}, {Blomqvist}, {Bolton}, {Borde}, {Bovy}, {Brandt}, {Brewington}, {Brinkmann}, {Brown}, {Brownstein}, {Bundy}, {Busca}, {Carithers}, {Carnero}, {Carr}, {Chen}, {Comparat}, {Connolly}, {Cope}, {Croft}, {Cuesta}, {da Costa}, {Davenport}, {Delubac}, {de Putter}, {Dhital}, {Ealet}, {Ebelke}, {Eisenstein}, {Escoffier}, {Fan}, {Filiz Ak}, {Finley}, {Font-Ribera}, {G{\'e}nova-Santos}, {Gunn}, {Guo}, {Haggard}, {Hall}, {Hamilton}, {Harris}, {Harris}, {Ho}, {Hogg}, {Holder}, {Honscheid}, {Huehnerhoff}, {Jordan}, {Jordan}, {Kauffmann}, {Kazin}, {Kirkby}, {Klaene}, {Kneib}, {Le Goff}, {Lee}, {Long}, {Loomis}, {Lundgren}, {Lupton}, {Maia}, {Makler}, {Malanushenko}, {Malanushenko}, {Mandelbaum}, {Manera}, {Maraston}, {Margala}, {Masters}, {McBride}, {McDonald}, {McGreer}, {McMahon}, {Mena}, {Miralda-Escud{\'e}},
  {Montero-Dorta}, {Montesano}, {Muna}, {Myers}, {Naugle}, {Nichol}, {Noterdaeme}, {Nuza}, {Olmstead}, {Oravetz}, {Oravetz}, {Owen}, {Padmanabhan}, {Palanque-Delabrouille}, {Pan}, {Parejko}, {P{\^a}ris}, {Percival}, {P{\'e}rez-Fournon}, {P{\'e}rez-R{\`a}fols}, {Petitjean}, {Pfaffenberger}, {Pforr}, {Pieri}, {Prada}, {Price-Whelan}, {Raddick}, {Rebolo}, {Rich}, {Richards}, {Rockosi}, {Roe}, {Ross}, {Ross}, {Rossi}, {Rubi{\~n}o-Martin}, {Samushia}, {S{\'a}nchez}, {Sayres}, {Schmidt}, {Schneider}, {Sc{\'o}ccola}, {Seo}, {Shelden}, {Sheldon}, {Shen}, {Shu}, {Slosar}, {Smee}, {Snedden}, {Stauffer}, {Steele}, {Strauss}, {Streblyanska}, {Suzuki}, {Swanson}, {Tal}, {Tanaka}, {Thomas}, {Tinker}, {Tojeiro}, {Tremonti}, {Vargas Maga{\~n}a}, {Verde}, {Viel}, {Wake}, {Watson}, {Weaver}, {Weinberg}, {Weiner}, {West}, {White}, {Wood-Vasey}, {Yeche}, {Zehavi}, {Zhao}, \& {Zheng}}]{dawson_13}
---. 2013{\natexlab{b}}, \aj, 145, 10, \dodoi{10.1088/0004-6256/145/1/10}

\bibitem[{{Dawson} {et~al.}(2016){Dawson}, {Kneib}, {Percival}, {Alam}, {Albareti}, {Anderson}, {Armengaud}, {Aubourg}, {Bailey}, {Bautista}, {Berlind}, {Bershady}, {Beutler}, {Bizyaev}, {Blanton}, {Blomqvist}, {Bolton}, {Bovy}, {Brandt}, {Brinkmann}, {Brownstein}, {Burtin}, {Busca}, {Cai}, {Chuang}, {Clerc}, {Comparat}, {Cope}, {Croft}, {Cruz-Gonzalez}, {da Costa}, {Cousinou}, {Darling}, {de la Macorra}, {de la Torre}, {Delubac}, {du Mas des Bourboux}, {Dwelly}, {Ealet}, {Eisenstein}, {Eracleous}, {Escoffier}, {Fan}, {Finoguenov}, {Font-Ribera}, {Frinchaboy}, {Gaulme}, {Georgakakis}, {Green}, {Guo}, {Guy}, {Ho}, {Holder}, {Huehnerhoff}, {Hutchinson}, {Jing}, {Jullo}, {Kamble}, {Kinemuchi}, {Kirkby}, {Kitaura}, {Klaene}, {Laher}, {Lang}, {Laurent}, {Le Goff}, {Li}, {Liang}, {Lima}, {Lin}, {Lin}, {Lin}, {Long}, {Lundgren}, {MacDonald}, {Geimba Maia}, {Malanushenko}, {Malanushenko}, {Mariappan}, {McBride}, {McGreer}, {M{\'e}nard}, {Merloni}, {Meza}, {Montero-Dorta}, {Muna}, {Myers}, {Nandra}, {Naugle},
  {Newman}, {Noterdaeme}, {Nugent}, {Ogando}, {Olmstead}, {Oravetz}, {Oravetz}, {Padmanabhan}, {Palanque-Delabrouille}, {Pan}, {Parejko}, {P{\^a}ris}, {Peacock}, {Petitjean}, {Pieri}, {Pisani}, {Prada}, {Prakash}, {Raichoor}, {Reid}, {Rich}, {Ridl}, {Rodriguez-Torres}, {Carnero Rosell}, {Ross}, {Rossi}, {Ruan}, {Salvato}, {Sayres}, {Schneider}, {Schlegel}, {Seljak}, {Seo}, {Sesar}, {Shandera}, {Shu}, {Slosar}, {Sobreira}, {Streblyanska}, {Suzuki}, {Taylor}, {Tao}, {Tinker}, {Tojeiro}, {Vargas-Maga{\~n}a}, {Wang}, {Weaver}, {Weinberg}, {White}, {Wood-Vasey}, {Yeche}, {Zhai}, {Zhao}, {Zhao}, {Zheng}, {Ben Zhu}, \& {Zou}}]{eboss}
{Dawson}, K.~S., {Kneib}, J.-P., {Percival}, W.~J., {et~al.} 2016, \aj, 151, 44, \dodoi{10.3847/0004-6256/151/2/44}

\bibitem[{{Delgado} {et~al.}(2022){Delgado}, {Wadekar}, {Hadzhiyska}, {Bose}, {Hernquist}, \& {Ho}}]{delgado_22}
{Delgado}, A.~M., {Wadekar}, D., {Hadzhiyska}, B., {et~al.} 2022, \mnras, 515, 2733, \dodoi{10.1093/mnras/stac1951}

\bibitem[{{DESI Collaboration} {et~al.}(2022){DESI Collaboration}, {Abareshi}, {Aguilar}, {Ahlen}, {Alam}, {Alexander}, {Alfarsy}, {Allen}, {Allende Prieto}, {Alves}, {Ameel}, {Armengaud}, {Asorey}, {Aviles}, {Bailey}, {Balaguera-Antol{\'\i}nez}, {Ballester}, {Baltay}, {Bault}, {Beltran}, {Benavides}, {BenZvi}, {Berti}, {Besuner}, {Beutler}, {Bianchi}, {Blake}, {Blanc}, {Blum}, {Bolton}, {Bose}, {Bramall}, {Brieden}, {Brodzeller}, {Brooks}, {Brownewell}, {Buckley-Geer}, {Cahn}, {Cai}, {Canning}, {Capasso}, {Carnero Rosell}, {Carton}, {Casas}, {Castander}, {Cervantes-Cota}, {Chabanier}, {Chaussidon}, {Chuang}, {Circosta}, {Cole}, {Cooper}, {da Costa}, {Cousinou}, {Cuceu}, {Davis}, {Dawson}, {de la Cruz-Noriega}, {de la Macorra}, {de Mattia}, {Della Costa}, {Demmer}, {Derwent}, {Dey}, {Dey}, {Dhungana}, {Ding}, {Dobson}, {Doel}, {Donald-McCann}, {Donaldson}, {Douglass}, {Duan}, {Dunlop}, {Edelstein}, {Eftekharzadeh}, {Eisenstein}, {Enriquez-Vargas}, {Escoffier}, {Evatt}, {Fagrelius}, {Fan}, {Fanning},
  {Fawcett}, {Ferraro}, {Ereza}, {Flaugher}, {Font-Ribera}, {Forero-Romero}, {Frenk}, {Fromenteau}, {G{\"a}nsicke}, {Garcia-Quintero}, {Garrison}, {Gazta{\~n}aga}, {Gerardi}, {Gil-Mar{\'\i}n}, {Gontcho a Gontcho}, {Gonzalez-Morales}, {Gonzalez-de-Rivera}, {Gonzalez-Perez}, {Gordon}, {Graur}, {Green}, {Grove}, {Gruen}, {Gutierrez}, {Guy}, {Hahn}, {Harris}, {Herrera}, {Herrera-Alcantar}, {Honscheid}, {Howlett}, {Huterer}, {Ir{\v{s}}i{\v{c}}}, {Ishak}, {Jelinsky}, {Jiang}, {Jimenez}, {Jing}, {Joyce}, {Jullo}, {Juneau}, {Kara{\c{c}}ayl{\i}}, {Karamanis}, {Karcher}, {Karim}, {Kehoe}, {Kent}, {Kirkby}, {Kisner}, {Kitaura}, {Koposov}, {Kov{\'a}cs}, {Kremin}, {Krolewski}, {L'Huillier}, {Lahav}, {Lambert}, {Lamman}, {Lan}, {Landriau}, {Lane}, {Lang}, {Lange}, {Lasker}, {Le Guillou}, {Leauthaud}, {Le Van Suu}, {Levi}, {Li}, {Magneville}, {Manera}, {Manser}, {Marshall}, {Martini}, {McCollam}, {McDonald}, {Meisner}, {Mena-Fern{\'a}ndez}, {Meneses-Rizo}, {Mezcua}, {Miller}, {Miquel}, {Montero-Camacho}, {Moon},
  {Moustakas}, {Mueller}, {Mu{\~n}oz-Guti{\'e}rrez}, {Myers}, {Nadathur}, {Najita}, {Napolitano}, {Neilsen}, {Newman}, {Nie}, {Ning}, {Niz}, {Norberg}, {Noriega}, {O'Brien}, {Obuljen}, {Palanque-Delabrouille}, {Palmese}, {Zhiwei}, {Pappalardo}, {PENG}, {Percival}, {Perruchot}, {Pogge}, {Poppett}, {Porredon}, {Prada}, {Prochaska}, {Pucha}, {P{\'e}rez-Fern{\'a}ndez}, {P{\'e}rez-R{\`a}fols}, {Rabinowitz}, {Raichoor}, {Ramirez-Solano}, {Ram{\'\i}rez-P{\'e}rez}, {Ravoux}, {Reil}, {Rezaie}, {Rocher}, {Rockosi}, {Roe}, {Roodman}, {Ross}, {Rossi}, {Ruggeri}, {Ruhlmann-Kleider}, {Sabiu}, {Safonova}, {Said}, {Saintonge}, {Salas Catonga}, {Samushia}, {Sanchez}, {Saulder}, {Schaan}, {Schlafly}, {Schlegel}, {Schmoll}, {Scholte}, {Schubnell}, {Secroun}, {Seo}, {Serrano}, {Sharples}, {Sholl}, {Silber}, {Silva}, {Sirk}, {Siudek}, {Smith}, {Sprayberry}, {Staten}, {Stupak}, {Tan}, {Tarl{\'e}}, {Tie}, {Tojeiro}, {Ure{\~n}a-L{\'o}pez}, {Valdes}, {Valenzuela}, {Valluri}, {Vargas-Maga{\~n}a}, {Verde}, {Walther}, {Wang}, {Wang},
  {Weaver}, {Weaverdyck}, {Wechsler}, {Wilson}, {Yang}, {Yu}, {Yuan}, {Y{\`e}che}, {Zhang}, {Zhang}, {Zhao}, {Zhou}, {Zhou}, {Zou}, {Zou}, {Zou}, {Zu}, \& {DESI Collaboration}}]{desi}
{DESI Collaboration}, {Abareshi}, B., {Aguilar}, J., {et~al.} 2022, \aj, 164, 207, \dodoi{10.3847/1538-3881/ac882b}

\bibitem[{{Diemer} {et~al.}(2023){Diemer}, {Behroozi}, \& {Mansfield}}]{diemer_23}
{Diemer}, B., {Behroozi}, P., \& {Mansfield}, P. 2023, arXiv e-prints, arXiv:2305.00993, \dodoi{10.48550/arXiv.2305.00993}

\bibitem[{{Durkan} {et~al.}(2019){Durkan}, {Bekasov}, {Murray}, \& {Papamakarios}}]{durkan_19}
{Durkan}, C., {Bekasov}, A., {Murray}, I., \& {Papamakarios}, G. 2019, arXiv e-prints, arXiv:1906.04032, \dodoi{10.48550/arXiv.1906.04032}

\bibitem[{Durkan {et~al.}(2020)Durkan, Bekasov, Murray, \& Papamakarios}]{nflows}
Durkan, C., Bekasov, A., Murray, I., \& Papamakarios, G. 2020, {nflows}: normalizing flows in {PyTorch}, v0.14,  Zenodo, \dodoi{10.5281/zenodo.4296287}

\bibitem[{{Eickenberg} {et~al.}(2022){Eickenberg}, {Allys}, {Moradinezhad Dizgah}, {Lemos}, {Massara}, {Abidi}, {Hahn}, {Hassan}, {Regaldo-Saint Blancard}, {Ho}, {Mallat}, {And{\'e}n}, \& {Villaescusa-Navarro}}]{eickenberg_22_wavelet}
{Eickenberg}, M., {Allys}, E., {Moradinezhad Dizgah}, A., {et~al.} 2022, arXiv e-prints, arXiv:2204.07646, \dodoi{10.48550/arXiv.2204.07646}

\bibitem[{{Eisenstein} {et~al.}(2001){Eisenstein}, {Annis}, {Gunn}, {Szalay}, {Connolly}, {Nichol}, {Bahcall}, {Bernardi}, {Burles}, {Castander}, {Fukugita}, {Hogg}, {Ivezi{\'c}}, {Knapp}, {Lupton}, {Narayanan}, {Postman}, {Reichart}, {Richmond}, {Schneider}, {Schlegel}, {Strauss}, {SubbaRao}, {Tucker}, {Vanden Berk}, {Vogeley}, {Weinberg}, \& {Yanny}}]{sdss_lrg_2001}
{Eisenstein}, D.~J., {Annis}, J., {Gunn}, J.~E., {et~al.} 2001, \aj, 122, 2267, \dodoi{10.1086/323717}

\bibitem[{{Foreman} {et~al.}(2020){Foreman}, {Coulton}, {Villaescusa-Navarro}, \& {Barreira}}]{foreman_2020}
{Foreman}, S., {Coulton}, W., {Villaescusa-Navarro}, F., \& {Barreira}, A. 2020, \mnras, 498, 2887, \dodoi{10.1093/mnras/staa2523}

\bibitem[{{Germain} {et~al.}(2015){Germain}, {Gregor}, {Murray}, \& {Larochelle}}]{made_2015}
{Germain}, M., {Gregor}, K., {Murray}, I., \& {Larochelle}, H. 2015, arXiv e-prints, arXiv:1502.03509, \dodoi{10.48550/arXiv.1502.03509}

\bibitem[{{Gill} {et~al.}(2005){Gill}, {Knebe}, \& {Gibson}}]{gill_2005}
{Gill}, S. P.~D., {Knebe}, A., \& {Gibson}, B.~K. 2005, \mnras, 356, 1327, \dodoi{10.1111/j.1365-2966.2004.08562.x}

\bibitem[{{Greenberg} {et~al.}(2019){Greenberg}, {Nonnenmacher}, \& {Macke}}]{greenberg_19}
{Greenberg}, D.~S., {Nonnenmacher}, M., \& {Macke}, J.~H. 2019, arXiv e-prints, arXiv:1905.07488, \dodoi{10.48550/arXiv.1905.07488}

\bibitem[{{Gualdi} {et~al.}(2021){Gualdi}, {Gil-Mar{\'\i}n}, \& {Verde}}]{gualdi_21}
{Gualdi}, D., {Gil-Mar{\'\i}n}, H., \& {Verde}, L. 2021, \jcap, 2021, 008, \dodoi{10.1088/1475-7516/2021/07/008}

\bibitem[{{Guo} {et~al.}(2015{\natexlab{a}}){Guo}, {Zheng}, {Zehavi}, {Behroozi}, {Chuang}, {Comparat}, {Favole}, {Gottloeber}, {Klypin}, {Prada}, {Weinberg}, \& {Yepes}}]{guo_15a}
{Guo}, H., {Zheng}, Z., {Zehavi}, I., {et~al.} 2015{\natexlab{a}}, \mnras, 453, 4368, \dodoi{10.1093/mnras/stv1966}

\bibitem[{{Guo} {et~al.}(2015{\natexlab{b}}){Guo}, {Zheng}, {Zehavi}, {Dawson}, {Skibba}, {Tinker}, {Weinberg}, {White}, \& {Schneider}}]{guo_15b}
---. 2015{\natexlab{b}}, \mnras, 446, 578, \dodoi{10.1093/mnras/stu2120}

\bibitem[{{Hadzhiyska} {et~al.}(2021){Hadzhiyska}, {Liu}, {Somerville}, {Gabrielpillai}, {Bose}, {Eisenstein}, \& {Hernquist}}]{hadzhiyska_21_gal_assembly}
{Hadzhiyska}, B., {Liu}, S., {Somerville}, R.~S., {et~al.} 2021, \mnras, 508, 698, \dodoi{10.1093/mnras/stab2564}

\bibitem[{{Hadzhiyska} {et~al.}(2022){Hadzhiyska}, {Eisenstein}, {Hernquist}, {Pakmor}, {Bose}, {Delgado}, {Contreras}, {Kannan}, {White}, {Springel}, {Frenk}, {Hern{\'a}ndez-Aguayo}, {Ferlito}, \& {Barrera}}]{hadzhisyksa_22_two_halo}
{Hadzhiyska}, B., {Eisenstein}, D., {Hernquist}, L., {et~al.} 2022, arXiv e-prints, arXiv:2210.10072, \dodoi{10.48550/arXiv.2210.10072}

\bibitem[{{Hadzhiyska} {et~al.}(2023){Hadzhiyska}, {Hernquist}, {Eisenstein}, {Delgado}, {Bose}, {Kannan}, {Pakmor}, {Springel}, {Contreras}, {Barrera}, {Ferlito}, {Hern{\'a}ndez-Aguayo}, {White}, \& {Frenk}}]{hadzhiyska_2022_one_halo}
{Hadzhiyska}, B., {Hernquist}, L., {Eisenstein}, D., {et~al.} 2023, \mnras, 524, 2524, \dodoi{10.1093/mnras/stad279}

\bibitem[{{Hahn} \& {Melchior}(2022)}]{sedflow_2022}
{Hahn}, C., \& {Melchior}, P. 2022, \apj, 938, 11, \dodoi{10.3847/1538-4357/ac7b84}

\bibitem[{{Hahn} \& {Villaescusa-Navarro}(2021{\natexlab{a}})}]{Molino}
{Hahn}, C., \& {Villaescusa-Navarro}, F. 2021{\natexlab{a}}, \jcap, 2021, 029, \dodoi{10.1088/1475-7516/2021/04/029}

\bibitem[{{Hahn} \& {Villaescusa-Navarro}(2021{\natexlab{b}})}]{hahn2021}
---. 2021{\natexlab{b}}, \jcap, 2021, 029, \dodoi{10.1088/1475-7516/2021/04/029}

\bibitem[{{Hahn} {et~al.}(2022){Hahn}, {Eickenberg}, {Ho}, {Hou}, {Lemos}, {Massara}, {Modi}, {Moradinezhad Dizgah}, {R{\'e}galdo-Saint Blancard}, \& {Abidi}}]{simbig2022_plk}
{Hahn}, C., {Eickenberg}, M., {Ho}, S., {et~al.} 2022, arXiv e-prints, arXiv:2211.00723, \dodoi{10.48550/arXiv.2211.00723}

\bibitem[{{Hahn} {et~al.}(2023{\natexlab{a}}){Hahn}, {Lemos}, {Parker}, {R{\'e}galdo-Saint Blancard}, {Eickenberg}, {Ho}, {Hou}, {Massara}, {Modi}, {Moradinezhad Dizgah}, \& {Spergel}}]{simbig_wave2}
{Hahn}, C., {Lemos}, P., {Parker}, L., {et~al.} 2023{\natexlab{a}}, arXiv e-prints, arXiv:2310.15246.
\newblock \doarXiv{2310.15246}

\bibitem[{{Hahn} {et~al.}(2023{\natexlab{b}}){Hahn}, {Eickenberg}, {Ho}, {Hou}, {Lemos}, {Massara}, {Modi}, {Moradinezhad Dizgah}, {Parker}, \& {R{\'e}galdo-Saint Blancard}}]{simbig_bk}
{Hahn}, C., {Eickenberg}, M., {Ho}, S., {et~al.} 2023{\natexlab{b}}, arXiv e-prints, arXiv:2310.15243.
\newblock \doarXiv{2310.15243}

\bibitem[{{Hahn} {et~al.}(2023{\natexlab{c}}){Hahn}, {Wilson}, {Ruiz-Macias}, {Cole}, {Weinberg}, {Moustakas}, {Kremin}, {Tinker}, {Smith}, {Wechsler}, {Ahlen}, {Alam}, {Bailey}, {Brooks}, {Cooper}, {Davis}, {Dawson}, {Dey}, {Dey}, {Eftekharzadeh}, {Eisenstein}, {Fanning}, {Forero-Romero}, {Frenk}, {Gazta{\~n}aga}, {Gontcho A Gontcho}, {Guy}, {Honscheid}, {Ishak}, {Juneau}, {Kehoe}, {Kisner}, {Lan}, {Landriau}, {Le Guillou}, {Levi}, {Magneville}, {Martini}, {Meisner}, {Myers}, {Nie}, {Norberg}, {Palanque-Delabrouille}, {Percival}, {Poppett}, {Prada}, {Raichoor}, {Ross}, {Safonova}, {Saulder}, {Schlafly}, {Schlegel}, {Sierra-Porta}, {Tarle}, {Weaver}, {Y{\`e}che}, {Zarrouk}, {Zhou}, {Zhou}, \& {Zou}}]{desi_bgs_target}
{Hahn}, C., {Wilson}, M.~J., {Ruiz-Macias}, O., {et~al.} 2023{\natexlab{c}}, \aj, 165, 253, \dodoi{10.3847/1538-3881/accff8}

\bibitem[{{Hahn} {et~al.}(2007){Hahn}, {Carollo}, {Porciani}, \& {Dekel}}]{hahn_2007}
{Hahn}, O., {Carollo}, C.~M., {Porciani}, C., \& {Dekel}, A. 2007, \mnras, 381, 41, \dodoi{10.1111/j.1365-2966.2007.12249.x}

\bibitem[{{Hahn} {et~al.}(2009){Hahn}, {Porciani}, {Dekel}, \& {Carollo}}]{hahn_2009}
{Hahn}, O., {Porciani}, C., {Dekel}, A., \& {Carollo}, C.~M. 2009, \mnras, 398, 1742, \dodoi{10.1111/j.1365-2966.2009.15271.x}

\bibitem[{{Hassan} {et~al.}(2022){Hassan}, {Villaescusa-Navarro}, {Wandelt}, {Spergel}, {Angl{\'e}s-Alc{\'a}zar}, {Genel}, {Cranmer}, {Bryan}, {Dav{\'e}}, {Somerville}, {Eickenberg}, {Narayanan}, {Ho}, \& {Andrianomena}}]{hiflow}
{Hassan}, S., {Villaescusa-Navarro}, F., {Wandelt}, B., {et~al.} 2022, \apj, 937, 83, \dodoi{10.3847/1538-4357/ac8b09}

\bibitem[{{Hearin} \& {Watson}(2013)}]{hearin_13_age}
{Hearin}, A.~P., \& {Watson}, D.~F. 2013, \mnras, 435, 1313, \dodoi{10.1093/mnras/stt1374}

\bibitem[{{Hearin} {et~al.}(2016){Hearin}, {Zentner}, {van den Bosch}, {Campbell}, \& {Tollerud}}]{hearin_16}
{Hearin}, A.~P., {Zentner}, A.~R., {van den Bosch}, F.~C., {Campbell}, D., \& {Tollerud}, E. 2016, \mnras, 460, 2552, \dodoi{10.1093/mnras/stw840}

\bibitem[{{Hellwing} {et~al.}(2016){Hellwing}, {Schaller}, {Frenk}, {Theuns}, {Schaye}, {Bower}, \& {Crain}}]{hellwing_2016}
{Hellwing}, W.~A., {Schaller}, M., {Frenk}, C.~S., {et~al.} 2016, \mnras, 461, L11, \dodoi{10.1093/mnrasl/slw081}

\bibitem[{{Hirschmann} {et~al.}(2014){Hirschmann}, {Dolag}, {Saro}, {Bachmann}, {Borgani}, \& {Burkert}}]{hirschmann2014_magneticum}
{Hirschmann}, M., {Dolag}, K., {Saro}, A., {et~al.} 2014, \mnras, 442, 2304, \dodoi{10.1093/mnras/stu1023}

\bibitem[{{Hou} {et~al.}(2023){Hou}, {Moradinezhad Dizgah}, {Hahn}, \& {Massara}}]{hou_2023}
{Hou}, J., {Moradinezhad Dizgah}, A., {Hahn}, C., \& {Massara}, E. 2023, \jcap, 2023, 045, \dodoi{10.1088/1475-7516/2023/03/045}

\bibitem[{{Huterer}(2023)}]{huterer2023_growth}
{Huterer}, D. 2023, \aapr, 31, 2, \dodoi{10.1007/s00159-023-00147-4}

\bibitem[{{Ivanov} \& {Philcox}(2023)}]{ivanov2023_H0chapter}
{Ivanov}, M.~M., \& {Philcox}, O. H.~E. 2023, arXiv e-prints, arXiv:2305.07977, \dodoi{10.48550/arXiv.2305.07977}

\bibitem[{{Ivanov} {et~al.}(2020){Ivanov}, {Simonovi{\'c}}, \& {Zaldarriaga}}]{ivanov_2020}
{Ivanov}, M.~M., {Simonovi{\'c}}, M., \& {Zaldarriaga}, M. 2020, \jcap, 2020, 042, \dodoi{10.1088/1475-7516/2020/05/042}

\bibitem[{{Jespersen} {et~al.}(2022){Jespersen}, {Cranmer}, {Melchior}, {Ho}, {Somerville}, \& {Gabrielpillai}}]{jespersen2022_gnn}
{Jespersen}, C.~K., {Cranmer}, M., {Melchior}, P., {et~al.} 2022, \apj, 941, 7, \dodoi{10.3847/1538-4357/ac9b18}

\bibitem[{Kingma \& Ba(2017)}]{kingma2017adam}
Kingma, D.~P., \& Ba, J. 2017, Adam: A Method for Stochastic Optimization.
\newblock \doarXiv{1412.6980}

\bibitem[{{Klypin} {et~al.}(1999){Klypin}, {Gottl{\"o}ber}, {Kravtsov}, \& {Khokhlov}}]{klypin_99_overmerging}
{Klypin}, A., {Gottl{\"o}ber}, S., {Kravtsov}, A.~V., \& {Khokhlov}, A.~M. 1999, \apj, 516, 530, \dodoi{10.1086/307122}

\bibitem[{{Klypin} {et~al.}(2016){Klypin}, {Yepes}, {Gottl{\"o}ber}, {Prada}, \& {He{\ss}}}]{klypin_16}
{Klypin}, A., {Yepes}, G., {Gottl{\"o}ber}, S., {Prada}, F., \& {He{\ss}}, S. 2016, \mnras, 457, 4340, \dodoi{10.1093/mnras/stw248}

\bibitem[{{Kobayashi} {et~al.}(2022){Kobayashi}, {Nishimichi}, {Takada}, \& {Miyatake}}]{kobayashi_23_sdss3}
{Kobayashi}, Y., {Nishimichi}, T., {Takada}, M., \& {Miyatake}, H. 2022, \prd, 105, 083517, \dodoi{10.1103/PhysRevD.105.083517}

\bibitem[{{Lange} {et~al.}(2023){Lange}, {Hearin}, {Leauthaud}, {van den Bosch}, {Xhakaj}, {Guo}, {Wechsler}, \& {DeRose}}]{lange2023}
{Lange}, J.~U., {Hearin}, A.~P., {Leauthaud}, A., {et~al.} 2023, \mnras, 520, 5373, \dodoi{10.1093/mnras/stad473}

\bibitem[{{Lange} {et~al.}(2019){Lange}, {van den Bosch}, {Zentner}, {Wang}, \& {Villarreal}}]{lange_19_sat_kin}
{Lange}, J.~U., {van den Bosch}, F.~C., {Zentner}, A.~R., {Wang}, K., \& {Villarreal}, A.~S. 2019, \mnras, 482, 4824, \dodoi{10.1093/mnras/sty2950}

\bibitem[{Laureijs {et~al.}(2011)Laureijs, Amiaux, Arduini, Augu{\`e}res, Brinchmann, Cole, Cropper, Dabin, Duvet, Ealet, Garilli, Gondoin, Guzzo, Hoar, Hoekstra, Holmes, Kitching, Maciaszek, Mellier, Pasian, Percival, Rhodes, Saavedra~Criado, Sauvage, Scaramella, Valenziano, Warren, Bender, Castander, Cimatti, Le~F{\`e}vre, {Kurki-Suonio}, Levi, Lilje, Meylan, Nichol, Pedersen, Popa, Rebolo~Lopez, Rix, Rottgering, Zeilinger, Grupp, Hudelot, Massey, Meneghetti, Miller, Paltani, {Paulin-Henriksson}, Pires, Saxton, Schrabback, Seidel, Walsh, Aghanim, Amendola, Bartlett, Baccigalupi, Beaulieu, Benabed, Cuby, Elbaz, Fosalba, Gavazzi, Helmi, Hook, Irwin, Kneib, Kunz, Mannucci, Moscardini, Tao, Teyssier, Weller, Zamorani, Zapatero~Osorio, Boulade, Foumond, Di~Giorgio, Guttridge, James, Kemp, Martignac, Spencer, Walton, Bl{\"u}mchen, Bonoli, Bortoletto, Cerna, Corcione, Fabron, Jahnke, Ligori, Madrid, Martin, Morgante, Pamplona, Prieto, Riva, Toledo, Trifoglio, Zerbi, Abdalla, Douspis, Grenet, Borgani, Bouwens,
  Courbin, Delouis, Dubath, Fontana, Frailis, Grazian, Koppenh{\"o}fer, Mansutti, Melchior, Mignoli, Mohr, Neissner, Noddle, Poncet, Scodeggio, Serrano, Shane, Starck, Surace, Taylor, {Verdoes-Kleijn}, Vuerli, Williams, Zacchei, Altieri, Escudero~Sanz, Kohley, Oosterbroek, Astier, Bacon, Bardelli, Baugh, Bellagamba, Benoist, Bianchi, Biviano, Branchini, Carbone, Cardone, Clements, Colombi, Conselice, Cresci, Deacon, Dunlop, Fedeli, Fontanot, Franzetti, Giocoli, {Garcia-Bellido}, Gow, Heavens, Hewett, Heymans, Holland, Huang, Ilbert, Joachimi, Jennins, Kerins, Kiessling, Kirk, Kotak, Krause, Lahav, {van Leeuwen}, Lesgourgues, Lombardi, Magliocchetti, Maguire, Majerotto, Maoli, Marulli, Maurogordato, McCracken, McLure, Melchiorri, Merson, Moresco, Nonino, Norberg, Peacock, Pello, Penny, Pettorino, Di~Porto, Pozzetti, Quercellini, Radovich, Rassat, Roche, Ronayette, Rossetti, Sartoris, Schneider, Semboloni, Serjeant, Simpson, Skordis, Smadja, Smartt, Spano, Spiro, Sullivan, Tilquin, Trotta, Verde, Wang,
  Williger, Zhao, Zoubian, \& Zucca}]{laureijs2011}
Laureijs, R., Amiaux, J., Arduini, S., {et~al.} 2011, arXiv e-prints, arXiv:1110.3193

\bibitem[{{Lemos} {et~al.}(2023{\natexlab{a}}){Lemos}, {Coogan}, {Hezaveh}, \& {Perreault-Levasseur}}]{lemos_23_tarp}
{Lemos}, P., {Coogan}, A., {Hezaveh}, Y., \& {Perreault-Levasseur}, L. 2023{\natexlab{a}}, 40th International Conference on Machine Learning, 202, 19256, \dodoi{10.48550/arXiv.2302.03026}

\bibitem[{{Lemos} {et~al.}(2023{\natexlab{b}}){Lemos}, {Parker}, {Hahn}, {Ho}, {Eickenberg}, {Hou}, {Massara}, {Modi}, {Moradinezhad Dizgah}, {Regaldo-Saint Blancard}, \& {Spergel}}]{simbig_cnn}
{Lemos}, P., {Parker}, L., {Hahn}, C., {et~al.} 2023{\natexlab{b}}, arXiv e-prints, arXiv:2310.15256.
\newblock \doarXiv{2310.15256}

\bibitem[{{Lovell} {et~al.}(2023){Lovell}, {Hassan}, {Villaescusa-Navarro}, {Genel}, {Hahn}, {Angles-Alcazar}, {Kwon}, {de Santi}, {Iyer}, {Fabbian}, \& {Bryan}}]{lovell_23}
{Lovell}, C.~C., {Hassan}, S., {Villaescusa-Navarro}, F., {et~al.} 2023, in Machine Learning for Astrophysics, 21, \dodoi{10.48550/arXiv.2307.06967}

\bibitem[{{Mansfield} \& {Avestruz}(2021)}]{mansfield_21}
{Mansfield}, P., \& {Avestruz}, C. 2021, \mnras, 500, 3309, \dodoi{10.1093/mnras/staa3388}

\bibitem[{{Mao} {et~al.}(2015){Mao}, {Williamson}, \& {Wechsler}}]{mao2015_concentration}
{Mao}, Y.-Y., {Williamson}, M., \& {Wechsler}, R.~H. 2015, \apj, 810, 21, \dodoi{10.1088/0004-637X/810/1/21}

\bibitem[{{Masjedi} {et~al.}(2006){Masjedi}, {Hogg}, {Cool}, {Eisenstein}, {Blanton}, {Zehavi}, {Berlind}, {Bell}, {Schneider}, {Warren}, \& {Brinkmann}}]{masjedi_06}
{Masjedi}, M., {Hogg}, D.~W., {Cool}, R.~J., {et~al.} 2006, \apj, 644, 54, \dodoi{10.1086/503536}

\bibitem[{Massara {et~al.}(2021)Massara, Villaescusa-Navarro, Ho, Dalal, \& Spergel}]{massara_2021}
Massara, E., Villaescusa-Navarro, F., Ho, S., Dalal, N., \& Spergel, D.~N. 2021, Phys. Rev. Lett., 126, 011301, \dodoi{10.1103/PhysRevLett.126.011301}

\bibitem[{{Massara} {et~al.}(2023){Massara}, {Villaescusa-Navarro}, {Hahn}, {Abidi}, {Eickenberg}, {Ho}, {Lemos}, {Dizgah}, \& {Blancard}}]{massara_2023}
{Massara}, E., {Villaescusa-Navarro}, F., {Hahn}, C., {et~al.} 2023, \apj, 951, 70, \dodoi{10.3847/1538-4357/acd44d}

\bibitem[{{McDonough} \& {Brainerd}(2022)}]{McDonough_2022}
{McDonough}, B., \& {Brainerd}, T.~G. 2022, \apj, 933, 161, \dodoi{10.3847/1538-4357/ac752d}

\bibitem[{{Mihos} \& {Hernquist}(1994)}]{mihos_94}
{Mihos}, J.~C., \& {Hernquist}, L. 1994, \apjl, 425, L13, \dodoi{10.1086/187299}

\bibitem[{{Moore} {et~al.}(1996){Moore}, {Katz}, \& {Lake}}]{moore_96}
{Moore}, B., {Katz}, N., \& {Lake}, G. 1996, \apj, 457, 455, \dodoi{10.1086/176745}

\bibitem[{{More} {et~al.}(2011){More}, {van den Bosch}, {Cacciato}, {Skibba}, {Mo}, \& {Yang}}]{more_11}
{More}, S., {van den Bosch}, F.~C., {Cacciato}, M., {et~al.} 2011, \mnras, 410, 210, \dodoi{10.1111/j.1365-2966.2010.17436.x}

\bibitem[{{Nagai} \& {Kravtsov}(2005)}]{nagai_05}
{Nagai}, D., \& {Kravtsov}, A.~V. 2005, \apj, 618, 557, \dodoi{10.1086/426016}

\bibitem[{{Navarro} {et~al.}(1996){Navarro}, {Frenk}, \& {White}}]{nfw}
{Navarro}, J.~F., {Frenk}, C.~S., \& {White}, S. D.~M. 1996, \apj, 462, 563, \dodoi{10.1086/177173}

\bibitem[{Nelson {et~al.}(2019)Nelson, Springel, Pillepich, Rodriguez-Gomez, Torrey, Genel, Vogelsberger, Pakmor, Marinacci, Weinberger, Kelley, Lovell, Diemer, \& Hernquist}]{nelson_2019}
Nelson, D., Springel, V., Pillepich, A., {et~al.} 2019, Computational Astrophysics and Cosmology, 6, 2, \dodoi{10.1186/s40668-019-0028-x}

\bibitem[{{Paillas} {et~al.}(2023){Paillas}, {Cuesta-Lazaro}, {Percival}, {Nadathur}, {Cai}, {Yuan}, {Beutler}, {de Mattia}, {Eisenstein}, {Forero-Sanchez}, {Padilla}, {Pinon}, {Ruhlmann-Kleider}, {S{\'a}nchez}, {Valogiannis}, \& {Zarrouk}}]{paillas2023}
{Paillas}, E., {Cuesta-Lazaro}, C., {Percival}, W.~J., {et~al.} 2023, arXiv e-prints, arXiv:2309.16541, \dodoi{10.48550/arXiv.2309.16541}

\bibitem[{{Pakmor} {et~al.}(2023){Pakmor}, {Springel}, {Coles}, {Guillet}, {Pfrommer}, {Bose}, {Barrera}, {Delgado}, {Ferlito}, {Frenk}, {Hadzhiyska}, {Hern{\'a}ndez-Aguayo}, {Hernquist}, {Kannan}, \& {White}}]{pakmor_mtng_23}
{Pakmor}, R., {Springel}, V., {Coles}, J.~P., {et~al.} 2023, \mnras, 524, 2539, \dodoi{10.1093/mnras/stac3620}

\bibitem[{{Papamakarios} {et~al.}(2017){Papamakarios}, {Pavlakou}, \& {Murray}}]{papam_17}
{Papamakarios}, G., {Pavlakou}, T., \& {Murray}, I. 2017, arXiv e-prints, arXiv:1705.07057, \dodoi{10.48550/arXiv.1705.07057}

\bibitem[{{Paranjape} {et~al.}(2018){Paranjape}, {Hahn}, \& {Sheth}}]{paranjape_18}
{Paranjape}, A., {Hahn}, O., \& {Sheth}, R.~K. 2018, \mnras, 476, 3631, \dodoi{10.1093/mnras/sty496}

\bibitem[{{Pe{\~n}arrubia} {et~al.}(2010){Pe{\~n}arrubia}, {Benson}, {Walker}, {Gilmore}, {McConnachie}, \& {Mayer}}]{per_10}
{Pe{\~n}arrubia}, J., {Benson}, A.~J., {Walker}, M.~G., {et~al.} 2010, \mnras, 406, 1290, \dodoi{10.1111/j.1365-2966.2010.16762.x}

\bibitem[{{Peacock} \& {Smith}(2000{\natexlab{a}})}]{peacock_00}
{Peacock}, J.~A., \& {Smith}, R.~E. 2000{\natexlab{a}}, \mnras, 318, 1144, \dodoi{10.1046/j.1365-8711.2000.03779.x}

\bibitem[{{Peacock} \& {Smith}(2000{\natexlab{b}})}]{peacock_2002}
---. 2000{\natexlab{b}}, \mnras, 318, 1144, \dodoi{10.1046/j.1365-8711.2000.03779.x}

\bibitem[{{Peacock} {et~al.}(2001){Peacock}, {Cole}, {Norberg}, {Baugh}, {Bland-Hawthorn}, {Bridges}, {Cannon}, {Colless}, {Collins}, {Couch}, {Dalton}, {Deeley}, {De Propris}, {Driver}, {Efstathiou}, {Ellis}, {Frenk}, {Glazebrook}, {Jackson}, {Lahav}, {Lewis}, {Lumsden}, {Maddox}, {Percival}, {Peterson}, {Price}, {Sutherland}, \& {Taylor}}]{peacock_01}
{Peacock}, J.~A., {Cole}, S., {Norberg}, P., {et~al.} 2001, \nat, 410, 169, \dodoi{10.1038/35065528}

\bibitem[{{Peebles}(1969)}]{peebles_1969}
{Peebles}, P.~J.~E. 1969, \apj, 155, 393, \dodoi{10.1086/149876}

\bibitem[{{Percival} {et~al.}(2001){Percival}, {Baugh}, {Bland-Hawthorn}, {Bridges}, {Cannon}, {Cole}, {Colless}, {Collins}, {Couch}, {Dalton}, {De Propris}, {Driver}, {Efstathiou}, {Ellis}, {Frenk}, {Glazebrook}, {Jackson}, {Lahav}, {Lewis}, {Lumsden}, {Maddox}, {Moody}, {Norberg}, {Peacock}, {Peterson}, {Sutherland}, \& {Taylor}}]{percival_01}
{Percival}, W.~J., {Baugh}, C.~M., {Bland-Hawthorn}, J., {et~al.} 2001, \mnras, 327, 1297, \dodoi{10.1046/j.1365-8711.2001.04827.x}

\bibitem[{{Pillepich} {et~al.}(2018){Pillepich}, {Springel}, {Nelson}, {Genel}, {Naiman}, {Pakmor}, {Hernquist}, {Torrey}, {Vogelsberger}, {Weinberger}, \& {Marinacci}}]{tng_pillepich_18}
{Pillepich}, A., {Springel}, V., {Nelson}, D., {et~al.} 2018, \mnras, 473, 4077, \dodoi{10.1093/mnras/stx2656}

\bibitem[{{Planck Collaboration} {et~al.}(2020){Planck Collaboration}, {Aghanim}, {Akrami}, {Ashdown}, {Aumont}, {Baccigalupi}, {Ballardini}, {Banday}, {Barreiro}, {Bartolo}, {Basak}, {Battye}, {Benabed}, {Bernard}, {Bersanelli}, {Bielewicz}, {Bock}, {Bond}, {Borrill}, {Bouchet}, {Boulanger}, {Bucher}, {Burigana}, {Butler}, {Calabrese}, {Cardoso}, {Carron}, {Challinor}, {Chiang}, {Chluba}, {Colombo}, {Combet}, {Contreras}, {Crill}, {Cuttaia}, {de Bernardis}, {de Zotti}, {Delabrouille}, {Delouis}, {Di Valentino}, {Diego}, {Dor{\'e}}, {Douspis}, {Ducout}, {Dupac}, {Dusini}, {Efstathiou}, {Elsner}, {En{\ss}lin}, {Eriksen}, {Fantaye}, {Farhang}, {Fergusson}, {Fernandez-Cobos}, {Finelli}, {Forastieri}, {Frailis}, {Fraisse}, {Franceschi}, {Frolov}, {Galeotta}, {Galli}, {Ganga}, {G{\'e}nova-Santos}, {Gerbino}, {Ghosh}, {Gonz{\'a}lez-Nuevo}, {G{\'o}rski}, {Gratton}, {Gruppuso}, {Gudmundsson}, {Hamann}, {Handley}, {Hansen}, {Herranz}, {Hildebrandt}, {Hivon}, {Huang}, {Jaffe}, {Jones}, {Karakci}, {Keih{\"a}nen},
  {Keskitalo}, {Kiiveri}, {Kim}, {Kisner}, {Knox}, {Krachmalnicoff}, {Kunz}, {Kurki-Suonio}, {Lagache}, {Lamarre}, {Lasenby}, {Lattanzi}, {Lawrence}, {Le Jeune}, {Lemos}, {Lesgourgues}, {Levrier}, {Lewis}, {Liguori}, {Lilje}, {Lilley}, {Lindholm}, {L{\'o}pez-Caniego}, {Lubin}, {Ma}, {Mac{\'\i}as-P{\'e}rez}, {Maggio}, {Maino}, {Mandolesi}, {Mangilli}, {Marcos-Caballero}, {Maris}, {Martin}, {Martinelli}, {Mart{\'\i}nez-Gonz{\'a}lez}, {Matarrese}, {Mauri}, {McEwen}, {Meinhold}, {Melchiorri}, {Mennella}, {Migliaccio}, {Millea}, {Mitra}, {Miville-Desch{\^e}nes}, {Molinari}, {Montier}, {Morgante}, {Moss}, {Natoli}, {N{\o}rgaard-Nielsen}, {Pagano}, {Paoletti}, {Partridge}, {Patanchon}, {Peiris}, {Perrotta}, {Pettorino}, {Piacentini}, {Polastri}, {Polenta}, {Puget}, {Rachen}, {Reinecke}, {Remazeilles}, {Renzi}, {Rocha}, {Rosset}, {Roudier}, {Rubi{\~n}o-Mart{\'\i}n}, {Ruiz-Granados}, {Salvati}, {Sandri}, {Savelainen}, {Scott}, {Shellard}, {Sirignano}, {Sirri}, {Spencer}, {Sunyaev}, {Suur-Uski}, {Tauber}, {Tavagnacco},
  {Tenti}, {Toffolatti}, {Tomasi}, {Trombetti}, {Valenziano}, {Valiviita}, {Van Tent}, {Vibert}, {Vielva}, {Villa}, {Vittorio}, {Wandelt}, {Wehus}, {White}, {White}, {Zacchei}, \& {Zonca}}]{planck_2018}
{Planck Collaboration}, {Aghanim}, N., {Akrami}, Y., {et~al.} 2020, \aap, 641, A6, \dodoi{10.1051/0004-6361/201833910}

\bibitem[{{Porth} {et~al.}(2023){Porth}, {Bernstein}, {Smith}, \& {Lee}}]{porth_23}
{Porth}, L., {Bernstein}, G.~M., {Smith}, R.~E., \& {Lee}, A.~J. 2023, \mnras, 518, 3344, \dodoi{10.1093/mnras/stac3225}

\bibitem[{{Power} {et~al.}(2003){Power}, {Navarro}, {Jenkins}, {Frenk}, {White}, {Springel}, {Stadel}, \& {Quinn}}]{power_03}
{Power}, C., {Navarro}, J.~F., {Jenkins}, A., {et~al.} 2003, \mnras, 338, 14, \dodoi{10.1046/j.1365-8711.2003.05925.x}

\bibitem[{{Reid} {et~al.}(2014){Reid}, {Seo}, {Leauthaud}, {Tinker}, \& {White}}]{reid_14}
{Reid}, B.~A., {Seo}, H.-J., {Leauthaud}, A., {Tinker}, J.~L., \& {White}, M. 2014, \mnras, 444, 476, \dodoi{10.1093/mnras/stu1391}

\bibitem[{{Reid} {et~al.}(2012){Reid}, {Samushia}, {White}, {Percival}, {Manera}, {Padmanabhan}, {Ross}, {S{\'a}nchez}, {Bailey}, {Bizyaev}, {Bolton}, {Brewington}, {Brinkmann}, {Brownstein}, {Cuesta}, {Eisenstein}, {Gunn}, {Honscheid}, {Malanushenko}, {Malanushenko}, {Maraston}, {McBride}, {Muna}, {Nichol}, {Oravetz}, {Pan}, {de Putter}, {Roe}, {Ross}, {Schlegel}, {Schneider}, {Seo}, {Shelden}, {Sheldon}, {Simmons}, {Skibba}, {Snedden}, {Swanson}, {Thomas}, {Tinker}, {Tojeiro}, {Verde}, {Wake}, {Weaver}, {Weinberg}, {Zehavi}, \& {Zhao}}]{reid_12_boss_hod}
{Reid}, B.~A., {Samushia}, L., {White}, M., {et~al.} 2012, \mnras, 426, 2719, \dodoi{10.1111/j.1365-2966.2012.21779.x}

\bibitem[{{Rocha} {et~al.}(2012){Rocha}, {Peter}, \& {Bullock}}]{rocha_12}
{Rocha}, M., {Peter}, A. H.~G., \& {Bullock}, J. 2012, \mnras, 425, 231, \dodoi{10.1111/j.1365-2966.2012.21432.x}

\bibitem[{{Rodr{\'\i}guez-Puebla} {et~al.}(2016){Rodr{\'\i}guez-Puebla}, {Behroozi}, {Primack}, {Klypin}, {Lee}, \& {Hellinger}}]{puebla_16}
{Rodr{\'\i}guez-Puebla}, A., {Behroozi}, P., {Primack}, J., {et~al.} 2016, \mnras, 462, 893, \dodoi{10.1093/mnras/stw1705}

\bibitem[{{Ruhe} {et~al.}(2022){Ruhe}, {Wong}, {Cranmer}, \& {Forr{\'e}}}]{gw_prop}
{Ruhe}, D., {Wong}, K., {Cranmer}, M., \& {Forr{\'e}}, P. 2022, arXiv e-prints, arXiv:2211.09008, \dodoi{10.48550/arXiv.2211.09008}

\bibitem[{{Sales} \& {Lambas}(2004)}]{sales_04}
{Sales}, L., \& {Lambas}, D.~G. 2004, \mnras, 348, 1236, \dodoi{10.1111/j.1365-2966.2004.07443.x}

\bibitem[{{Schaye} {et~al.}(2015){Schaye}, {Crain}, {Bower}, {Furlong}, {Schaller}, {Theuns}, {Dalla Vecchia}, {Frenk}, {McCarthy}, {Helly}, {Jenkins}, {Rosas-Guevara}, {White}, {Baes}, {Booth}, {Camps}, {Navarro}, {Qu}, {Rahmati}, {Sawala}, {Thomas}, \& {Trayford}}]{eagle_15}
{Schaye}, J., {Crain}, R.~A., {Bower}, R.~G., {et~al.} 2015, \mnras, 446, 521, \dodoi{10.1093/mnras/stu2058}

\bibitem[{{Schneider} {et~al.}(2019){Schneider}, {Teyssier}, {Stadel}, {Chisari}, {Le Brun}, {Amara}, \& {Refregier}}]{schneider2019_baryons}
{Schneider}, A., {Teyssier}, R., {Stadel}, J., {et~al.} 2019, \jcap, 2019, 020, \dodoi{10.1088/1475-7516/2019/03/020}

\bibitem[{{Scoccimarro} {et~al.}(2001){Scoccimarro}, {Sheth}, {Hui}, \& {Jain}}]{scoccimarro_2001}
{Scoccimarro}, R., {Sheth}, R.~K., {Hui}, L., \& {Jain}, B. 2001, \apj, 546, 20, \dodoi{10.1086/318261}

\bibitem[{{Sefusatti} \& {Scoccimarro}(2005)}]{sefusatti2005}
{Sefusatti}, E., \& {Scoccimarro}, R. 2005, \prd, 71, 063001, \dodoi{10.1103/PhysRevD.71.063001}

\bibitem[{{Seljak}(2000)}]{seljak_2000}
{Seljak}, U. 2000, \mnras, 318, 203, \dodoi{10.1046/j.1365-8711.2000.03715.x}

\bibitem[{{Shen} {et~al.}(2006){Shen}, {Abel}, {Mo}, \& {Sheth}}]{shen_2006}
{Shen}, J., {Abel}, T., {Mo}, H.~J., \& {Sheth}, R.~K. 2006, \apj, 645, 783, \dodoi{10.1086/504513}

\bibitem[{{Sheth} \& {Tormen}(2004)}]{sheth_tormen_2004}
{Sheth}, R.~K., \& {Tormen}, G. 2004, \mnras, 350, 1385, \dodoi{10.1111/j.1365-2966.2004.07733.x}

\bibitem[{Spergel {et~al.}(2015)Spergel, Gehrels, Baltay, Bennett, Breckinridge, Donahue, Dressler, Gaudi, Greene, Guyon, Hirata, Kalirai, Kasdin, Macintosh, Moos, Perlmutter, Postman, Rauscher, Rhodes, Wang, Weinberg, Benford, Hudson, Jeong, Mellier, Traub, Yamada, Capak, Colbert, Masters, Penny, Savransky, Stern, Zimmerman, Barry, Bartusek, Carpenter, Cheng, Content, Dekens, Demers, Grady, Jackson, Kuan, Kruk, Melton, Nemati, Parvin, Poberezhskiy, Peddie, Ruffa, Wallace, Whipple, Wollack, \& Zhao}]{spergel2015}
Spergel, D., Gehrels, N., Baltay, C., {et~al.} 2015, Wide-{{Field InfrarRed Survey Telescope-Astrophysics Focused Telescope Assets WFIRST-AFTA}} 2015 {{Report}}

\bibitem[{{Springel} {et~al.}(2018){Springel}, {Pakmor}, {Pillepich}, {Weinberger}, {Nelson}, {Hernquist}, {Vogelsberger}, {Genel}, {Torrey}, {Marinacci}, \& {Naiman}}]{springel_18}
{Springel}, V., {Pakmor}, R., {Pillepich}, A., {et~al.} 2018, \mnras, 475, 676, \dodoi{10.1093/mnras/stx3304}

\bibitem[{{Steinmetz} \& {Bartelmann}(1995)}]{spin_dist}
{Steinmetz}, M., \& {Bartelmann}, M. 1995, \mnras, 272, 570, \dodoi{10.1093/mnras/272.3.570}

\bibitem[{{Storey-Fisher} {et~al.}(2022){Storey-Fisher}, {Tinker}, {Zhai}, {DeRose}, {Wechsler}, \& {Banerjee}}]{storey-fisher_23_aemulus_VI}
{Storey-Fisher}, K., {Tinker}, J., {Zhai}, Z., {et~al.} 2022, arXiv e-prints, arXiv:2210.03203, \dodoi{10.48550/arXiv.2210.03203}

\bibitem[{Tabak \& Turner(2013)}]{tabak_13}
Tabak, E.~G., \& Turner, C.~V. 2013, Communications on Pure and Applied Mathematics, 66, 145, \dodoi{https://doi.org/10.1002/cpa.21423}

\bibitem[{Tabak \& Vanden-Eijnden(2010)}]{tabak_vanden_10}
Tabak, E.~G., \& Vanden-Eijnden, E. 2010, Communications in Mathematical Sciences, 8, 217

\bibitem[{Takada {et~al.}(2014)Takada, Ellis, Chiba, Greene, Aihara, Arimoto, Bundy, Cohen, Dor{\'e}, Graves, Gunn, Heckman, Hirata, Ho, Kneib, Le~F{\`e}vre, Lin, More, Murayama, Nagao, Ouchi, Seiffert, Silverman, Sodr{\'e}, Spergel, Strauss, Sugai, Suto, Takami, \& Wyse}]{takada2014}
Takada, M., Ellis, R.~S., Chiba, M., {et~al.} 2014, Publications of the Astronomical Society of Japan, 66, R1, \dodoi{10.1093/pasj/pst019}

\bibitem[{{Tal} {et~al.}(2012){Tal}, {Wake}, \& {van Dokkum}}]{tal_2012}
{Tal}, T., {Wake}, D.~A., \& {van Dokkum}, P.~G. 2012, \apjl, 751, L5, \dodoi{10.1088/2041-8205/751/1/L5}

\bibitem[{Tegmark {et~al.}(2004)Tegmark, Strauss, Blanton, Abazajian, Dodelson, Sandvik, Wang, Weinberg, Zehavi, Bahcall, Hoyle, Schlegel, Scoccimarro, Vogeley, Berlind, Budavari, Connolly, Eisenstein, Finkbeiner, Frieman, Gunn, Hui, Jain, Johnston, Kent, Lin, Nakajima, Nichol, Ostriker, Pope, Scranton, Seljak, Sheth, Stebbins, Szalay, Szapudi, Xu, Annis, Brinkmann, Burles, Castander, Csabai, Loveday, Doi, Fukugita, Gillespie, Hennessy, Hogg, Ivezi\ifmmode~\acute{c}\else \'{c}\fi{}, Knapp, Lamb, Lee, Lupton, McKay, Kunszt, Munn, O'Connell, Peoples, Pier, Richmond, Rockosi, Schneider, Stoughton, Tucker, Vanden~Berk, Yanny, \& York}]{tegmark_04}
Tegmark, M., Strauss, M.~A., Blanton, M.~R., {et~al.} 2004, Phys. Rev. D, 69, 103501, \dodoi{10.1103/PhysRevD.69.103501}

\bibitem[{Tejero-Cantero {et~al.}(2020)Tejero-Cantero, Boelts, Deistler, Lueckmann, Durkan, Gonçalves, Greenberg, \& Macke}]{sbi_2020}
Tejero-Cantero, A., Boelts, J., Deistler, M., {et~al.} 2020, Journal of Open Source Software, 5, 2505, \dodoi{10.21105/joss.02505}

\bibitem[{{Teyssier, R.}(2002)}]{teyssier02_ramses}
{Teyssier, R.} 2002, A\&A, 385, 337, \dodoi{10.1051/0004-6361:20011817}

\bibitem[{{Thiele} {et~al.}(2023){Thiele}, {Massara}, {Pisani}, {Hahn}, {Spergel}, {Ho}, \& {Wandelt}}]{thiele2023}
{Thiele}, L., {Massara}, E., {Pisani}, A., {et~al.} 2023, arXiv e-prints, arXiv:2307.07555, \dodoi{10.48550/arXiv.2307.07555}

\bibitem[{{Tinker} {et~al.}(2012){Tinker}, {Sheldon}, {Wechsler}, {Becker}, {Rozo}, {Zu}, {Weinberg}, {Zehavi}, {Blanton}, {Busha}, \& {Koester}}]{tinker_12}
{Tinker}, J.~L., {Sheldon}, E.~S., {Wechsler}, R.~H., {et~al.} 2012, \apj, 745, 16, \dodoi{10.1088/0004-637X/745/1/16}

\bibitem[{{Valogiannis} \& {Dvorkin}(2022)}]{valogiannis_22_wst}
{Valogiannis}, G., \& {Dvorkin}, C. 2022, \prd, 105, 103534, \dodoi{10.1103/PhysRevD.105.103534}

\bibitem[{{Valogiannis} {et~al.}(2023){Valogiannis}, {Yuan}, \& {Dvorkin}}]{valogiannis2023_wst}
{Valogiannis}, G., {Yuan}, S., \& {Dvorkin}, C. 2023, arXiv e-prints, arXiv:2310.16116, \dodoi{10.48550/arXiv.2310.16116}

\bibitem[{{van Daalen} {et~al.}(2020){van Daalen}, {McCarthy}, \& {Schaye}}]{van_daalen_2020_gal_form_power_spectrum}
{van Daalen}, M.~P., {McCarthy}, I.~G., \& {Schaye}, J. 2020, \mnras, 491, 2424, \dodoi{10.1093/mnras/stz3199}

\bibitem[{{van Daalen} {et~al.}(2011){van Daalen}, {Schaye}, {Booth}, \& {Dalla Vecchia}}]{van_daalen_2011_agn_small}
{van Daalen}, M.~P., {Schaye}, J., {Booth}, C.~M., \& {Dalla Vecchia}, C. 2011, \mnras, 415, 3649, \dodoi{10.1111/j.1365-2966.2011.18981.x}

\bibitem[{{van den Bosch}(2017)}]{vdB_17}
{van den Bosch}, F.~C. 2017, \mnras, 468, 885, \dodoi{10.1093/mnras/stx520}

\bibitem[{{van den Bosch} {et~al.}(2013){van den Bosch}, {More}, {Cacciato}, {Mo}, \& {Yang}}]{vdB_13_small_scale}
{van den Bosch}, F.~C., {More}, S., {Cacciato}, M., {Mo}, H., \& {Yang}, X. 2013, \mnras, 430, 725, \dodoi{10.1093/mnras/sts006}

\bibitem[{{van den Bosch} \& {Ogiya}(2018)}]{vdB_18}
{van den Bosch}, F.~C., \& {Ogiya}, G. 2018, \mnras, 475, 4066, \dodoi{10.1093/mnras/sty084}

\bibitem[{{van den Bosch} {et~al.}(2018){van den Bosch}, {Ogiya}, {Hahn}, \& {Burkert}}]{vdB_18_2}
{van den Bosch}, F.~C., {Ogiya}, G., {Hahn}, O., \& {Burkert}, A. 2018, \mnras, 474, 3043, \dodoi{10.1093/mnras/stx2956}

\bibitem[{{van den Bosch} {et~al.}(2005){van den Bosch}, {Weinmann}, {Yang}, {Mo}, {Li}, \& {Jing}}]{vdb_05_ps}
{van den Bosch}, F.~C., {Weinmann}, S.~M., {Yang}, X., {et~al.} 2005, \mnras, 361, 1203, \dodoi{10.1111/j.1365-2966.2005.09260.x}

\bibitem[{{van Kampen}(1995)}]{van_kampen_95}
{van Kampen}, E. 1995, \mnras, 273, 295, \dodoi{10.1093/mnras/273.2.295}

\bibitem[{{van Kampen}(2000)}]{kampen_2000}
---. 2000, arXiv e-prints, astro, \dodoi{10.48550/arXiv.astro-ph/0002027}

\bibitem[{{van Kampen} {et~al.}(1999){van Kampen}, {Jimenez}, \& {Peacock}}]{vKampen_99}
{van Kampen}, E., {Jimenez}, R., \& {Peacock}, J.~A. 1999, \mnras, 310, 43, \dodoi{10.1046/j.1365-8711.1999.02955.x}

\bibitem[{{Verde} {et~al.}(2002){Verde}, {Heavens}, {Percival}, {Matarrese}, {Baugh}, {Bland-Hawthorn}, {Bridges}, {Cannon}, {Cole}, {Colless}, {Collins}, {Couch}, {Dalton}, {De Propris}, {Driver}, {Efstathiou}, {Ellis}, {Frenk}, {Glazebrook}, {Jackson}, {Lahav}, {Lewis}, {Lumsden}, {Maddox}, {Madgwick}, {Norberg}, {Peacock}, {Peterson}, {Sutherland}, \& {Taylor}}]{verde2002}
{Verde}, L., {Heavens}, A.~F., {Percival}, W.~J., {et~al.} 2002, \mnras, 335, 432, \dodoi{10.1046/j.1365-8711.2002.05620.x}

\bibitem[{{Villaescusa-Navarro} {et~al.}(2021){Villaescusa-Navarro}, {Angl{\'e}s-Alc{\'a}zar}, {Genel}, {Spergel}, {Somerville}, {Dave}, {Pillepich}, {Hernquist}, {Nelson}, {Torrey}, {Narayanan}, {Li}, {Philcox}, {La Torre}, {Maria Delgado}, {Ho}, {Hassan}, {Burkhart}, {Wadekar}, {Battaglia}, {Contardo}, \& {Bryan}}]{camels_2021}
{Villaescusa-Navarro}, F., {Angl{\'e}s-Alc{\'a}zar}, D., {Genel}, S., {et~al.} 2021, \apj, 915, 71, \dodoi{10.3847/1538-4357/abf7ba}

\bibitem[{{Villaescusa-Navarro} {et~al.}(2022){Villaescusa-Navarro}, {Genel}, {Angl{\'e}s-Alc{\'a}zar}, {Perez}, {Villanueva-Domingo}, {Wadekar}, {Shao}, {Mohammad}, {Hassan}, {Moser}, {Lau}, {Machado Poletti Valle}, {Nicola}, {Thiele}, {Jo}, {Philcox}, {Oppenheimer}, {Tillman}, {Hahn}, {Kaushal}, {Pisani}, {Gebhardt}, {Delgado}, {Caliendo}, {Kreisch}, {Wong}, {Coulton}, {Eickenberg}, {Parimbelli}, {Ni}, {Steinwandel}, {La Torre}, {Dave}, {Battaglia}, {Nagai}, {Spergel}, {Hernquist}, {Burkhart}, {Narayanan}, {Wandelt}, {Somerville}, {Bryan}, {Viel}, {Li}, {Irsic}, {Kraljic}, \& {Vogelsberger}}]{camels_2022}
{Villaescusa-Navarro}, F., {Genel}, S., {Angl{\'e}s-Alc{\'a}zar}, D., {et~al.} 2022, arXiv e-prints, arXiv:2201.01300.
\newblock \doarXiv{2201.01300}

\bibitem[{{Villanueva-Domingo} {et~al.}(2022){Villanueva-Domingo}, {Villaescusa-Navarro}, {Angl{\'e}s-Alc{\'a}zar}, {Genel}, {Marinacci}, {Spergel}, {Hernquist}, {Vogelsberger}, {Dave}, \& {Narayanan}}]{villanueva-domingo2022_gnn}
{Villanueva-Domingo}, P., {Villaescusa-Navarro}, F., {Angl{\'e}s-Alc{\'a}zar}, D., {et~al.} 2022, \apj, 935, 30, \dodoi{10.3847/1538-4357/ac7aa3}

\bibitem[{Wang {et~al.}(2022)Wang, Zhai, Alavi, Massara, Pisani, Benson, Hirata, Samushia, Weinberg, Colbert, Dor{\'e}, Eifler, Heinrich, Ho, Krause, Padmanabhan, Spergel, \& Teplitz}]{wang2022a}
Wang, Y., Zhai, Z., Alavi, A., {et~al.} 2022, The Astrophysical Journal, 928, 1, \dodoi{10.3847/1538-4357/ac4973}

\bibitem[{{Watson} {et~al.}(2012{\natexlab{a}}){Watson}, {Berlind}, {McBride}, {Hogg}, \& {Jiang}}]{watson_12}
{Watson}, D.~F., {Berlind}, A.~A., {McBride}, C.~K., {Hogg}, D.~W., \& {Jiang}, T. 2012{\natexlab{a}}, \apj, 749, 83, \dodoi{10.1088/0004-637X/749/1/83}

\bibitem[{{Watson} {et~al.}(2012{\natexlab{b}}){Watson}, {Berlind}, \& {Zentner}}]{watson_hod_mass_loss}
{Watson}, D.~F., {Berlind}, A.~A., \& {Zentner}, A.~R. 2012{\natexlab{b}}, \apj, 754, 90, \dodoi{10.1088/0004-637X/754/2/90}

\bibitem[{{Wechsler} \& {Tinker}(2018)}]{wechsler_tinker_18}
{Wechsler}, R.~H., \& {Tinker}, J.~L. 2018, \araa, 56, 435, \dodoi{10.1146/annurev-astro-081817-051756}

\bibitem[{{Weinberger} {et~al.}(2017){Weinberger}, {Springel}, {Hernquist}, {Pillepich}, {Marinacci}, {Pakmor}, {Nelson}, {Genel}, {Vogelsberger}, {Naiman}, \& {Torrey}}]{tng_weinberger_17}
{Weinberger}, R., {Springel}, V., {Hernquist}, L., {et~al.} 2017, \mnras, 465, 3291, \dodoi{10.1093/mnras/stw2944}

\bibitem[{{White} {et~al.}(2007){White}, {Zheng}, {Brown}, {Dey}, \& {Jannuzi}}]{white_07}
{White}, M., {Zheng}, Z., {Brown}, M. J.~I., {Dey}, A., \& {Jannuzi}, B.~T. 2007, \apjl, 655, L69, \dodoi{10.1086/512015}

\bibitem[{{White} {et~al.}(2011){White}, {Blanton}, {Bolton}, {Schlegel}, {Tinker}, {Berlind}, {da Costa}, {Kazin}, {Lin}, {Maia}, {McBride}, {Padmanabhan}, {Parejko}, {Percival}, {Prada}, {Ramos}, {Sheldon}, {de Simoni}, {Skibba}, {Thomas}, {Wake}, {Zehavi}, {Zheng}, {Nichol}, {Schneider}, {Strauss}, {Weaver}, \& {Weinberg}}]{white_11_hod_clustering}
{White}, M., {Blanton}, M., {Bolton}, A., {et~al.} 2011, \apj, 728, 126, \dodoi{10.1088/0004-637X/728/2/126}

\bibitem[{{Wright} {et~al.}(2022){Wright}, {Lagos}, {Power}, {Stevens}, {Cortese}, \& {Poulton}}]{wright_22}
{Wright}, R.~J., {Lagos}, C. d.~P., {Power}, C., {et~al.} 2022, \mnras, 516, 2891, \dodoi{10.1093/mnras/stac2042}

\bibitem[{{Wu} \& {Kragh Jespersen}(2023)}]{gnn_wu_23}
{Wu}, J.~F., \& {Kragh Jespersen}, C. 2023, arXiv e-prints, arXiv:2306.12327, \dodoi{10.48550/arXiv.2306.12327}

\bibitem[{{Xu} {et~al.}(2021){Xu}, {Zehavi}, \& {Contreras}}]{xu_2021}
{Xu}, X., {Zehavi}, I., \& {Contreras}, S. 2021, \mnras, 502, 3242, \dodoi{10.1093/mnras/stab100}

\bibitem[{{Yang} {et~al.}(2003){Yang}, {Mo}, \& {van den Bosch}}]{yang_2003_clf}
{Yang}, X., {Mo}, H.~J., \& {van den Bosch}, F.~C. 2003, \mnras, 339, 1057, \dodoi{10.1046/j.1365-8711.2003.06254.x}

\bibitem[{{York} {et~al.}(2000){York}, {Adelman}, {Anderson}, {Anderson}, {Annis}, {Bahcall}, {Bakken}, {Barkhouser}, {Bastian}, {Berman}, {Boroski}, {Bracker}, {Briegel}, {Briggs}, {Brinkmann}, {Brunner}, {Burles}, {Carey}, {Carr}, {Castander}, {Chen}, {Colestock}, {Connolly}, {Crocker}, {Csabai}, {Czarapata}, {Davis}, {Doi}, {Dombeck}, {Eisenstein}, {Ellman}, {Elms}, {Evans}, {Fan}, {Federwitz}, {Fiscelli}, {Friedman}, {Frieman}, {Fukugita}, {Gillespie}, {Gunn}, {Gurbani}, {de Haas}, {Haldeman}, {Harris}, {Hayes}, {Heckman}, {Hennessy}, {Hindsley}, {Holm}, {Holmgren}, {Huang}, {Hull}, {Husby}, {Ichikawa}, {Ichikawa}, {Ivezi{\'c}}, {Kent}, {Kim}, {Kinney}, {Klaene}, {Kleinman}, {Kleinman}, {Knapp}, {Korienek}, {Kron}, {Kunszt}, {Lamb}, {Lee}, {Leger}, {Limmongkol}, {Lindenmeyer}, {Long}, {Loomis}, {Loveday}, {Lucinio}, {Lupton}, {MacKinnon}, {Mannery}, {Mantsch}, {Margon}, {McGehee}, {McKay}, {Meiksin}, {Merelli}, {Monet}, {Munn}, {Narayanan}, {Nash}, {Neilsen}, {Neswold}, {Newberg}, {Nichol}, {Nicinski},
  {Nonino}, {Okada}, {Okamura}, {Ostriker}, {Owen}, {Pauls}, {Peoples}, {Peterson}, {Petravick}, {Pier}, {Pope}, {Pordes}, {Prosapio}, {Rechenmacher}, {Quinn}, {Richards}, {Richmond}, {Rivetta}, {Rockosi}, {Ruthmansdorfer}, {Sandford}, {Schlegel}, {Schneider}, {Sekiguchi}, {Sergey}, {Shimasaku}, {Siegmund}, {Smee}, {Smith}, {Snedden}, {Stone}, {Stoughton}, {Strauss}, {Stubbs}, {SubbaRao}, {Szalay}, {Szapudi}, {Szokoly}, {Thakar}, {Tremonti}, {Tucker}, {Uomoto}, {Vanden Berk}, {Vogeley}, {Waddell}, {Wang}, {Watanabe}, {Weinberg}, {Yanny}, {Yasuda}, \& {SDSS Collaboration}}]{sdss}
{York}, D.~G., {Adelman}, J., {Anderson}, John~E., J., {et~al.} 2000, \aj, 120, 1579, \dodoi{10.1086/301513}

\bibitem[{{Yuan} {et~al.}(2022{\natexlab{a}}){Yuan}, {Garrison}, {Eisenstein}, \& {Wechsler}}]{yuan_22_fs8}
{Yuan}, S., {Garrison}, L.~H., {Eisenstein}, D.~J., \& {Wechsler}, R.~H. 2022{\natexlab{a}}, \mnras, 515, 871, \dodoi{10.1093/mnras/stac1830}

\bibitem[{{Yuan} {et~al.}(2022{\natexlab{b}}){Yuan}, {Garrison}, {Eisenstein}, \& {Wechsler}}]{yuan2022}
---. 2022{\natexlab{b}}, \mnras, 515, 871, \dodoi{10.1093/mnras/stac1830}

\bibitem[{{Yuan} {et~al.}(2022{\natexlab{c}}){Yuan}, {Hadzhiyska}, {Bose}, \& {Eisenstein}}]{yuan_22_lrg_elg}
{Yuan}, S., {Hadzhiyska}, B., {Bose}, S., \& {Eisenstein}, D.~J. 2022{\natexlab{c}}, \mnras, 512, 5793, \dodoi{10.1093/mnras/stac830}

\bibitem[{{Zehavi} {et~al.}(2018){Zehavi}, {Contreras}, {Padilla}, {Smith}, {Baugh}, \& {Norberg}}]{zehavi_18}
{Zehavi}, I., {Contreras}, S., {Padilla}, N., {et~al.} 2018, \apj, 853, 84, \dodoi{10.3847/1538-4357/aaa54a}

\bibitem[{{Zehavi} {et~al.}(2011){Zehavi}, {Zheng}, {Weinberg}, {Blanton}, {Bahcall}, {Berlind}, {Brinkmann}, {Frieman}, {Gunn}, {Lupton}, {Nichol}, {Percival}, {Schneider}, {Skibba}, {Strauss}, {Tegmark}, \& {York}}]{zehavi_11}
{Zehavi}, I., {Zheng}, Z., {Weinberg}, D.~H., {et~al.} 2011, \apj, 736, 59, \dodoi{10.1088/0004-637X/736/1/59}

\bibitem[{{Zentner} {et~al.}(2014){Zentner}, {Hearin}, \& {van den Bosch}}]{zentner_14}
{Zentner}, A.~R., {Hearin}, A.~P., \& {van den Bosch}, F.~C. 2014, \mnras, 443, 3044, \dodoi{10.1093/mnras/stu1383}

\bibitem[{{Zentner} {et~al.}(2005){Zentner}, {Kravtsov}, {Gnedin}, \& {Klypin}}]{zentner_05}
{Zentner}, A.~R., {Kravtsov}, A.~V., {Gnedin}, O.~Y., \& {Klypin}, A.~A. 2005, \apj, 629, 219, \dodoi{10.1086/431355}

\bibitem[{{Zhai} {et~al.}(2019){Zhai}, {Tinker}, {Becker}, {DeRose}, {Mao}, {McClintock}, {McLaughlin}, {Rozo}, \& {Wechsler}}]{zhai2019_aemulus}
{Zhai}, Z., {Tinker}, J.~L., {Becker}, M.~R., {et~al.} 2019, \apj, 874, 95, \dodoi{10.3847/1538-4357/ab0d7b}

\bibitem[{{Zhai} {et~al.}(2023){Zhai}, {Tinker}, {Banerjee}, {DeRose}, {Guo}, {Mao}, {McLaughlin}, {Storey-Fisher}, \& {Wechsler}}]{zhai_23_aemulus}
{Zhai}, Z., {Tinker}, J.~L., {Banerjee}, A., {et~al.} 2023, \apj, 948, 99, \dodoi{10.3847/1538-4357/acc65b}

\bibitem[{{Zhang} {et~al.}(2019){Zhang}, {Liao}, {Li}, \& {Gao}}]{zhang_19_opt}
{Zhang}, T., {Liao}, S., {Li}, M., \& {Gao}, L. 2019, \mnras, 487, 1227, \dodoi{10.1093/mnras/stz1370}

\bibitem[{{Zheng} {et~al.}(2007){Zheng}, {Coil}, \& {Zehavi}}]{zheng+_2007}
{Zheng}, Z., {Coil}, A.~L., \& {Zehavi}, I. 2007, \apj, 667, 760, \dodoi{10.1086/521074}

\bibitem[{{Zheng} \& {Weinberg}(2007)}]{zheng_07_degeneracy}
{Zheng}, Z., \& {Weinberg}, D.~H. 2007, \apj, 659, 1, \dodoi{10.1086/512151}

\bibitem[{{Zheng} {et~al.}(2009){Zheng}, {Zehavi}, {Eisenstein}, {Weinberg}, \& {Jing}}]{lrg_09}
{Zheng}, Z., {Zehavi}, I., {Eisenstein}, D.~J., {Weinberg}, D.~H., \& {Jing}, Y.~P. 2009, \apj, 707, 554, \dodoi{10.1088/0004-637X/707/1/554}

\bibitem[{{Zheng} {et~al.}(2005){Zheng}, {Berlind}, {Weinberg}, {Benson}, {Baugh}, {Cole}, {Dav{\'e}}, {Frenk}, {Katz}, \& {Lacey}}]{zheng_2005}
{Zheng}, Z., {Berlind}, A.~A., {Weinberg}, D.~H., {et~al.} 2005, \apj, 633, 791, \dodoi{10.1086/466510}

\bibitem[{{Zhou} {et~al.}(2023){Zhou}, {Dey}, {Newman}, {Eisenstein}, {Dawson}, {Bailey}, {Berti}, {Guy}, {Lan}, {Zou}, {Aguilar}, {Ahlen}, {Alam}, {Brooks}, {de la Macorra}, {Dey}, {Dhungana}, {Fanning}, {Font-Ribera}, {Gontcho}, {Honscheid}, {Ishak}, {Kisner}, {Kov{\'a}cs}, {Kremin}, {Landriau}, {Levi}, {Magneville}, {Manera}, {Martini}, {Meisner}, {Miquel}, {Moustakas}, {Myers}, {Nie}, {Palanque-Delabrouille}, {Percival}, {Poppett}, {Prada}, {Raichoor}, {Ross}, {Schlafly}, {Schlegel}, {Schubnell}, {Tarl{\'e}}, {Weaver}, {Wechsler}, {Y{\'e}che}, \& {Zhou}}]{desi_lrg_23}
{Zhou}, R., {Dey}, B., {Newman}, J.~A., {et~al.} 2023, \aj, 165, 58, \dodoi{10.3847/1538-3881/aca5fb}

\bibitem[{{Zitrin} {et~al.}(2012){Zitrin}, {Bartelmann}, {Umetsu}, {Oguri}, \& {Broadhurst}}]{zitrin_12}
{Zitrin}, A., {Bartelmann}, M., {Umetsu}, K., {Oguri}, M., \& {Broadhurst}, T. 2012, \mnras, 426, 2944, \dodoi{10.1111/j.1365-2966.2012.21886.x}

\bibitem[{{Zu} \& {Weinberg}(2013)}]{zw_13}
{Zu}, Y., \& {Weinberg}, D.~H. 2013, \mnras, 431, 3319, \dodoi{10.1093/mnras/stt411}

\bibitem[{{Zu} {et~al.}(2014){Zu}, {Weinberg}, {Jennings}, {Li}, \& {Wyman}}]{zw_14}
{Zu}, Y., {Weinberg}, D.~H., {Jennings}, E., {Li}, B., \& {Wyman}, M. 2014, \mnras, 445, 1885, \dodoi{10.1093/mnras/stu1739}

\end{thebibliography}
\bibliographystyle{aasjournal}

\end{document}